\documentclass[allclo,onecollarge,square, comma, sort&compress,natbib]{svjour2}

\usepackage{amsmath}
\usepackage{amsfonts}
\usepackage{amssymb}
\usepackage{bm}

\usepackage{graphicx}

\usepackage{dcolumn}
\usepackage{color}
\usepackage{soul}

\definecolor{purple}{rgb}{0.5,0,0.5}
\definecolor{blue}{rgb}{0.0,0,0.9}

\usepackage[mathscr,scaled=1.15]{urwchancal}
\DeclareFontFamily{OT1}{pzc}{}
\DeclareFontShape{OT1}{pzc}{m}{it}%
{<-> s * [1.15] pzcmi7t}{}
\DeclareMathAlphabet{\mathpzc}{OT1}{pzc}{m}{it}

\newcommand{\sfrac}[2]{\mbox{\footnotesize $\displaystyle \frac{#1}{#2}$}}

\newcommand{\lsim}{\mathrel{\rlap{\lower4pt\hbox{\hskip0pt$\sim$}}
\raise1pt\hbox{$<$}}}           
\newcommand{\gsim}{\mathrel{\rlap{\lower4pt\hbox{\hskip0pt$\sim$}}
\raise1pt\hbox{$>$}}}           

\bibpunct{[}{]}{,}{n}{,}{,} 

\begin{document}

\title{Nucleon and \mbox{\boldmath $\Delta$} elastic and transition form factors}

\authorrunning{Jorge Segovia \emph{et al}.}
\titlerunning{Nucleon and \mbox{\boldmath $\Delta$} elastic and transition form factors}

\author{Jorge Segovia
\and
        Ian C.~Clo\"et
\and
        Craig~D.~Roberts
\and
    Sebastian~M.~Schmidt
}

\institute{
Jorge Segovia \and Ian C.~Clo\"et \and Craig D.~Roberts
\at
Physics Division, Argonne National Laboratory, Argonne
Illinois 60439, USA
\and
Sebastian M. Schmidt
\at
Institute for Advanced Simulation, Forschungszentrum J\"ulich and JARA, D-52425 J\"ulich, Germany
}

\date{3 September 2014}

\maketitle

\begin{abstract}
We present a unified study of nucleon and $\Delta$ elastic and transition form factors, and compare predictions made using a framework built upon a Faddeev equation kernel and interaction vertices that possess QCD-like momentum dependence with results obtained using a symmetry-preserving treatment of a vector$\,\otimes\,$vector contact-interaction.  The comparison emphasises that experiments are sensitive to the momentum dependence of the running couplings and masses in the strong interaction sector of the Standard Model and highlights that the key to describing hadron properties is a veracious expression of dynamical chiral symmetry breaking in the bound-state problem. Amongst the results we describe, the following are of particular interest:
$G_E^p(Q^2)/G_M^p(Q^2)$ possesses a zero at $Q^2=9.5\,$GeV$^2$; any change in the interaction which shifts a zero in the proton ratio to larger $Q^2$ relocates a zero in $G_E^n(Q^2)/G_M^n(Q^2)$ to smaller $Q^2$; there is likely a value of momentum transfer above which $G_E^n>G_E^p$; and the presence of strong diquark correlations within the nucleon is sufficient to understand empirical extractions of the flavour-separated form factors.
Regarding the $\Delta(1232)$-baryon, we find that, \emph{inter alia}: the electric monopole form factor exhibits a zero; the electric quadrupole form factor is negative, large in magnitude, and sensitive to the nature and strength of correlations in the $\Delta(1232)$ Faddeev amplitude; and the magnetic octupole form factor is negative so long as rest-frame $P$- and $D$-wave correlations are included.
In connection with the $N \to \Delta$ transition, the momentum-dependence of the magnetic transition form factor, $G_M^\ast$, matches that of $G_M^n$ once the momentum transfer is high enough to pierce the meson-cloud; and the electric quadrupole ratio is a keen measure of diquark and orbital angular momentum correlations, the zero in which is obscured by meson-cloud effects on the domain currently accessible to experiment.
Importantly, within each framework, identical propagators and vertices are sufficient to describe all properties discussed herein.
Our analysis and predictions should therefore serve as motivation for measurement of elastic and transition form factors involving the nucleon and its resonances at high photon virtualities using modern electron-beam facilities.
%
%
\end{abstract}


\maketitle

\section{Introduction}
\label{sec:Introduction}
Nonperturbative quantum chromodynamics (QCD) poses significant challenges.  Primary amongst them is a need to chart the behaviour of QCD's running coupling and masses into the domain of infrared momenta.  Contemporary theory is incapable of solving this problem alone but a collaboration with experiment holds a promise for progress.  This effort can benefit substantially by exposing the structure of nucleon excited states and measuring the associated transition form factors at large momentum transfers \cite{Aznauryan:2012baS}.  Large momenta are needed in order to pierce the meson-cloud that, often to a significant extent, screens the dressed-quark core of all baryons \cite{Suzuki:2009nj,Roberts:2011rr,Kamano:2013iva}; and it is via the $Q^2$ evolution of form factors that one gains access to the running of QCD's coupling and masses from the infrared into the ultraviolet \cite{Cloet:2013gva,Chang:2013nia}.

It is within the context just described that the present study is undertaken but it may also be viewed as a reevaluation and significant extension of Ref.\,\cite{Cloet:2008re}, which surveyed nucleon electromagnetic form factors.  Herein we employ the same approach in a simultaneous treatment of elastic and transition form factors involving the nucleon and/or $\Delta(1232)$-baryon; and it will be seen that a unified description can successfully be achieved with just three minor modifications of the elements in Ref.\,\cite{Cloet:2008re}.  Importantly, identical propagators and vertices are sufficient to describe all properties discussed herein.
A unified QCD-based description of elastic and transition form factors involving the nucleon and its resonances has acquired additional significance owing to substantial progress in the extraction of transition electrocouplings, $g_{{\rm v}NN^\ast}$, from meson electroproduction data, obtained primarily with the CLAS detector at the Thomas Jefferson National Accelerator Facility (JLab).  The electrocouplings of all low-lying $N^\ast$ states with mass less-than $1.6\,$GeV have been determined via independent analyses of $\pi^+ n$, $\pi^0p$ and $\pi^+ \pi^- p$ exclusive channels \cite{Beringer:1900zz,Mokeev:2012vsa}; and preliminary results for the $g_{{\rm v}NN^\ast}$ electrocouplings of most high-lying $N^\ast$ states with masses below $1.8\,$GeV have also been obtained from CLAS meson electroproduction data \cite{Aznauryan:2012baS,Mokeev:2013kka}.

In order to address the issue of charting the behaviour of the running coupling and masses in the strong interaction sector of the Standard Model, throughout this work we make comparisons between the predictions of the QCD-based framework in Ref.\,\cite{Cloet:2008re} and results obtained using a confining, symmetry-preserving treatment of a vector$\,\otimes\,$vector contact interaction in a widely-used leading-order (rainbow-ladder) truncation of QCD's Dyson-Schwinger equations (DSEs) \cite{Chang:2011vu,Bashir:2012fs,Cloet:2013jya}.  This is a pertinent comparison because the contact-interaction (CI) framework has judiciously been applied to a large body of hadron phenomena \cite{GutierrezGuerrero:2010md,%
Roberts:2010rn,Roberts:2011cf,Roberts:2011wy,Roberts:2011rr,%
Wilson:2011aa,Chen:2012qr,Chen:2012txa,Segovia:2013rca,Segovia:2013uga} and by contrasting the results obtained for the same observables one can expose those quantities which are most sensitive to the momentum dependence of elementary quantities in QCD.
Of particular relevance herein is a comparison with Refs.\,\cite{Wilson:2011aa,Segovia:2013rca,Segovia:2013uga}, which together provide a complete body of CI results for the quantities we study.  Those analyses explain that the CI framework produces hard form factors, curtails some quark orbital angular momentum correlations within a baryon, and suppresses two-loop diagrams in the elastic and transition electromagnetic currents, defects that are rectified in our QCD-based approach.

This manuscript is arranged as follows.  In Sect.\,\ref{sec:BaryonModel} we present a short survey of our framework, both the Faddeev equation treatment of the baryon dressed-quark cores and the currents that describe the interaction of a photon with a baryon composed from such consistently dressed constituents.  Additional material is expressed in appendices.
The relationship between the Faddeev amplitudes and currents and the elastic and transition form factors is detailed in Sect.\,\ref{sec:BEMCurrents}; and we discuss nucleon elastic form factors in Sect.\,\ref{sec:BEMFFs}, confirming the predictions of Ref.\,\cite{Cloet:2008re} and elucidating novel results for a range of quantities including the neutron's electric form factor.

The $\Delta(1232)$-baryon's elastic form factors are canvassed in Sect.\,\ref{subsec:FFdelta}; but some background material might be useful here in case it should appear odd that we have spent effort on computing a wide range of quantities that are unlikely ever to be measured because $\Delta$-baryons are unstable.  Consider, therefore, that $\Delta(1232)$-baryons are the lightest baryon resonances, with a mass just 30\% heavier than the nucleon; and despite possessing a width of $0.12\,$GeV, the $\Delta$-resonances are well isolated from other nucleon excitations.  Much of the interest in $\Delta$-resonances originates in the fact that, at the simplest level, the $\Delta^{+}$ and $\Delta^{0}$ can respectively be viewed as spin- and isospin-flip excitations of the proton and neutron.\footnote{As we shall see, however, this apparently elementary connection obscures a deeper truth; namely, the structure of the $\Delta$-baryon's dressed-quark-core is actually far simpler than that of the nucleons.}  In addition, the strong $\Delta \to \pi N$ coupling entails that the $\Delta(1232)$-resonance is an important platform for developing and honing an understanding of the role a meson cloud plays in baryon physics \cite{Sato:2000jf,JuliaDiaz:2006xt} so that this may be separated from effects more properly attributable to a baryon's dressed-quark core \cite{Alkofer:2004yf,Eichmann:2008ae,Eichmann:2008ef,Cloet:2008wg,Cloet:2008re}.

Since the $\Delta(1232)$ is a $J=\frac{3}{2}$ state, a complete description of its electromagnetic structure requires four form factors \cite{Scadron:1968zz}: electric charge, $G_{E0}$; magnetic dipole $G_{M1}$; electric quadrupole, $G_{E2}$; and magnetic octupole $G_{M3}$.  The first two listed here are, respectively, the analogues of those form factors which describe the momentum-space distribution of the nucleon's charge and magnetisation.  The remaining two may be associated with shape deformation of the $\Delta$-baryon because they are identically zero within any quark-model framework in which $SU(6)$ spin-flavour symmetry is only broken by electromagnetism and the associated current transforms according to the adjoint representation of the symmetry group \cite{Beg:1964nm,Buchmann:1996bd}.  At zero momentum transfer the form factors can be used to define dimensionless $\Delta$-baryon multipole moments: electric charge, $e_{\Delta}=G_{E0}(0)$; magnetic moment $\hat\mu_{\Delta}=G_{M1}(0)$; electric quadrupole moment $\hat{\cal Q}_{\Delta}=G_{E2}(0)$; and magnetic octupole moment $\hat{\cal O}_{\Delta}=G_{M2}(0)$.

Whilst it is relatively straightforward to extend a theoretical framework applicable to the nucleon so that it may be employed to describe the $\Delta(1232)$, the state thus obtained is commonly stable; i.e., one obtains the zero-width dressed-quark-core of the $\Delta$.  Since the width-to-mass ratio is small, this is a reasonable approximation, when interpreted judiciously, just as it is for the $\rho$-meson \cite{Pichowsky:1999mu,Maris:1999nt,Jarecke:2002xd}.  Empirically, on the other hand, one must deal with the very short $\Delta$-lifetime: $\tau_\Delta \sim 10^{-16} \tau_{\pi^+}$, and therefore little is experimentally known about the electromagnetic properties of $\Delta(1232)$-baryons.  Information has been obtained through analysis of the $\pi p \to \pi p \gamma$ and $\gamma p \to p \pi^0 \gamma^\prime$ reactions, so that Ref.\,\cite{Beringer:1900zz} reports $\mu_{\Delta^{++}} = 3.7\,$--$\,7.5\,\mu_N$ and
$\mu_{\Delta^{+}}=2.7^{+1.0}_{-1.3}{\rm (stat)}\pm1.5{\rm (syst)}\pm3.0{\rm (theor)}\,\mu_{N}$, where $\mu_N$ is the nuclear magneton.

Notwithstanding these features, in continuum QCD one computes $\Delta$-baryon elastic form factors largely because their $Q^2=0$ values are required in order to normalise the $\Delta$-baryon's Faddeev amplitude.  That normalisation is necessary if one wishes to analyse the $\gamma^\ast N \to \Delta$ transition, which is empirically accessible, owing to the appearance in recent times of intense, energetic electron-beam facilities: data are now available on $0 \leq Q^2 \lesssim 8\,$GeV$^2$ \cite{Aznauryan:2011ub,Aznauryan:2011qj}.  This transition is described by three form factors \cite{Jones:1972ky}: magnetic-dipole, $G_{M}^{\ast}$; electric quadrupole, $G_{E}^{\ast}$; and Coulomb (longitudinal) quadrupole, $G_{C}^{\ast}$.  Our analysis of these form factors is described in Sect.\,\ref{subsec:FFnucdel}.

We provide a summary and perspective in Sect.\,\ref{sec:summary}.


\section{Faddeev Equations for the Nucleon and \mbox{\boldmath $\Delta$}}
\label{sec:BaryonModel}
In quantum field theory a baryon appears as a pole in a six-point quark Green
function.  The pole's residue is proportional to the baryon's Faddeev amplitude, which is obtained from a Poincar\'e covariant Faddeev equation that sums all possible quantum field theoretical exchanges and interactions that can take place between three dressed-quarks.  Canonical normalisation of the Faddeev amplitude guarantees unit residue for the $s$-channel poles in the $J^P = \frac{1}{2}^+$, $\frac{3}{2}^+$ three-quark vacuum polarisation diagrams associated with the nucleon and $\Delta$, respectively, and entails the appropriate electric charge.

\begin{figure}[t!]
\begin{center}
\includegraphics[clip,width=0.66\textwidth]{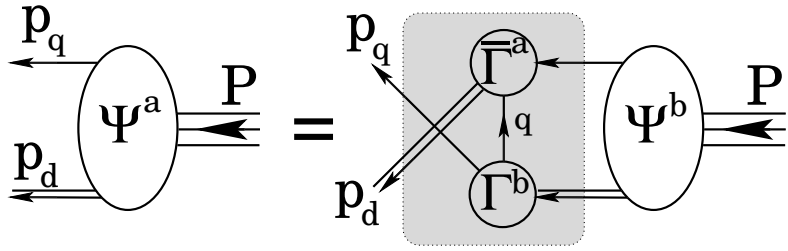}
\caption{\label{faddeevfigure} Poincar\'e covariant Faddeev equation,
Eq.\,(\protect\ref{FEone}), employed herein to calculate nucleon and $\Delta$ properties. $\Psi$ in Eq.\,(\protect\ref{PsiNucleon}) is the amplitude for a baryon of total momentum $P= p_q + p_d$.  It expresses the relative momentum correlation between the dressed-quark and -diquarks within the nucleon.  The shaded region demarcates the kernel of the Faddeev equation, Sec.\,\protect\ref{completing}, in which: the \emph{single line} denotes the dressed-quark propagator, Sec.\,\protect\ref{subsubsec:S}; $\Gamma$ is the diquark Bethe-Salpeter-like amplitude, Sec.\,\protect\ref{qqBSA}; and the \emph{double line} is the diquark propagator, Sec.\,\protect\ref{qqprop}.}
\end{center}
\end{figure}

A tractable truncation of the Faddeev equation is based~\cite{Cahill:1988dx} on the observation that an interaction which describes mesons also generates diquark correlations in the colour-$\bar 3$ channel~\cite{Cahill:1987qr}.  The dominant correlations for ground state octet and decuplet baryons are scalar ($0^+$) and axial-vector ($1^+$) diquarks because, for example, they have the correct parity and the associated mass-scales are smaller than the baryons' masses \cite{Chen:2012qr,Burden:1996nh,Maris:2002yu}: for systems constituted from $u$,$d$-quarks (in GeV)
\begin{equation}
\label{diquarkmass}
m_{[ud]_{0^+}} = 0.7 - 0.8
 \,,\;
m_{(uu)_{1^+}}=m_{(ud)_{1^+}}=m_{(dd)_{1^+}}=0.9 - 1.1\,.
\end{equation}
The kernel of the Faddeev equation is completed by specifying that the quarks are dressed, with two of the three dressed-quarks correlated always as a colour-$\bar 3$ diquark.  As illustrated in Fig~\ref{faddeevfigure}, binding is then effected by the iterated exchange of roles between the bystander and diquark-participant quarks.

We employ the Faddeev equations used in Refs.\,\cite{Alkofer:2004yf,Cloet:2008wg,Cloet:2008re}, which are reproduced in App.\,\ref{app:FE} for completeness.  With all the elements specified as described therein, the equations can be solved to obtain the nucleon and $\Delta$ masses and amplitudes.  Owing to Eq.\,(\ref{DQPropConstr}), in this calculation the masses of the scalar and axial-vector diquarks are the only variable parameters.  The mass of the axial-vector correlation is chosen in order to obtain a desired mass for the $\Delta$ and the scalar diquark's mass is subsequently set by requiring a particular nucleon mass.  In this, too, we follow Refs.\,\cite{Alkofer:2004yf,Cloet:2008wg,Cloet:2008re} and choose diquark masses so as to obtain $m_N=1.18\,$GeV and $m_\Delta=1.33\,$GeV, which are greater than those determined empirically \cite{Beringer:1900zz}: $m_N^{\rm exp}=0.94\,$GeV and $m_{\Delta}^{\rm exp} = 1.23\,{\rm GeV}$.  This is appropriate because our Faddeev equation kernels omit resonant contributions; i.e., do not contain effects that may phenomenologically be associated with a meson cloud.

In practical calculations, meson-cloud effects divide into two distinct classes. \label{page:pionloops} The first (type-1) is within the gap equation, where pseudoscalar-meson loop corrections to the dressed-quark-gluon vertex
act uniformly to reduce the infrared mass-scale associated with the mass-function of a dressed-quark \cite{Blaschke:1995gr,Fischer:2007ze,Eichmann:2008ae,Cloet:2008fw,
Chang:2009ae}.  This effect can be pictured as a single quark emitting and reabsorbing a pseudoscalar meson and can be mimicked by simply choosing the parameters in the gap equation's kernel so as to obtain a dressed-quark mass that is characterised by an energy-scale of roughly $400\,$MeV.  Such an approach has implicitly been widely employed with phenomenological success \cite{Roberts:2000aa,Maris:2003vk,Roberts:2007jh,%
Chang:2011vu,Bashir:2012fs,Cloet:2013jya}; and we use it herein.

\begin{table}[t!]
\begin{center}
\caption{\label{tableNmass} Mass-scale parameters (in GeV) for the scalar and
axial-vector diquark correlations, fixed by fitting nucleon and $\Delta$ masses
offset to allow for ``meson cloud'' contributions \protect\cite{Hecht:2002ej}. We also list $\omega_{J^{P}}= m_{J^{P}}/\surd 2$, the width-parameter in the
$(qq)_{J^P}$ Bethe-Salpeter amplitude, Eqs.\,(\protect\ref{Gamma0p}) \& (\protect\ref{Gamma1p}):  its inverse is an indication of the diquark's matter
radius.  Row~2 illustrates effects of omitting the $1^+$-diquark correlation: the $\Delta$ cannot be formed and $m_N$ is significantly increased.  Evidently, the $1^+$-diquark provides significant attraction in the Faddeev equation's kernel.  (Dimensioned quantities in GeV, unless otherwise indicated.}
\begin{tabular*}{1.0\textwidth}{
c@{\extracolsep{0ptplus1fil}}c@{\extracolsep{0ptplus1fil}}
c@{\extracolsep{0ptplus1fil}}
c@{\extracolsep{0ptplus1fil}}c@{\extracolsep{0ptplus1fil}}c@{\extracolsep{0ptplus1fil
}}}
\hline
$m_N$ & $m_{\Delta}$~ & $m_{0^{+}}$ & $m_{1^{+}}$~ &
$\omega_{0^{+}} $ & $\omega_{1^{+}}$ \\
\hline
1.18 & 1.33~ & 0.796 & 0.893 & 0.56=1/(0.35\,{\rm fm}) & 0.63=1/(0.31\,{\rm fm}) \\
1.46 &  & 0.796 &  & 0.56=1/(0.35\,{\rm fm}) &  \\
\hline
\end{tabular*}
\end{center}
\end{table}

The second sort of correction (type-2) arises in connection with bound-states and may be likened to adding pseudoscalar meson exchange between dressed-quarks within the bound-state \cite{Roberts:1988yz,Hollenberg:1992nj,Alkofer:1993gu,Mitchell:1996dn,%
Ishii:1998tw,Pichowsky:1999mu,Hecht:2002ej,Sanchis-Alepuz:2014wea}, as opposed to type-1; i.e., emission and absorption of a meson by the same quark.  The type-2 contribution, depicted explicitly in Fig.\,1 of Ref.\,\cite{Ishii:1998tw}, is that computed in typical evaluations of meson-loop corrections to hadron observables based on a point-hadron Lagrangian \cite{Hecht:2002ej} and they are the corrections that should be added to calculated baryon masses.

These observations underpin a view that bound-state kernels which omit type-2
meson-cloud corrections should produce dressed-quark-core masses for hadron
ground-states that are larger than the empirical values.  This is true in practice, as emphasised by agreement between Faddeev equation predictions such as ours \cite{Roberts:2011cf,Chen:2012qr,Wang:2013wk} and the bare masses fitted in dynamical coupled-channels computations \cite{Aznauryan:2012baS,Suzuki:2009nj,Doring:2010ap}.
Moreover, as we shall again see herein, this perspective also has implications for the description of elastic and transition form factors \cite{Alkofer:2004yf,Eichmann:2008ef,Cloet:2008re,Cloet:2008wg,%
Segovia:2013rca,Wilson:2011aa}.


\section{Electromagnetic Currents}
\label{sec:BEMCurrents}

\subsection{Nucleon Elastic Form Factors}
\label{subsec:BEMnucleon}
The nucleon's electromagnetic current is
\begin{eqnarray}
J_\mu(K,Q) & = & ie\,\bar u(P_{f})\, \Lambda_\mu(K,Q) \,u(P_{i})
= i e \,\bar u(P_{f})\,\left( \gamma_{\mu} F_{1}(Q^{2}) +
\frac{1}{2m_{N}}\, \sigma_{\mu\nu}\,Q_\nu\,F_2(Q^2)\right) u(P_{i})\,,
\label{JnucleonB}
\end{eqnarray}
where $P_{i}$ ($P_{f}$) is the momentum of the incoming (outgoing) nucleon, $Q=
P_{f} - P_{i}$ is the photon momentum, $K=(P_{i}+P_{f})/2$ is the total momentum of the system, $K\cdot Q=0$ and $K^2 = - m_N^2 (1+\tau_N)$, $\tau_N = Q^2/(4 m_N^2)$ for elastic scattering; and $F_1$ and $F_2$ are, respectively, the Dirac and Pauli form factors, from which one obtains the nucleon's electric and magnetic (Sachs) form factors
\begin{equation}
\label{GEpeq}
G_E(Q^2)  =  F_1(Q^2) - \frac{Q^2}{4 m_{N}^2} F_2(Q^2)\,, \quad
G_M(Q^2)  =  F_1(Q^2) + F_2(Q^2)\,.
\end{equation}
The Sachs form factors can be obtained directly from the current using any sensible projection operators, for example, with the trace over spinor indices:
\begin{equation}
\label{ProjectNucleon}
G_{E} = \frac{m_{N}}{2K^{2}} \, K_{\mu}\, {\rm tr} J_{\mu}\,, \quad
G_{M} = \frac{m_{N}^{2}}{Q^{2}} \left(\delta_{\mu\nu} - \frac{K_{\mu}K_{\nu}}{K^{2}} \right) \,  {\rm tr} \,i \gamma_{\mu} J_{\nu}\,.
\end{equation}

Nucleon static electromagnetic properties are associated with the behaviour of these form factors in the neighbourhood $Q^2\simeq 0$.  The charges are given by $G_{E}(0)$ and the magnetic moments are
\begin{equation}
\label{momdef}
\mu_N = G_M^N(0) = F_1^N(0) + F_2^N(0) \Rightarrow
\mu_n = F_2^n(0) =: \kappa_n \,, \quad
\mu_p = 1 + F_2^p(0) =: 1 + \kappa_p\,,
\end{equation}
where $\kappa_N$, $N=n,p$, are the anomalous magnetic moments.

\subsection{\mbox{\boldmath $\Delta$} Elastic Form Factors}
\label{subsec:BEMdelta}
The matrix element of the electromagnetic current operator between $J=3/2$ states can be expressed through four form factors: Coulomb monopole (E0); magnetic dipole (M1); electric quadrupole (E2); and magnetic octupole (M3).  In order to construct those form factors, one may first write the $\Delta\gamma\Delta$ vertex as \cite{Nicmorus:2010sd}:
\begin{equation}
\Lambda_{\mu,\lambda\omega}(K,Q) = \Lambda_{+}(P_{f})R_{\lambda\alpha}(P_{f})
\Gamma_{\mu,\alpha\beta}(K,Q) \Lambda_{+}(P_{i})R_{\beta\omega}(P_{i}),
\label{eq:DEMcurrent}
\end{equation}
where the positive-energy spinor projector $\Lambda_{+}(P)$ and Rarita-Schwinger projection operator $R_{\mu\nu}(P)$ are defined in App.\,\ref{App:EM}, and
\begin{equation}
\Gamma_{\mu,\alpha\beta}(K,Q) =
\left[(F_{1}^{\ast}+F_{2}^{\ast})i\gamma_{\mu}-\frac{F_{2}^{\ast}}{m_{\Delta}}K_{\mu}
\right]\delta_{\alpha\beta}
-\left[(F_{3}^{\ast}+F_{4}^{\ast})i\gamma_{\mu}-\frac{F_{4}^{\ast}}{m_{\Delta}}K_{
\mu }\right]\frac{Q_{\alpha}Q_{\beta}}{4m_{\Delta}^{2}}.
\label{eq:Gammamualbe}
\end{equation}
This vertex also involves two momenta, expressed through the ingoing, $P_{i}$, and outgoing, $P_{f}$, $\Delta$-baryon momenta, or by the average momentum $K=(P_{i}+P_{f})/2$ and the photon momentum $Q=P_{f}-P_{i}$.  Once more, since the scattering is elastic and $P_{i}^{2}=P_{f}^{2}=-m_{\Delta}^{2}$, $K\cdot Q = 0$ and $K^{2} = -m_{\Delta}^{2}(1+\tau_\Delta)$, $\tau_\Delta=Q^{2}/(4m_{\Delta}^{2})$.  It follows that the Poincar\'e invariant form factors which constitute the vertex depend only on the photon momentum-transfer $Q^{2}$: $\{F_i^\ast=F_i^\ast(Q^2), i=1,2,3,4\}$.  The multipole form factors are constructed as follows:
\begin{subequations}
\label{DefsDeltaElastic}
\begin{eqnarray}
G_{E0}(Q^{2}) &=& \left(1+\frac{2\tau_\Delta}{3}\right)(F_{1}^{\ast}-\tau_\Delta F_{2}^{\ast}) -
\frac{\tau_\Delta}{3} (1+\tau_\Delta) (F_{3}^{\ast}-\tau_\Delta F_{4}^{\ast}), \\
G_{M1}(Q^{2}) &=& \left(1+\frac{4\tau_\Delta}{5}\right)(F_{1}^{\ast}+F_{2}^{\ast}) -
\frac{2\tau_\Delta}{5} (1+\tau_\Delta) (F_{3}^{\ast}+F_{4}^{\ast}), \\
G_{E2}(Q^{2}) &=& (F_{1}^{\ast}-\tau_\Delta F_{2}^{\ast}) -
\frac{1}{2} (1+\tau_\Delta) (F_{3}^{\ast}-\tau_\Delta F_{4}^{\ast}), \\
G_{M3}(Q^{2}) &= &(F_{1}^{\ast}+F_{2}^{\ast}) -
\frac{1}{2} (1+\tau_\Delta) (F_{3}^{\ast}+ F_{4}^{\ast})\,,
\end{eqnarray}
\end{subequations}
and their $Q^2=0$ values define dimensionless multipole moments:
\begin{equation}
e_{\Delta} = G_{E0}(0)\,, \quad
\hat \mu_{\Delta} = G_{M1}(0)\,, \quad
\hat {\cal Q}_{\Delta} = G_{E2}(0) \,, \quad
\hat {\cal O}_{\Delta} = G_{M3}(0)\,.
\end{equation}

Given the vertex, $\Gamma_{\mu,\alpha\beta}(K,Q)$, one may obtain the multipole form factors using any four reasonable projection operators; e.g., with
\cite{Nicmorus:2010sd}
\begin{equation}
\mathpzc{p}_{1} = \check{P}_{\mu}\check{P}_{\lambda}\check{P}_{\omega}{\rm tr} \Lambda_{\mu,\lambda\omega}\,, \quad
\mathpzc{p}_{2} = \check{P}_{\mu} {\rm tr} \Lambda_{\mu,\lambda\lambda}\,,
\quad
\mathpzc{p}_{3} = \check{P}_{\lambda}\check{P}_{\omega} {\rm
tr} \Lambda_{\mu,\lambda\omega}\gamma^{\perp}_{\mu}\,, \quad
\mathpzc{p}_{4} = {\rm tr}
\Lambda_{\mu,\lambda\lambda}\gamma^{\perp}_{\mu}\,,
\label{eq:scalars}
\end{equation}
where $P\cdot \gamma^{\perp}=0$ and $\check{P}^{2}=+1$, one has
\begin{equation}
\label{DEprojections}
\begin{array}{ll}
\displaystyle G_{E0} = \frac{\mathpzc{p}_{2}-2\mathpzc{p}_{1}}{4i\sqrt{1+\tau_\Delta}}\,, &
\displaystyle G_{M1} = \frac{9i(\mathpzc{p}_{4}-2\mathpzc{p}_{3})}{40\tau_\Delta}\,, \\
\displaystyle G_{E2} =
\frac{3\left[\mathpzc{p}_{1}\left(3+2\tau_\Delta\right)
-\mathpzc{p}_{2}\tau_\Delta\right]}{8i\tau_\Delta^{2}\sqrt{1+\tau_\Delta}}\,, &
\displaystyle G_{M3} =
\frac{3i\left[\mathpzc{p}_{3}(5+4\tau_\Delta)-2\mathpzc{p}_{4} \tau_\Delta\right]}{32\tau_\Delta^{3}}\,.
\end{array}
\end{equation}

\subsection{\mbox{\boldmath $\gamma + N \to \Delta$} Transition Form Factors}
\label{subsec:BEMnucdel}
The $\gamma^{\ast} N \to \Delta$ transition is described by three
Poincar\'e-invariant form factors~\cite{Jones:1972ky}: magnetic-dipole,
$G_{M}^{\ast}$; electric quadrupole, $G_{E}^{\ast}$; and Coulomb (longitudinal)
quadrupole, $G_{C}^{\ast}$. They arise through consideration of the $N\to \Delta$
transition current:
\begin{equation}
J_{\mu\lambda}(K,Q) =
\Lambda_{+}(P_{f})R_{\lambda\alpha}(P_{f})i\gamma_{5}\Gamma_{\alpha\mu}(K,Q)\Lambda_
{+}(P_{i}),
\label{eq:JTransition}
\end{equation}
where: $P_{i}$, $P_{f}$ are, respectively, the incoming nucleon and outgoing $\Delta$
momenta, with $P_{i}^{2}=-m_{N}^{2}$, $P_{f}^{2}=-m_{\Delta}^{2}$; the incoming
photon momentum is $Q_\mu=(P_{f}-P_{i})_\mu$ and $K=(P_{i}+P_{f})/2$; and
$\Lambda_{+}(P_{i})$, $\Lambda_{+}(P_{f})$ are, respectively, positive-energy
projection operators for the nucleon and $\Delta$, with the Rarita-Schwinger tensor
projector $R_{\lambda\alpha}(P_f)$ arising in the latter connection.

In order to succinctly express $\Gamma_{\alpha\mu}(K,Q)$, we define
\begin{equation}
\check K_{\mu}^{\perp} = {\cal T}_{\mu\nu}^{Q} \check{K}_{\nu}
= (\delta_{\mu\nu} - \check{Q}_{\mu} \check{Q}_{\nu}) \check{K}_{\nu},
\end{equation}
with $\check{K}^{2} = 1 = \check{Q}^{2}$, in which case
\begin{equation}
\Gamma_{\alpha\mu}(K,Q) =
\mathpzc{k}
\left[\frac{\lambda_m}{2\lambda_{+}}(G_{M}^{\ast}-G_{E}^{\ast})\gamma_{5}
\varepsilon_{\alpha\mu\gamma\delta} \check K_{\gamma}\check{Q}_{\delta}  -
G_{E}^{\ast} {\cal T}_{\alpha\gamma}^{Q} {\cal T}_{\gamma\mu}^{K}
- \frac{i\varsigma}{\lambda_m}G_{C}^{\ast}\check{Q}_{\alpha} \check
K^\perp_{\mu}\right],
\label{eq:Gamma2Transition}
\end{equation}
where
$\mathpzc{k} = \sqrt{(3/2)}(1+m_\Delta/m_N)$,
$\varsigma = Q^{2}/[2\Sigma_{\Delta N}]$,
$\lambda_\pm = \varsigma + t_\pm/[2 \Sigma_{\Delta N}]$
with $t_\pm = (m_\Delta \pm m_N)^2$,
$\lambda_m = \sqrt{\lambda_+ \lambda_-}$,
$\Sigma_{\Delta N} = m_\Delta^2 + m_N^2$, $\Delta_{\Delta N} = m_\Delta^2 - m_N^2$.

With a concrete expression for the current in hand, one may obtain the form factors using any three sensibly chosen projection operations; e.g.,
with~\cite{Eichmann:2011aa}
\begin{equation}
\mathpzc{t}_{1} = \mathpzc{n}
\frac{\sqrt{\varsigma(1+2\mathpzc{d})}}{\mathpzc{d}-\varsigma}
{\cal T}^{K}_{\mu\nu}\check K^\perp_{\lambda} {\rm tr}
\gamma_{5}J_{\mu\lambda}\gamma_{\nu}\,,
\mathpzc{t}_{2} = \mathpzc{n} \frac{\lambda_{+}}{\lambda_m} {\cal T}^{K}_{\mu\lambda}
{\rm tr} \gamma_{5} J_{\mu \lambda}\,,
\mathpzc{t}_{3} =  3 \mathpzc{n}
\frac{\lambda_+}{\lambda_m}\frac{(1+2\mathpzc{d})}{\mathpzc{d}-\varsigma} \check
K^\perp_{\mu}\check K^\perp_{\lambda} {\rm tr}\gamma_{5}J_{\mu\lambda} \,,
\end{equation}
where $\mathpzc{d}=\Delta_{\Delta N}/[2 \Sigma_{\Delta N}]$,
$\mathpzc{n}= \sqrt{1-4\mathpzc{d}^{2}}/[4i\mathpzc{k}\lambda_m]$), then
\begin{equation}
\label{GMGEGC}
G_{M}^{\ast} = 3
\left[ \mathpzc{t}_{2}+\mathpzc{t}_{1}\right]\,, \;
G_{E}^{\ast} = \mathpzc{t}_{2}-\mathpzc{t}_{1}\,, \;
G_{C}^{\ast} = \mathpzc{t}_{3}.
\end{equation}

The following ratios are often considered in connection with the $\gamma N\to \Delta$ transition:
\begin{equation}
\label{eqREMSM}
R_{\rm EM} = -\frac{G_E^{\ast}}{G_M^{\ast}}, \quad
R_{\rm SM} = - \frac{|\vec{Q}|}{2 m_\Delta} \frac{G_C^{\ast}}{G_M^{\ast}}
=  - \frac{\lambda_m}{m_\Delta} \frac{G_C^{\ast}}{G_M^{\ast}}\,,
\end{equation}
because they can be read as measures of deformation in one or both of the hadrons involved since they are identically zero in $SU(6)$-symmetric constituent-quark models.


\section{Nucleon Elastic Form Factors}
\label{sec:BEMFFs}
In order to calculate the electromagnetic form factors specified above one must know the manner by which the baryon described in Sec.\,\ref{sec:BaryonModel} couples to a photon.  That is described in Ref.\,\cite{Oettel:1999gc} and detailed in App.\,\ref{NPVertex}.  As apparent in that Appendix, the current depends on the electromagnetic properties of the diquark correlations.  Estimates exist of the size of diquark correlations \cite{Bloch:1999ke,Maris:2004bp,Alexandrou:2006cq,Babich:2007ah,Roberts:2011wy} and one typically finds
\begin{equation}
r_{\{uu\}_{1^+}} \approx 1.1 r_{[ud]_{0^+}},\;
r_{[ud]_{0^+}} \approx 1.1 r_\pi\,,
\end{equation}
where $r_\pi$ is the pion charge radius.  It is thus evident that diquark correlations within a baryon are not pointlike.  Hence, with increasing $Q^2$, interaction diagrams in which the photon resolves a diquark's substructure must be suppressed with respect to contributions from diagrams that describe either a photon interacting with a bystander or an exchanged quark.  These latter two are the only hard interactions with dressed-quarks allowed in a baryon.

Motivated by these considerations, we employ electromagnetic form factors for the diquark correlations, which are expressed in Eqs.\,(\ref{Fqqform}), (\ref{Gamma0plus}), (\ref{AnsatzF1}), (\ref{calTvalue}).  A one-parameter monopole is appropriate because the correlations involve two quarks: the parameter is a length-scale which characterises the associated interaction radius.  We use the same single scale for all diquark form factors; viz., scalar diquark, axial-vector diquark and scalar$\,\leftrightarrow\,$axial-vector transition:
\begin{equation}
\label{radiusqq}
r_{qq} = 0.8\,{\rm fm} \,.
\end{equation}
This value was used elsewhere \cite{Cloet:2008re}; and within any reasonable estimate of the theoretical error in our formulation, it is meaningless to distinguish 10\% differences between radii for the different form factors.  Moreover, 10\% changes in the common value of $r_{qq}$ have no material effect.



\begin{figure}[t]
\begin{center}
\begin{tabular}{cc}
\includegraphics[clip,width=0.47\linewidth]{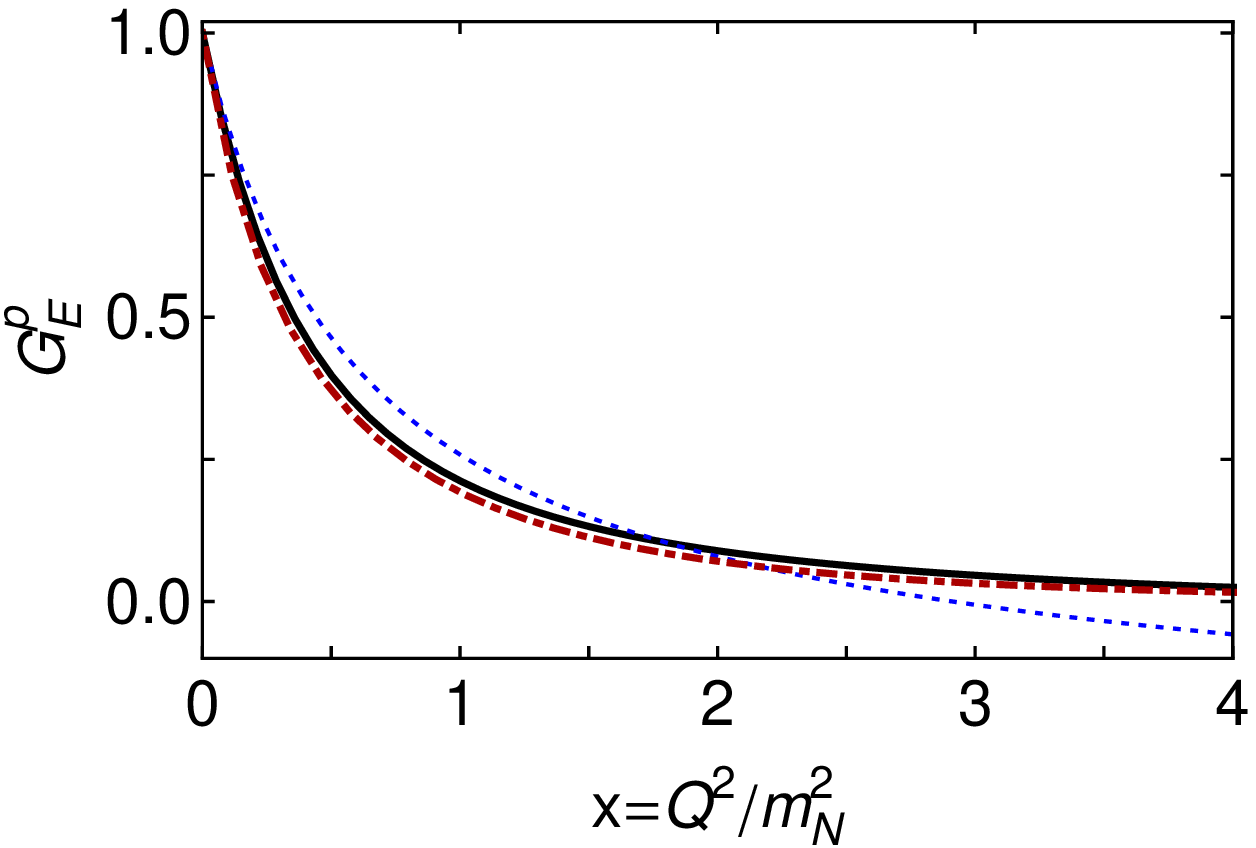}\vspace*
{-1ex } &
\includegraphics[clip,width=0.47\linewidth]{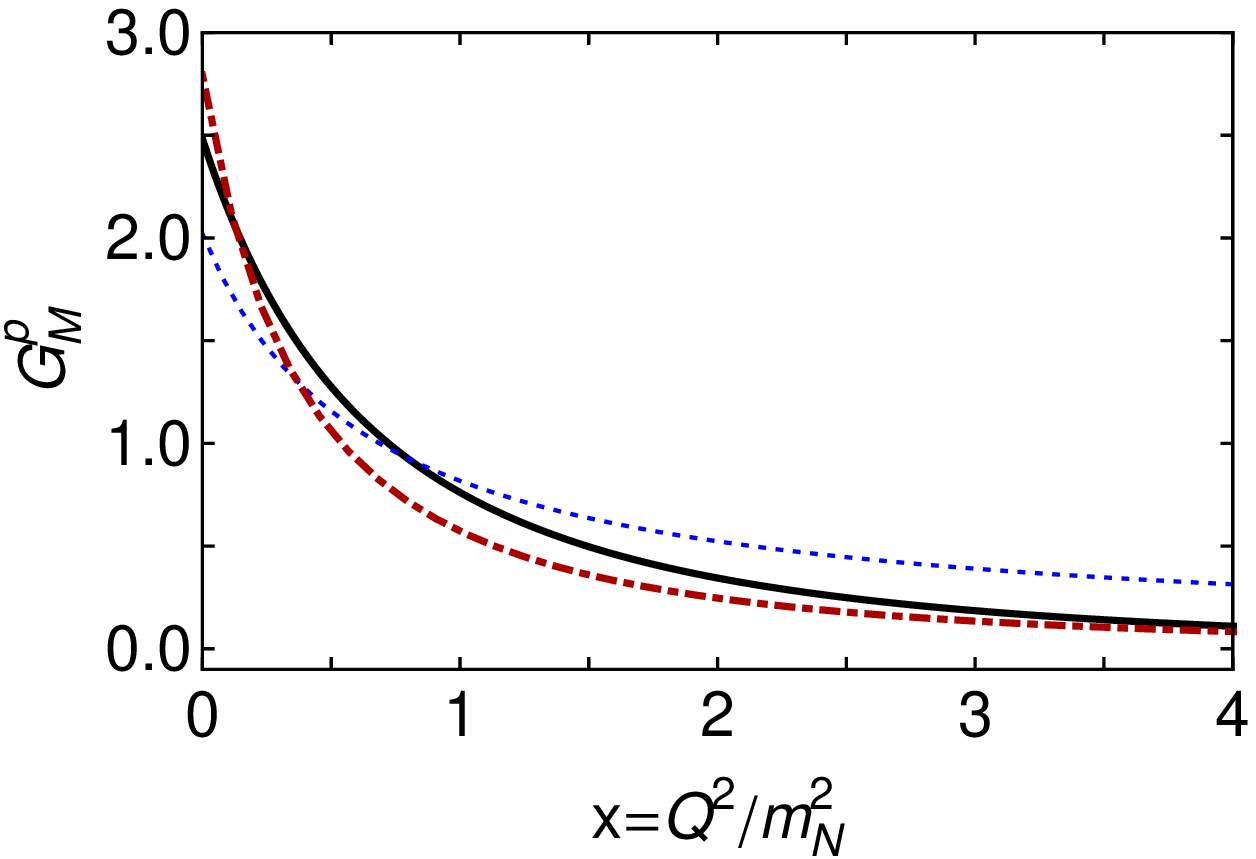}\vspace*
{-1ex}
\end{tabular}
\begin{tabular}{cc}
\includegraphics[clip,width=0.47\linewidth]{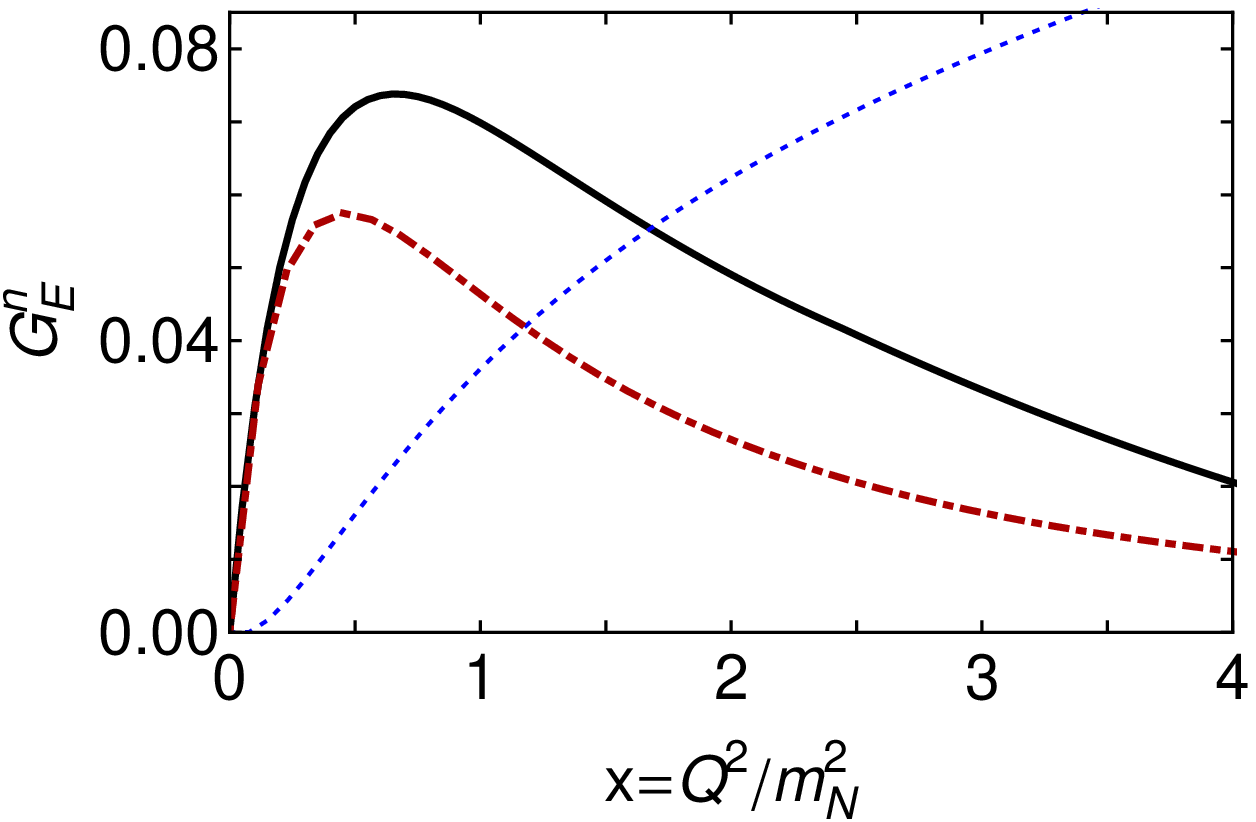}\vspace*
{-1ex } &
\includegraphics[clip,width=0.47\linewidth]{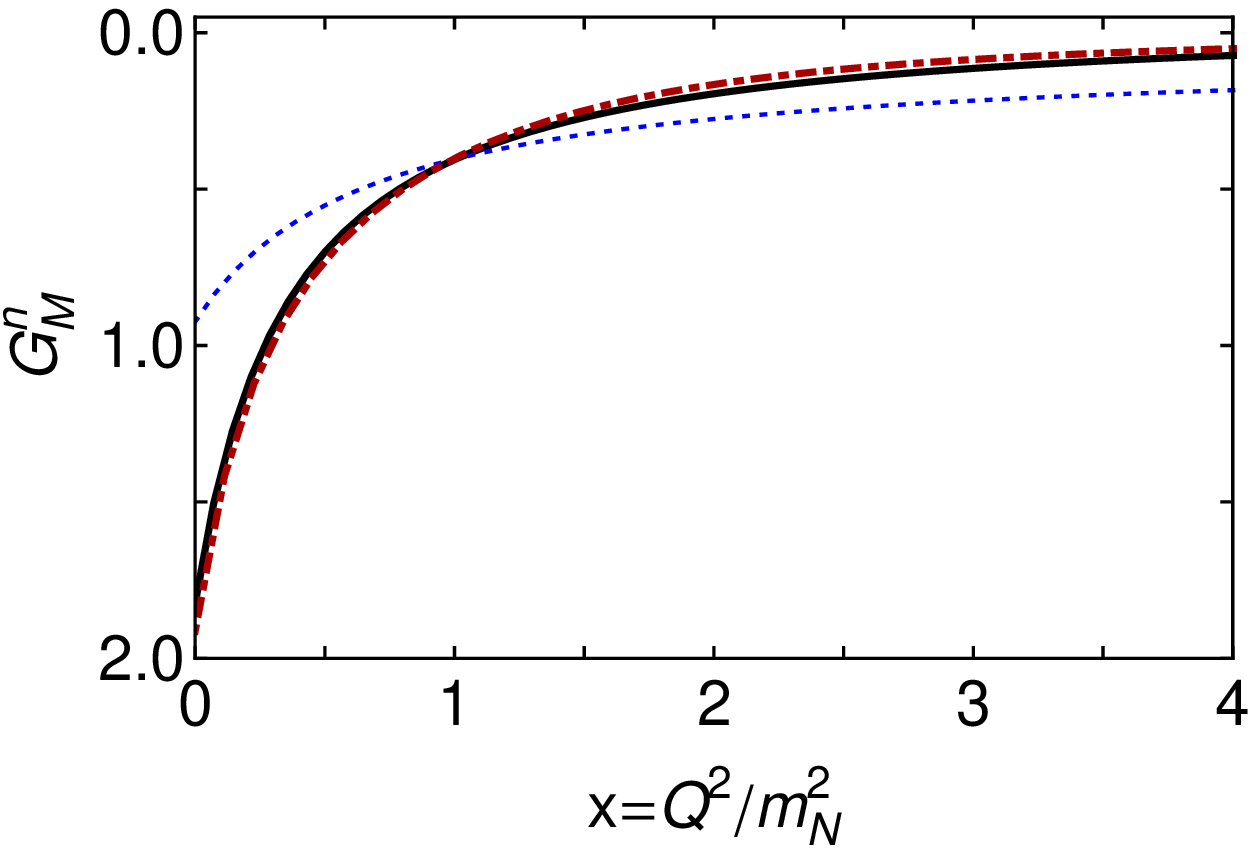}\vspace*
{-1ex}
\end{tabular}
\caption{\label{fig:FFNucleon1}
Proton (top) and neutron (bottom) electromagnetic form factors.
In both rows: {\it left panel} -- Sachs electric; {\it right panel} -- Sachs magnetic.
Curves in all panels: {\it solid, black} -- result obtained herein, using a Faddeev equation kernel and interaction vertices that possess QCD-like momentum dependence; viz., a QCD-kindred framework; {\it dotted, blue} -- result obtained with a symmetry preserving treatment of a contact interaction (CI framework) \cite{Wilson:2011aa}; {\it dot-dashed,red} -- 2004 parametrisation of experimental data \cite{Kelly:2004hm}.}
\end{center}
\end{figure}

The explicit form of the nucleon current is
\begin{equation}
J_\mu(P_f,P_i) =  \sum_{n=1}^6
\int\frac{d^4p}{(2\pi)^4} \frac{d^4k}{(2\pi)^4} \;
\bar\Psi(-p;P_f) \, J_\mu^n(p,P_f,k,P_i) \, \Psi(k;P_i)\,,
\label{JNucleonExplicit}
\end{equation}
where $\Psi$ is the nucleon Faddeev amplitude and the sum ranges over each one of the six diagrams depicted in Fig.\,\ref{vertex} and detailed in App.\,\ref{NPVertex}.  The nucleon form factors are obtained from Eq.\,\eqref{JNucleonExplicit} via the projections in Eq.\,\eqref{ProjectNucleon}.

\subsection{Sachs Form Factors}
In Fig.\,\ref{fig:FFNucleon1} we depict the dressed-quark core Sachs electric and magnetic form factors for the proton and neutron computed using a Faddeev equation kernel and interaction vertices that possess QCD-like momentum dependence.  We will hereafter describe this pairing as the QCD-kindred framework.  The form factors Fig.\,\ref{fig:FFNucleon1} produce the following values for nucleon static properties:
\begin{equation}
\label{nucleonstatic}
\begin{array}{c}
\begin{array}{l|ccccc}
{\rm Th} &
r_E^2 & r_E^2 M_N^2 & r_M^2 & r_M^2 M_N^2 & \mu\\ \hline
p &
\phantom{-}(0.61\,{\rm fm})^2 & \phantom{-}(4.31)^2 &
(0.53\,{\rm fm})^2 & (3.16)^2 & \phantom{-}2.50 \\
n & -(0.28\,{\rm fm})^2 & -(1.99)^2 & (0.70\,{\rm fm})^2 & (4.95)^2 & -1.83
\end{array},\\
\begin{array}{l|ccccc}
{\rm Exp} & r_E^2 & r_E^2 M_N^2 & r_M^2 & r_M^2 M_N^2 & \mu \\ \hline
p &
\phantom{-}(0.88\,{\rm fm})^2 & \phantom{-}(4.19)^2 &
(0.84\,{\rm fm})^2 & (4.00)^2 & \phantom{-}2.79 \\
n & -(0.34\,{\rm fm})^2 & -(1.62)^2 & (0.89\,{\rm fm})^2& (4.24)^2 & -1.91
\end{array},
\end{array}
\end{equation}
where the experimental values are drawn from Ref.\,\cite{Beringer:1900zz} and we quote a proton charge radius determined from electron scattering.\footnote{Meson cloud corrections to these dressed-quark core results are canvassed in Ref.\,\cite{Cloet:2008wg}; and a discussion of the challenge posed by muonic hydrogen measurements may be found in Ref.\,\cite{Pohl:2013yb}.}
The figure also displays results obtained with a symmetry preserving treatment of a contact interaction, which we will subsequently designate as the CI-framework.

It is apparent in Fig.\,\ref{fig:FFNucleon1} that the QCD-kindred results are in fair agreement with experiment, which is represented by the year-2004 parametrisation in Ref.\,\cite{Kelly:2004hm}.  (Comparisons made with a more recent parametrisation \cite{Bradford:2006yz} are not materially different.)  No parameters were tuned in order to achieve this outcome.  The most notable mismatch appears to be in our description of the neutron electric form factor at low $Q^2$.  However appearances are somewhat deceiving in this case because, on the low-$Q^2$ domain: $G_E^n$ is small and hence slight differences appear large; and $G_E^n$ is much affected by subdominant effects that we have neglected, such as so-called meson-cloud contributions.  On the other hand, as was previously observed \cite{Wilson:2011aa}, form factors obtained via a symmetry-preserving DSE treatment of a contact-interaction are typically too hard.  The defects of a contact-interaction are expressed with greatest force in the neutron electric form factor.

A comparison between the curves in Fig.\,\ref{fig:FFNucleon1} shows clearly that readily accessible observables are typically very sensitive to the nature of the interaction that produces the amplitudes and propagators which describe bound-states and their dressed constituents.  This fact will be reinforced by subsequent figures.  Thus, as explained and illustrated in a series of contributions 
\cite{GutierrezGuerrero:2010md,%
Roberts:2010rn,Roberts:2011cf,Roberts:2011wy,Roberts:2011rr,%
Wilson:2011aa,Chen:2012qr,Chen:2012txa,Segovia:2013rca,Segovia:2013uga}
experiment and theory together can be used effectively to chart the evolution of QCD's $\beta$-function.

\begin{figure}[t]
\begin{center}
\begin{tabular}{cc}
\includegraphics[clip,width=0.47\linewidth]{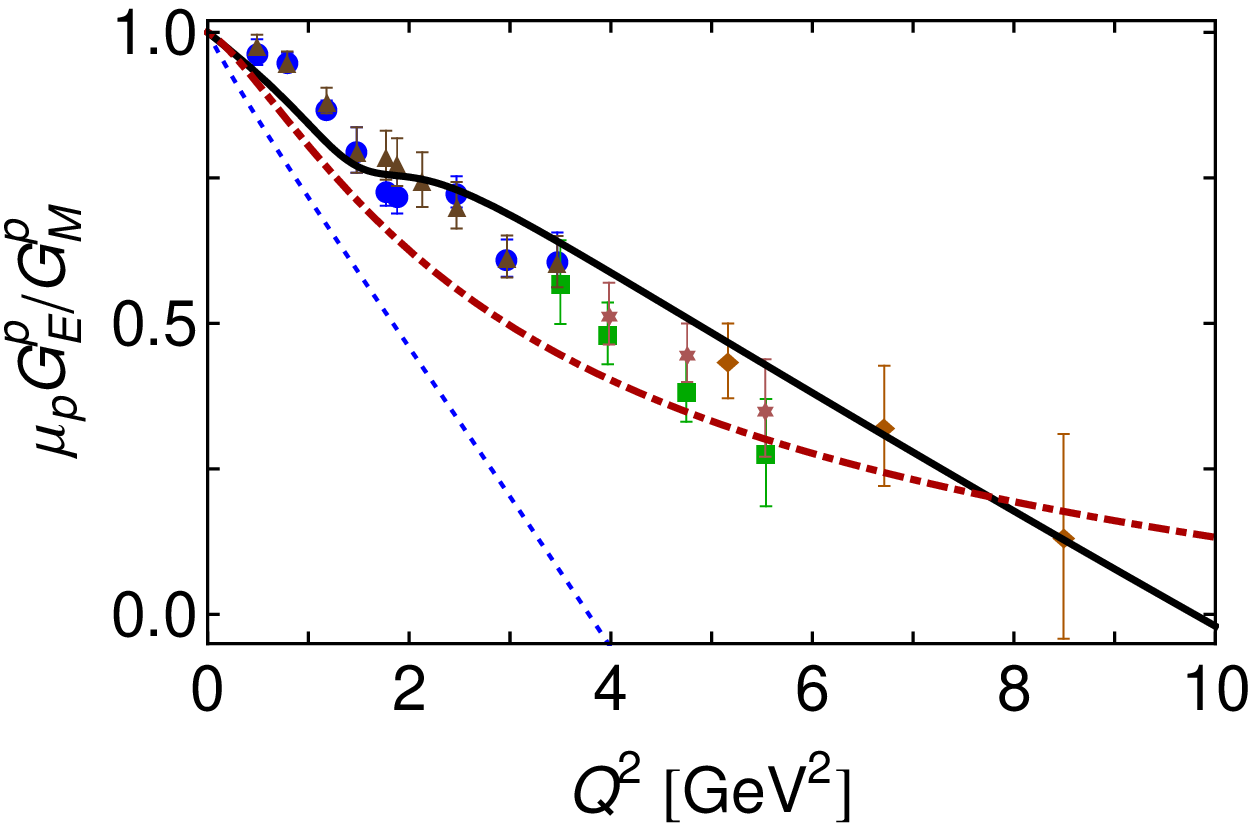}\vspace*
{-1ex } &
\includegraphics[clip,width=0.47\linewidth]{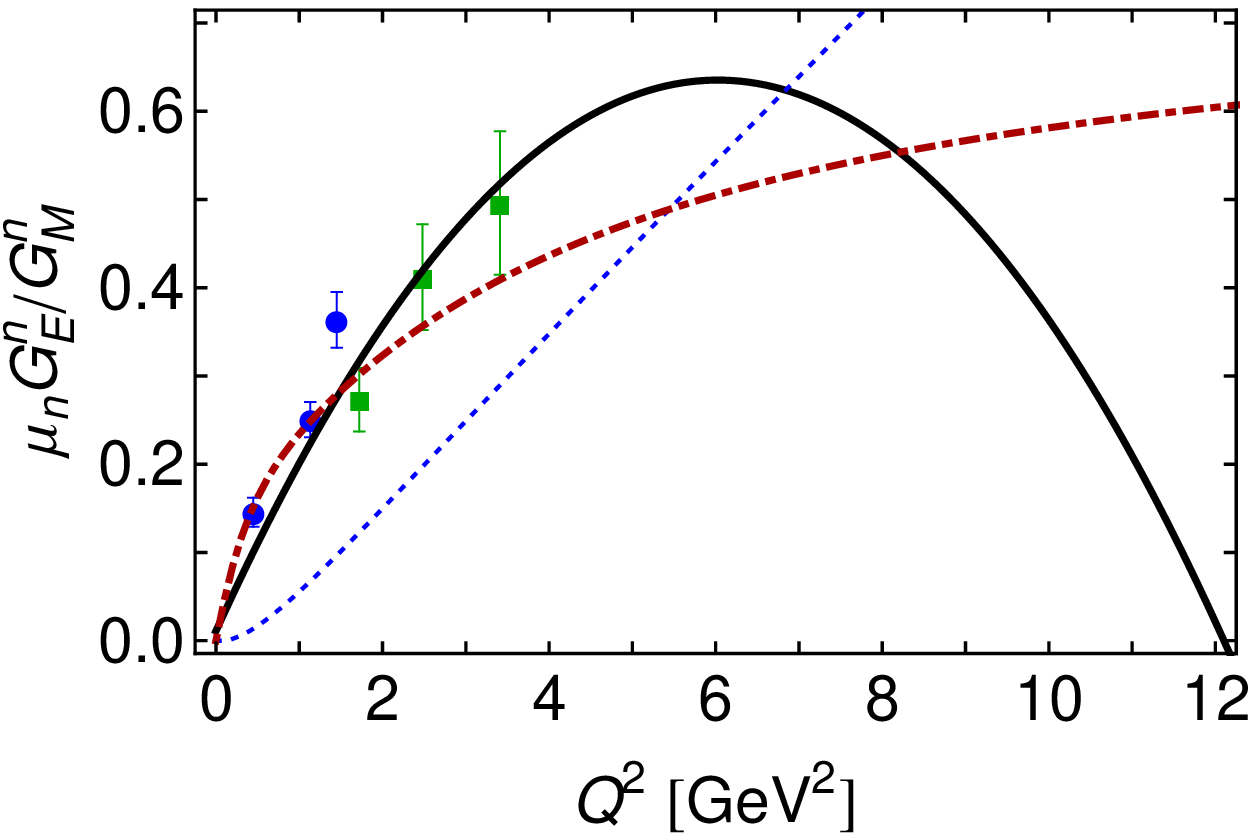}\vspace*
{-1ex}
\end{tabular}
\caption{\label{fig:FFNucleon2}
\emph{Left panel}: normalised ratio of proton electric and magnetic form factors.
Curves: {\it solid, black} -- result obtained herein, using our QCD-kindred framework; {\it Dashed, blue} -- CI result \cite{Wilson:2011aa}; and {\it dot-dashed, red} -- ratio inferred from 2004 parametrisation of experimental data \cite{Kelly:2004hm}.
Data:
blue circles \cite{Jones:1999rz};
green squares \cite{Gayou:2001qd};
brown triangles \cite{Punjabi:2005wqS};
purple asterisk \cite{Puckett:2010ac};
and orange diamonds \cite{Puckett:2011xgS}.
\emph{Right panel}: normalised ratio of neutron electric and magnetic form factors.  Curves: same as in left panel.
Data: blue circles \cite{Madey:2003av};
and green squares \cite{Riordan:2010id}.}
\end{center}
\end{figure}

In Fig.\,\ref{fig:FFNucleon2} we depict the unit-normalised ratio of Sachs electric and magnetic form factors for the proton and neutron.
Let us first consider the proton's ratio (left panel).  Both the CI and QCD-kindred frameworks predict a zero in $G_E^p/G_M^p$.  The comparison with extant experimental results indicates that the contact-interaction is not a useful tool on $Q^2\gtrsim M(0)^2$, where $M(p^2)$ is the dressed-quark mass function explained in App.\,\ref{subsubsec:S}.  The result obtained with the QCD-kindred framework, on the other hand, agrees with available data and predicts a zero in this ratio at\footnote{The zero is herein shifted to larger $Q^2$ by 20\% compared with the result in Ref.\,\cite{Cloet:2008re} because we use a larger value for the magnetic moment of the axial-vector diquark: $\mu_{1^+}=1.0$ instead of $0.37$.  The new value of $\mu_{1^+}=1.0$ was obtained by requiring a unified description of nucleon and $\Delta$ elastic and transition form factors, which introduced constraints not available to Ref.\,\cite{Cloet:2008re}.}
$Q^2\approx 9.5\,$GeV$^2$.  We note that owing to the presence of strong diquark correlations, the singly-represented $d$-quark is usually sequestered inside a soft (scalar) diquark correlation.  The appearance of a zero is therefore driven primarily by the contribution to $G_E^p$ from the doubly-represented $u$-quark \cite{Cloet:2013jya}, which is four times more likely than the $d$-quark to be involved in a hard interaction.

As explained elsewhere \cite{Wilson:2011aa,Cloet:2013gva}, the behaviour of the dressed-quark contributions to the proton's electric form factor on $Q^2\gtrsim 5\,$GeV$^2$, and hence $G_E^p$ itself, are particularly sensitive to the rate at which the dressed-quark mass runs from the nonperturbative into the perturbative domain of QCD.  This is readily explicated using the information in the left panel of Fig.\,\ref{fig:FFNucleon2}.

The contact-interaction produces a momentum-independent dressed-quark mass; and in this counterpoint to QCD the dressed-quarks produce hard Dirac and Pauli form factors, which yield a ratio $\mu_p G_E/G_M$ that possesses a zero at $Q^2\lesssim 4\,$GeV$^2$.
Alternatively, the dressed-quark mass function used herein is large at infrared momenta and approaches the current-quark mass as the momentum of the dressed-quark increases.  Such is the behaviour in QCD: dressed-quarks are massive in the infrared but become parton-like in the ultraviolet, characterised thereupon by a mass function that is modulated by the current-quark mass.  Hence, the proton's dressed-quarks possess constituent-quark-like masses at small momenta and thus have a large anomalous magnetic moment on this domain.  As the momentum transfer grows, the structure of the integrands in the computation of the elastic form factors ensures that the dressed-quark mass functions are increasingly sampled within the domain upon which the transition from nonperturbative to perturbative behaviour takes place.  This corresponds empirically to momentum transfers $Q^2 \gtrsim 5\,$GeV$^2$.  The rate at which the transition occurs determines how quickly the dressed-quarks become parton-like; i.e., how rapidly they are unclothed and come to behave as light fermion degrees of freedom.  Since light-quarks must have a small anomalous magnetic moment \cite{Chang:2010hb}, then this transition entails that the proton Pauli form factor generated dynamically therewith drops to zero.  This produces an interplay between the Dirac and Pauli form factors which, via Eq.\,\eqref{GEpeq}, entails that a momentum-dependent mass-function must beget a zero at larger values of $Q^2$ than is obtained with momentum-independent dressed-quark masses.

Our model for the dressed-quark mass function is characterised by a particular rate of transition between the nonperturbative and perturbative domains.  If one were to increase that rate, then the transformation to partonlike quarks would become more rapid and hence the proton's Pauli form factor would drop even more rapidly to zero.  In this case the quark angular momentum correlations, expressed by the diquark structure, remain but the individual dressed-quark magnetic moments diminish markedly.  Thus a more rapid transition pushes the zero in $\mu_p G_{Ep}/G_{Mp}$ to larger values of $Q^2$.  There is a rate of transformation beyond which the zero disappears completely \cite{Cloet:2013gva}: a theory in which the mass-function rapidly becomes partonic produces no zero at all.

It follows that the possible existence and location of the zero in the ratio of proton elastic form factors [$\mu_p G_{E}^p(Q^2)/G_{M}^p(Q^2)$] are a fairly direct measure of the nature of the quark-quark interaction in the Standard Model.  Like the dilation of the meson valence-quark parton distribution amplitudes \cite{Chang:2013pqS,Cloet:2013tta,Segovia:2013ecaS,Gao:2014bca,Shi:2014uwa}, they are a cumulative gauge of the momentum dependence of the interaction, the transition between the associated theory's nonperturbative and perturbative domains, and the width of that domain.  Hence, in extending experimental measurements of this ratio, and thereby the proton's charge form factor, to larger momentum transfers; i.e., in reliably determining the proton's charge distribution, there is an extraordinary opportunity for a constructive dialogue between experiment and theory.  That feedback will assist greatly with contemporary efforts to reveal the character of the strongly interacting part of the Standard Model and its emergent phenomena.

\begin{figure}[t]
\begin{center}
\begin{tabular}{cc}
\includegraphics[clip,width=0.47\linewidth]{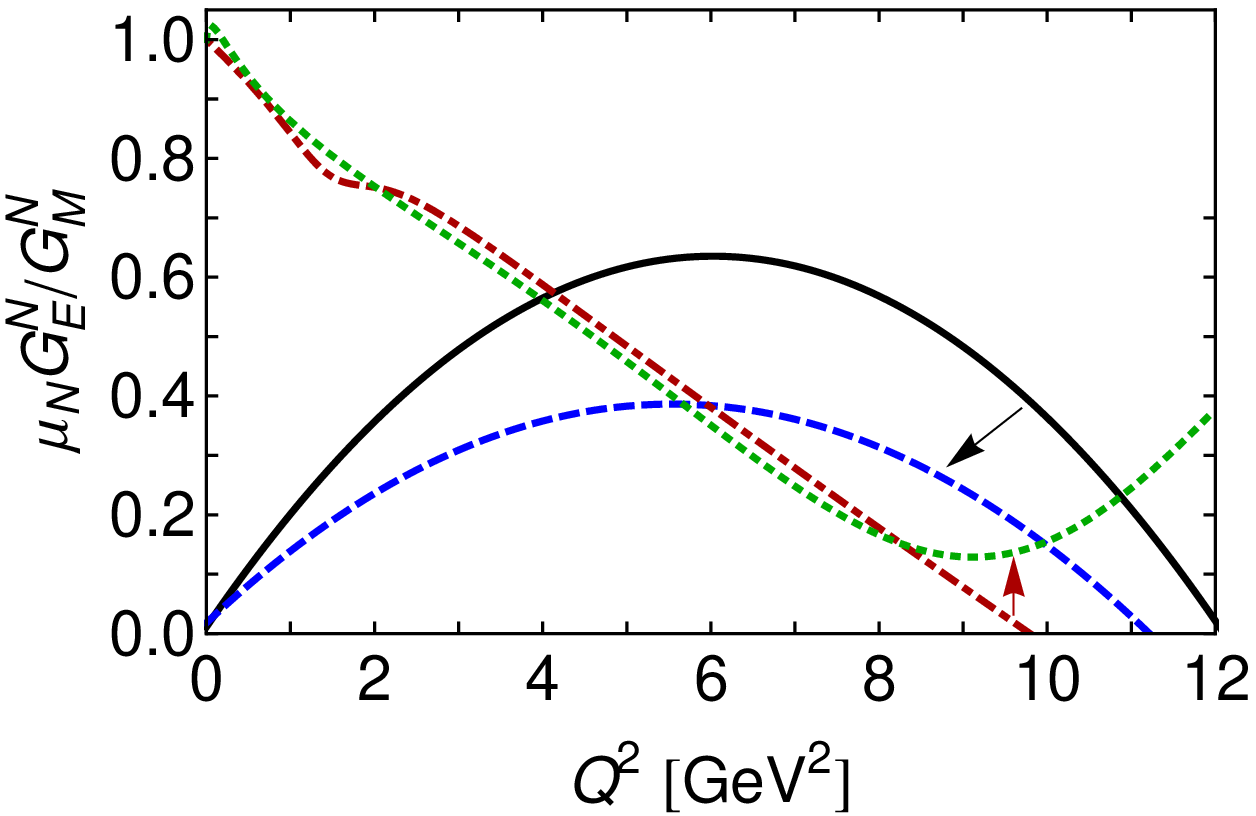}\vspace*
{-1ex } &
\includegraphics[clip,width=0.47\linewidth]{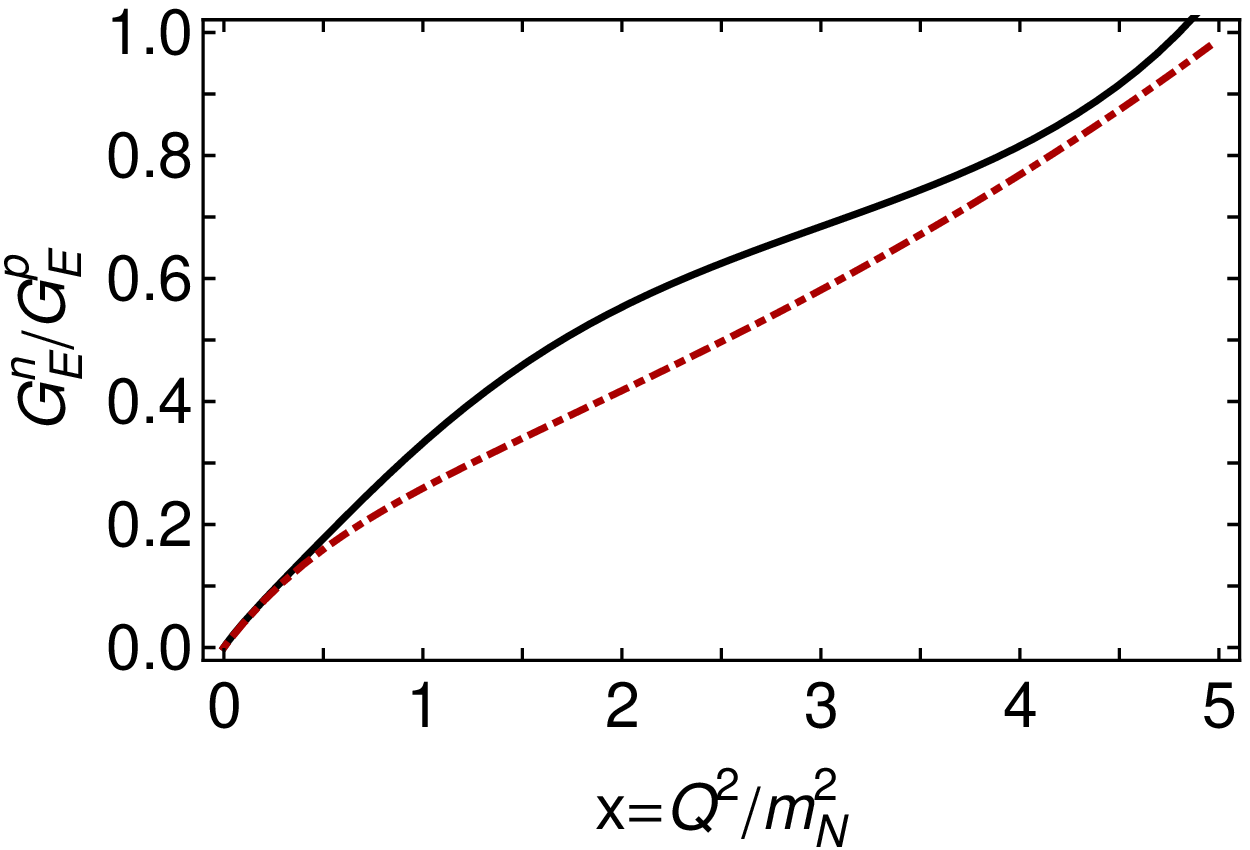} 
\end{tabular}
\caption{\label{fig:GEnGEp}
\emph{Left panel}.  Ratio $\mu_N G_E^N/G_M^N$ for $N=\,$neutron, proton:
{\it solid, black} -- neutron result obtained herein, using QCD-like momentum-dependent quark dressing; {\it dashed, blue} -- neutron result obtained with such dressing but an accelerated rate of transition from dressed-quark to parton; and \emph{dot-dashed, red} -- proton result obtained herein, \emph{dotted-green} -- proton result with accelerated transition.  (The arrows highlight the response to accelerating the dressed-quark$\,\to\,$ parton transformation.)
\emph{Right panel} -- Ratio of neutron and proton Sachs electric form factors:
{\it solid-black} -- result obtained herein; and {\it dot-dashed-red} -- ratio inferred from 2004 parametrisation of experimental data \cite{Kelly:2004hm}.}
\end{center}
\end{figure}

The right panel of Fig.\,\ref{fig:FFNucleon2} displays the ratio $\mu_n G_E^n/G_M^n$, which also exhibits a zero but at $Q^2\approx 12\,$GeV$^2$; i.e., shifted to a 25\% larger value of $Q^2$ compared with the zero in the proton ratio.
The properties of the dressed-quark propagators and bound-state amplitudes which influence the appearance of a zero in $\mu_n G_E^n/G_M^n$ are qualitatively the same as those described in connection with $\mu_p G_{E}^p/G_{M}^p$.  However, owing to the different electric charge weightings of the quark contributions in the neutron, the quantitative effect is opposite to that for the proton.  Namely, when the transformation from dressed-quark to parton is accelerated as described in Ref.\,\cite{Cloet:2013gva}, the zero occurs at smaller $Q^2$.  This is illustrated in the left panel of Fig.\,\ref{fig:GEnGEp}.  On the other hand, as indicated by the dotted curve in the lower-left panel of Fig.\,\ref{fig:FFNucleon1}, one typically finds that a contact interaction produces no zero in the neutron ratio \cite{Wilson:2011aa,Cloet:2014rja}.

In order to elucidate the origin of these features, let us first capitalise on the fact that the $s$-quark contributes very little to nucleon electromagnetic form factors \cite{Aniol:2005zf,Armstrong:2005hs,Young:2006jc} and therefore write
\begin{equation}
\label{eqFlavourSep}
G_{E}^p = e_u G_E^{p,u} - |e_d| G_E^{p,d} \,,\quad
G_{E}^n = e_u G_E^{n,u} - |e_d| G_E^{n,d}\,,
\end{equation}
where the isolated terms denote the contribution from each quark flavour.  Consider next that charge symmetry is almost exact in QCD, so that
\begin{equation}
G_E^{n,d} = G_E^{p,u}\,, \quad G_E^{n,u} = G_E^{p,d}\,,
\end{equation}
and hence, to a very good level of approximation,
\begin{equation}
G_{E}^n = e_u G_E^{n,u} - |e_d| G_E^{n,d} =  e_u G_E^{p,d} - |e_d| G_E^{p,u}\,.
\label{GEnCS}
\end{equation}

Now, with a zero in $G_{E}^p$ at $Q^2 \approx 9.5\,$GeV$^2=:s_z$, one has $G_E^{p,d}(s_z) = 2 \,G_E^{p,u}(s_z)$ and hence $G_{E}^n(s_z) = G_E^{p,u}(s_z)>0$, where the last result is evident in Fig.\,7.3 of Ref.\,\cite{Cloet:2013jya}, which shows that although the behaviour of $G_E^{p,u}$ and  $G_{E}^p$ is qualitatively similar, the zero in $G_E^{p,u}$ occurs at a larger value of $Q^2$ than that in $G_{E}^p$ itself.  Under these conditions, any zero in $G_{E}^n$ must occur at a larger value of $Q^2$ than the zero in $G_E^p$: compare the dot-dashed and solid curves in Fig.\,\ref{fig:GEnGEp}

This relative ordering of zeros can change, however, because, in contrast to $G_E^{p,u}$, $G_E^{p,d}$ evolves more slowly with changes in the rate at which the dressed-quark mass function transits from the nonperturbative to the perturbative domain, something which is also apparent in Fig.\,7.3 of Ref.\,\cite{Cloet:2013jya}.  As explained above, this inertia owes to the $d$-quark being preferentially sequestered inside a soft (scalar) diquark correlation.  Subject to these insights, consider Eq.\,\eqref{GEnCS}: with the location of a zero in $G_E^{p,d}$ shifting slowly to larger values of $Q^2$ but that in $G_E^{p,u}$ moving rapidly, one is subtracting from $G_E^{p,d}(Q^2)$ a function whose domain of positive support is becoming increasingly large.  That operation will typically shift the zero in $G_E^n$ to smaller values of $Q^2$ and eventually enable a zero in $G_E^n$ even when that in the $G_E^p$ has disappeared.

The right panel in Fig.\,\ref{fig:GEnGEp} displays a curious effect arising from the faster-than-dipole decrease of the proton's electric form factor (and possible appearance of a zero); namely, there will likely be a domain of $Q^2$ upon which the magnitude of the neutron's electric form factor exceeds that of the proton's.  This being the case, then at some value of momentum transfer the electric form factor of the neutral composite fermion becomes larger than that of its positively charged counterpart.  That occurs at $Q^2 = 4.8 M_N^2$ in our QCD-kindred analysis.

\subsection{Flavour Separation}
\begin{figure}[t]
\begin{center}
\includegraphics[clip,width=0.47\linewidth]{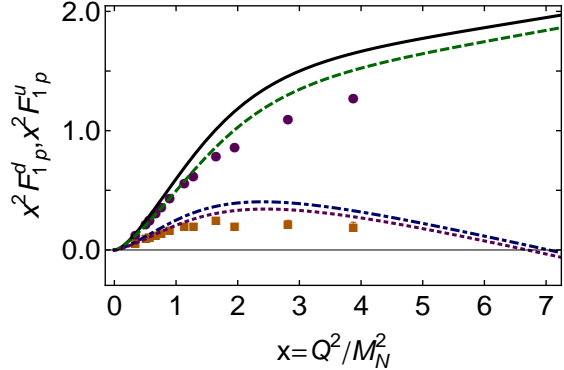}\vspace*
{-1ex }
\caption{\label{fig:flavoursepF1}
Flavour separation of the proton's Dirac form factor as a function of $x=Q^2/m_N^2$: normalisation: $F_{1p}^u(0)=2$, $F_{1p}^d(0)=1$.
\emph{Solid, black} -- $u$-quark obtained using the QCD-kindred framework and the dressed-quark anomalous magnetic moment (DqAMM) described in association with Eq.\,\eqref{eq:DQAMM}; and \emph{dashed, green} -- $u$-quark, DqAMM removed.  \emph{Dot-dashed, blue} -- $d$-quark with DqAMM; and \emph{dotted, purple} -- $d$-quark, DqAMM removed.
The data are from Refs.\,\protect\cite{Zhu:2001md,Bermuth:2003qhS,Warren:2003ma,Glazier:2004ny,
Plaster:2005cx,Riordan:2010id,Cates:2011pz}: $u$-quark, circles; and $d$-quark, squares.
%
}
\end{center}
\end{figure}

\begin{figure}[t]
\begin{center}
\begin{tabular}{cc}
\includegraphics[clip,width=0.47\linewidth]{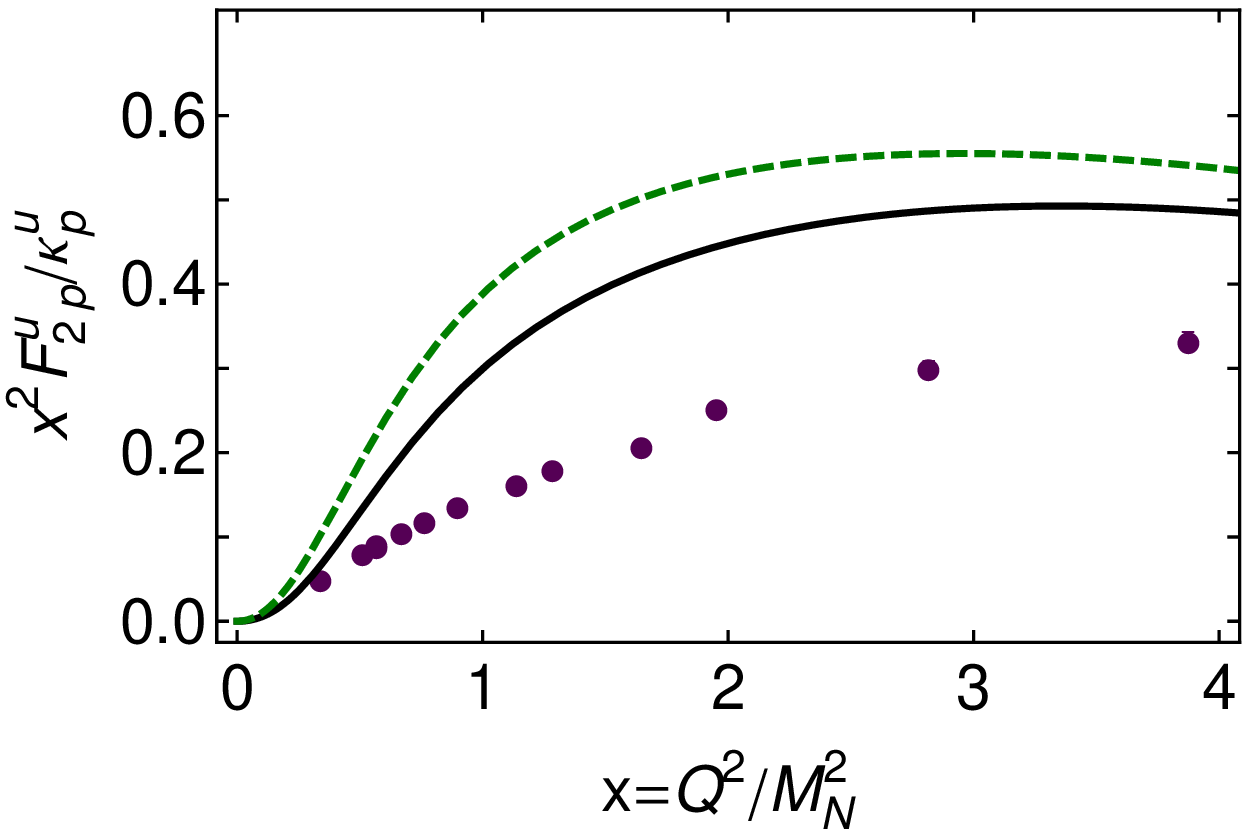} \vspace*{-1ex}
& \includegraphics[clip,width=0.47\linewidth]{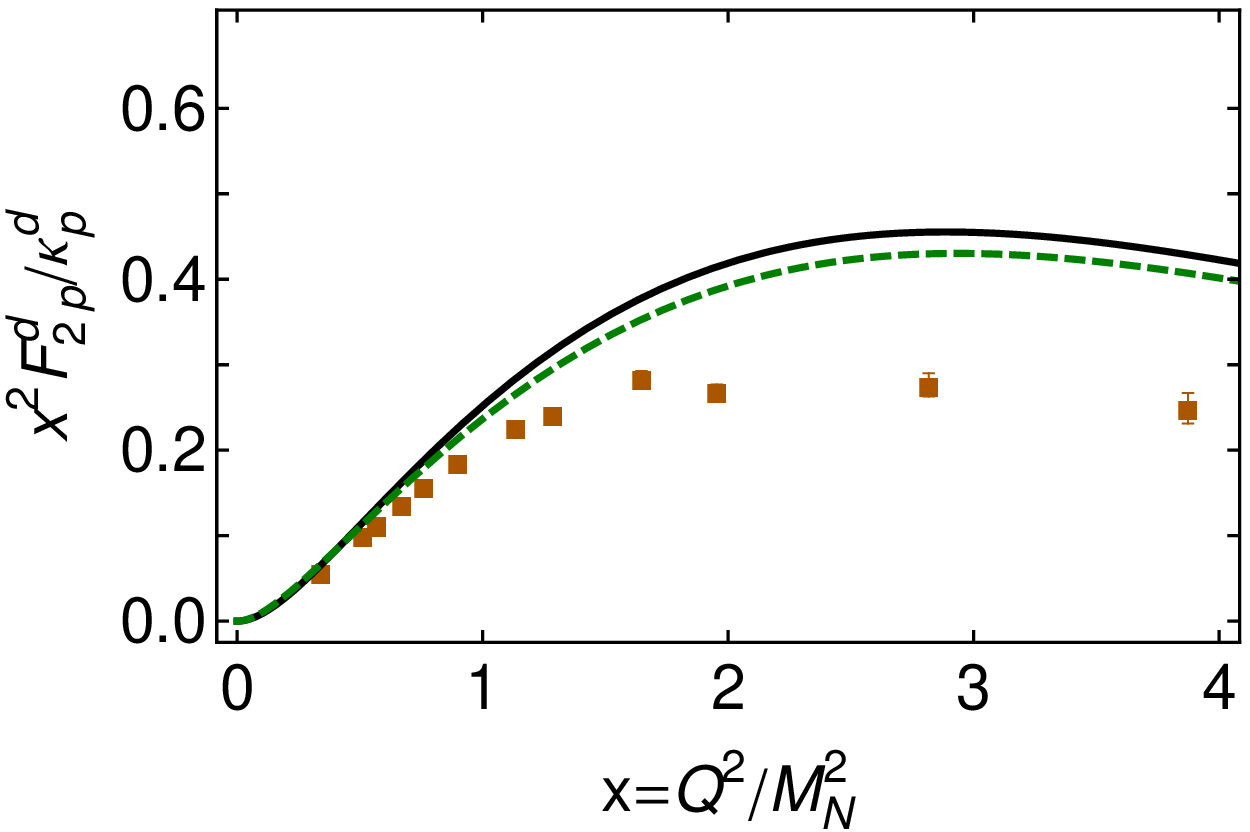} 
\end{tabular}
\caption{\label{fig:flavoursepF2}
Flavour separation of the proton's Pauli form factor, as a function of $x=Q^2/m_N^2$: $u$-quark, left panel; and $d$-quark, right panel.
\emph{Solid, black} -- result obtained using the QCD-kindred framework and the DqAMM described in association with Eq.\,\eqref{eq:DQAMM}; and \emph{dashed, green} -- DqAMM removed.
The data are from Refs.\,\protect\cite{Zhu:2001md,Bermuth:2003qhS,Warren:2003ma,Glazier:2004ny,
Plaster:2005cx,Riordan:2010id,Cates:2011pz}: $u$-quark, circles; and $d$-quark, squares.}
\end{center}
\end{figure}

With precise measurement of the neutron's electric form factor to $Q^2=3.4\,$GeV$^2$ \cite{Riordan:2010id} it has become possible to separate the $u$- and $d$-quark contributions to the nucleon elastic form factors on a sizeable domain \cite{Cates:2011pz}.  This is illustrated in Figs.\,\ref{fig:flavoursepF1} and \ref{fig:flavoursepF2}.  Plainly, in neither the data nor the calculations is the scaling behaviour anticipated from perturbative QCD evident on the momentum domain depicted.  This fact is emphasised by the zero in $F_{1p}^d$, which is also present when a contact interaction is employed \cite{Wilson:2011aa}.  The zero is explained by the presence of diquark correlations in the nucleon.  It was observed in Ref.\,\cite{Cloet:2008re} that the proton's singly-represented $d$-quark is more likely to be struck in association with an axial-vector diquark correlation than with a scalar diquark [see Table~\ref{tab:probabilities} herein], and form factor contributions involving an axial-vector diquark are soft.  On the other hand, the doubly-represented $u$-quark is predominantly linked with harder scalar-diquark contributions.  This interference produces the zero in the Dirac form factor of the $d$-quark in the proton.  The location of the zero depends on the relative probability of finding $1^+$ and $0^+$ diquarks in the proton: with all other things held fixed, the zero moves to smaller values of $x$ with increasing probability for the appearance of an axial-vector diquark \cite{Wilson:2011aa}.

We have included two curves for each form factor depicted Figs.\,\ref{fig:flavoursepF1} and \ref{fig:flavoursepF2}; namely, results obtained with a dressed-quark anomalous magnetic moment (DqAMM) [described in association with Eq.\,\eqref{eq:DQAMM}] and results obtained without the DqAMM.  There is no qualitative difference between the results.  Thus, whilst inclusion of mechanisms that can generate or modify a DqAMM, such as meson-cloud effects, may quantitatively affect the agreement between our prediction and data, they are not the key to explaining the data's basic features.  Instead, as observed above and elsewhere \cite{Roberts:2010hu,Cloet:2011qu,Wilson:2011aa}, the presence of strong diquark correlations within the nucleon is sufficient to explain the empirically verified behaviour of the flavour-separated form factors.

\subsection{Valence-quark Distributions at $x_B=1$}
At this point it is possible to exploit a connection between the $Q^2=0$ values of elastic form factors and the dimensionless structure functions of deep inelastic scattering at Bjorken-$x=:x_B=1$ in order to make predictions for some important properties of the nucleon.  Whilst all familiar parton distribution functions (PDFs) vanish at $x_B=1$, ratios of any two need not; and, under DGLAP evolution, the value of such a ratio is invariant \cite{Holt:2010vj}.  Thus, e.g., with $d_v(x_B)$, $u_v(x_B)$ the proton's $d$, $u$ valence-quark PDFs, the value of $\lim_{x_B\to 1} d_v(x_B)/u_v(x_B)$ is an unambiguous, scale invariant, nonperturbative feature of QCD.  It is therefore a keen discriminator between frameworks that claim to explain nucleon structure.  Furthermore, $x_B=1$ corresponds strictly to the situation in which the invariant mass of the hadronic final state is precisely that of the target; viz., elastic scattering.  The structure functions inferred experimentally on the neighborhood $x_B\simeq 1$ are therefore determined theoretically by the target's elastic form factors.\footnote{The nucleon resonance programme with the CLAS12 detector at JLab will furnish electrocouplings of all prominent $N^\ast$ states with masses less-than 2\,GeV on $Q^2 \leq 5\,$GeV$^2$ \cite{Aznauryan:2012baS}.  The associated analysis will also deliver resonance amplitudes for most excited nucleon states, so that resonant contributions to electron scattering processes may be described by using these experimental results, potentially providing new opportunities for the exploration of inclusive structure functions in the resonance region on $x_B\simeq 1$.}

These observations were exploited in Refs.\,\cite{Wilson:2011aa,Roberts:2013mja} in order to deduce a collection of simple formulae, expressed in terms of the diquark appearance and mixing probabilities listed in Table~\ref{tab:probabilities}, from which one may compute ratios of spin-averaged and longitudinal-spin-dependent $u$- and $d$-quark parton distribution functions on the domain $x_B\simeq 1$:
\begin{subequations}
\label{PDFratios}
\begin{eqnarray}
\label{dvuvF1result}
\left. \frac{d_v(x_B)}{u_v(x_B)}\right|_{x_B\simeq 1} &=& \frac{P_{1}^{p,d}}{P_{1}^{p,u}} =
\frac{\frac{2}{3} P_1^{p,a} + \frac{1}{3} P_1^{p,m}}
{P_1^{p,s}+\frac{1}{3} P_1^{p,a} + \frac{2}{3} P_1^{p,m}},\\
%
%
A_1^p &=& \frac{
\frac{4}{9} (P_{u\uparrow}-P_{u\downarrow}) + \frac{1}{9}(P_{d\uparrow}-P_{d\downarrow})}
{\frac{4}{9} (P_{u\uparrow}+P_{u\downarrow}) + \frac{1}{9}(P_{d\uparrow}+P_{d\downarrow})}\,,
%
\quad A_1^n = \frac{
\frac{4}{9} (P_{d\uparrow}-P_{d\downarrow}) + \frac{1}{9}(P_{u\uparrow}-P_{u\downarrow})}
{\frac{4}{9} (P_{d\uparrow}+P_{d\downarrow}) + \frac{1}{9}(P_{u\uparrow}+P_{u\downarrow})}\,,
\end{eqnarray}
\end{subequations}
where
\begin{equation}
\label{proball}
\begin{array}{l}
P_{u_\uparrow} =
P^{p,s}_{u_\uparrow} + \frac{1}{9} P_1^{p,a} + \frac{1}{3} P_1^{p,m}
= \psi_{L=0}^2 + 2 \psi_{L=0} \psi_{L=1}+ \frac{1}{3} \psi_{L=1}^2 + \frac{1}{9} P_1^{p,a} + \frac{1}{3} P_1^{p,m}, \\[2ex]
P_{u_\downarrow} = P^{p,s}_{u_\downarrow} + \frac{2}{9} P_1^{p,a} + \frac{1}{3} P_1^{p,m} =
\frac{2}{3} \psi_{L=1}^2 + \frac{2}{9} P_1^{p,a} + \frac{1}{3} P_1^{p,m},\\[2ex]
P_{d_\uparrow} = \frac{2}{9} P_1^{p,a} + \frac{1}{6} P_1^{p,m}, \quad
P_{d_\downarrow} = \frac{4}{9} P_1^{p,a} + \frac{1}{6} P_1^{p,m}\,.
\end{array}
\end{equation}
The first line of Eq.\,\eqref{proball} can be understood once one recalls that $P_1^{p,s}$ is the probability for finding a $u$-quark bystander in association with a scalar $[ud]$-diquark correlation in the proton.  Owing to Poincar\'e covariance, this term expresses a sum of quark-diquark angular momentum $L^{u[ud]}=0$ and $L^{u[ud]}=1$ correlations within the nucleon.  With $L^{u[ud]}=0$, the bystander quark carries all the nucleon's spin.  On the other hand, the $L^{q[ud]}=1$ correlation contributes to both the parallel and antiparallel alignment probabilities of the bystander quark: $2 [ud]_{L_z^{u[ud]}=1} u_{\downarrow} \oplus [ud]_{L_z^{u[ud]}=0} u_{\uparrow}$.  The relative strength of these terms is fixed by solving the Faddeev equation and expressed thereafter in the Faddeev amplitude: $\Psi_{0^+} \sim \psi_{L=0} + \psi_{L=1}$, so that, converting the amplitude to probabilities,
\begin{equation}
\label{Pscalar}
\begin{array}{ll}
P_1^{p,s} = P^{p,s}_{u_\uparrow} + P^{p,s}_{u_\downarrow}, & \\
P^{p,s}_{u_\uparrow} = \psi_{L=0}^2 + 2 \psi_{L=0} \psi_{L=1}+ \frac{1}{3} \psi_{L=1}^2, &
P^{p,s}_{u_\downarrow} =  \frac{2}{3} \psi_{L=1}^2.\;
\end{array}
\end{equation}
With the Faddeev equation used herein, based upon a momentum-dependent dressed-quark mass function, $\psi_{L=0}=0.55$, $\psi_{L=1}=0.22$ \cite{Cloet:2008re,Cloet:2007pi}.  On the other hand, in the analysis of Ref.\,\cite{Wilson:2011aa} one finds $\psi_{L=0}=0.88$, $\psi_{L=1}=0$ because that treatment of the contact interaction produces a momentum-independent nucleon Faddeev amplitude.

\begin{table}[t]
\begin{center}
\caption{\label{tab:a}
Selected predictions for the $x_B=1$ value of the indicated quantities.
The DSE results are computed using the formulae in Eqs.\,\eqref{PDFratios}\,--\,\eqref{Pscalar}.
The next four rows are, respectively, results drawn from Refs.\,\protect\cite{Close:1988br,Cloet:2005pp,Hughes:1999wr,Isgur:1998yb}.
The last row, labeled ``pQCD,'' expresses predictions made in Refs.\,\protect\cite{Farrar:1975yb,Brodsky:1994kg}, which are based on an SU$(6)$ spin-flavour wave function for the proton's valence-quarks and assume helicity conservation in their interaction with hard-photons.
}
\begin{tabular*}
{\hsize}
{
l|@{\extracolsep{0ptplus1fil}}
l@{\extracolsep{0ptplus1fil}}
l@{\extracolsep{0ptplus1fil}}
l@{\extracolsep{0ptplus1fil}}
l@{\extracolsep{0ptplus1fil}}
l@{\extracolsep{0ptplus1fil}}
l@{\extracolsep{0ptplus1fil}}
l@{\extracolsep{0ptplus1fil}}}\hline
    & $\frac{F_2^n}{F_2^p}$ & $\frac{d}{u}$ & $\frac{\Delta d}{\Delta u}$
    & $\frac{\Delta u}{u}$ & $\frac{\Delta d}{d}$ & $A_1^n$ & $A_1^p$\\\hline
%
Herein
& $0.50$ & $0.29$ & $-0.12$ & 0.67 & $-0.29$ & 0.16 & 0.61 \\[0.5ex]
DSE-realistic \cite{Cloet:2008re}
& $0.49$ & $0.28$ & $-0.11$ & 0.65 & $-0.26$ & 0.17 & 0.59 \\[0.5ex]
DSE-contact \cite{Roberts:2011wy}
& $0.41$ & $0.18$ & $-0.07$ & 0.88 & $-0.33$ & 0.34 & 0.88\\\hline
$0_{[ud]}^+$ & $\frac{1}{4}$ & 0 & 0 & 1 & 0 & 1 & 1 \\[0.5ex]
NJL & $0.43$ & $0.20$ & $-0.06$ & 0.80 & $-0.25$ & 0.35& 0.77 \\[0.5ex]
SU$(6)$ & $\frac{2}{3}$ & $\frac{1}{2}$ & $-\frac{1}{4}$ & $\frac{2}{3}$ & $-\frac{1}{3}$ & 0 & $\frac{5}{9}$ \\[0.5ex]
CQM & $\frac{1}{4}$ & 0 & 0 & 1 & $-\frac{1}{3}$ & 1 & 1 \\[0.5ex]
pQCD & $\frac{3}{7}$ & $\frac{1}{5}$ & $\frac{1}{5}$ & 1 & 1 & 1 & 1 \\\hline
\end{tabular*}
\end{center}
\end{table}

Equations\,\eqref{PDFratios}\,--\,\eqref{Pscalar} yield the predictions listed in Table~\ref{tab:a}.  In completing the table we used additional simple identities; e.g.,
\begin{equation}
\frac{F_2^n}{F_2^p} = \frac{1+4 d/u}{4-d/u}\,,\;
u=P_{u\uparrow}+P_{u\downarrow}\,\; \delta u=P_{u\uparrow}-P_{u\downarrow}\,,\;\mbox{etc.}
\end{equation}
Given that the Faddeev equation used herein is precisely the same as that in Ref.\,\cite{Cloet:2008re} and the current is only slightly modified, it is natural that there is no meaningful difference between our results and those produced by the solutions in Ref.\,\cite{Cloet:2008re}.

Table~\ref{tab:a} highlights the fact that no single ratio is capable of completely distinguishing between distinct pictures of nucleon structure.  On the other hand, it shows that a comparison between experiment and different predictions for the combination of \emph{all} tabulated quantities provides a very effective means of discriminating between competing descriptions.  Thus, as emphasised in Ref.\,\cite{Roberts:2013mja},  empirical results for unpolarised distributions and longitudinal spin asymmetries on $x_B\simeq 1$ will add greatly to our capacity for identifying the correct dynamical explanation of nucleon structure at accessible energy scales.

\begin{figure}[t]
\begin{center}
\begin{tabular}{cc}
\includegraphics[clip,width=0.46\linewidth]{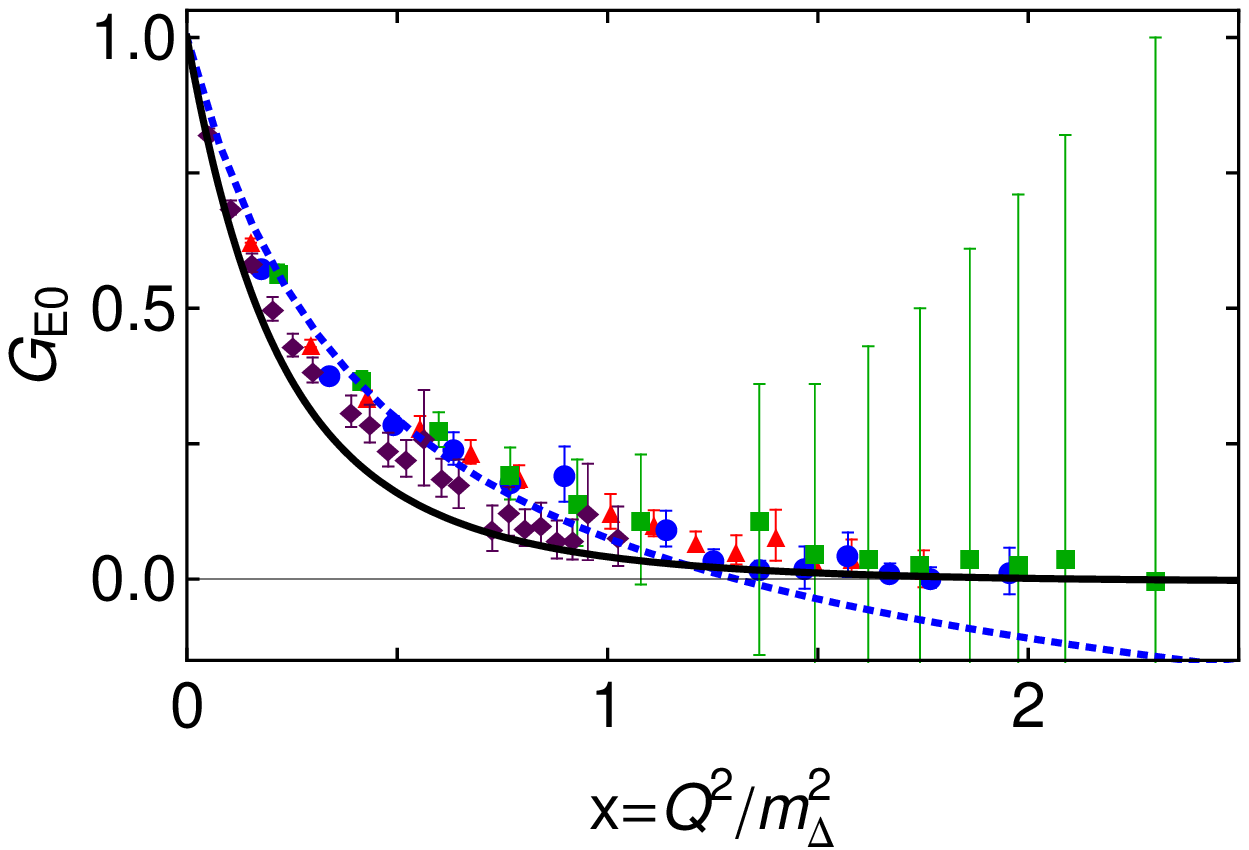}\vspace*
{-1ex } &
\includegraphics[clip,width=0.44\linewidth]{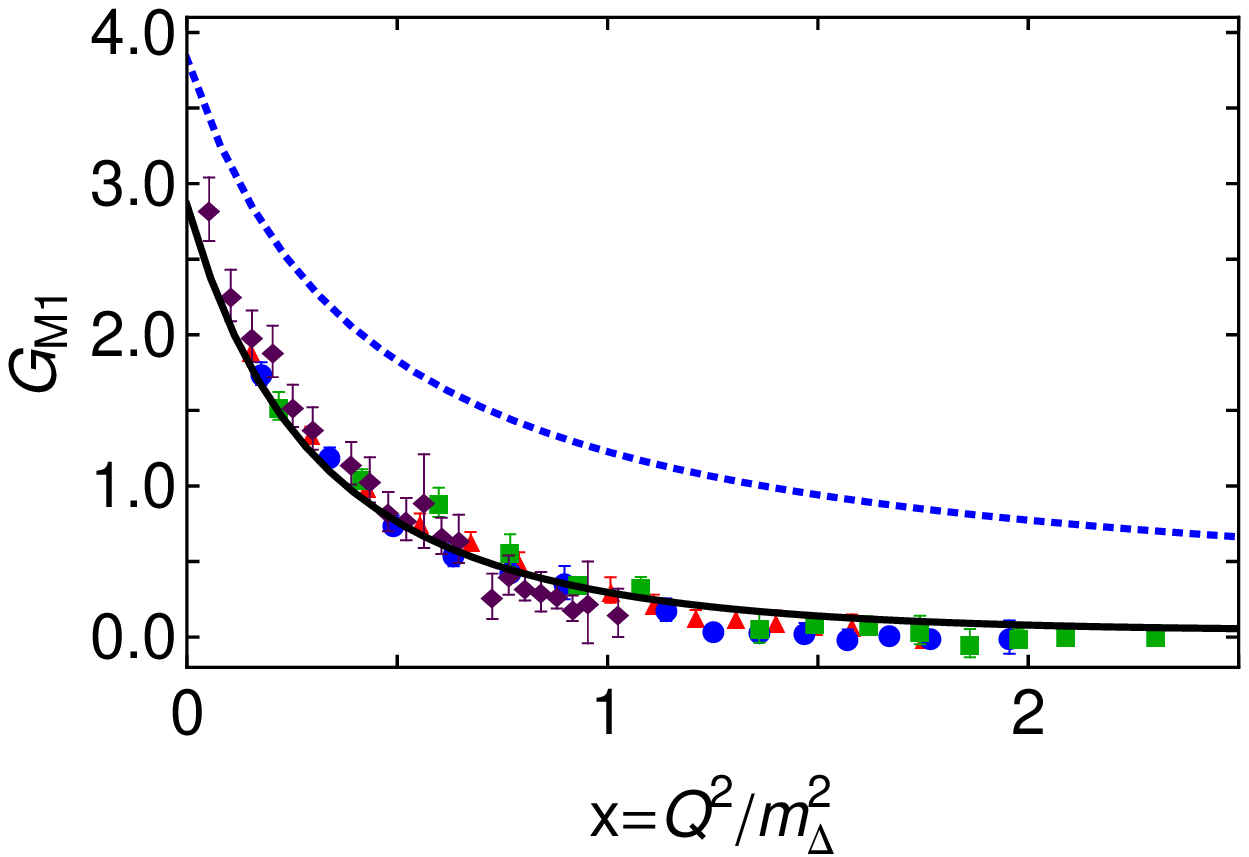}\vspace*
{-1ex} \\
\includegraphics[clip,width=0.44\linewidth]{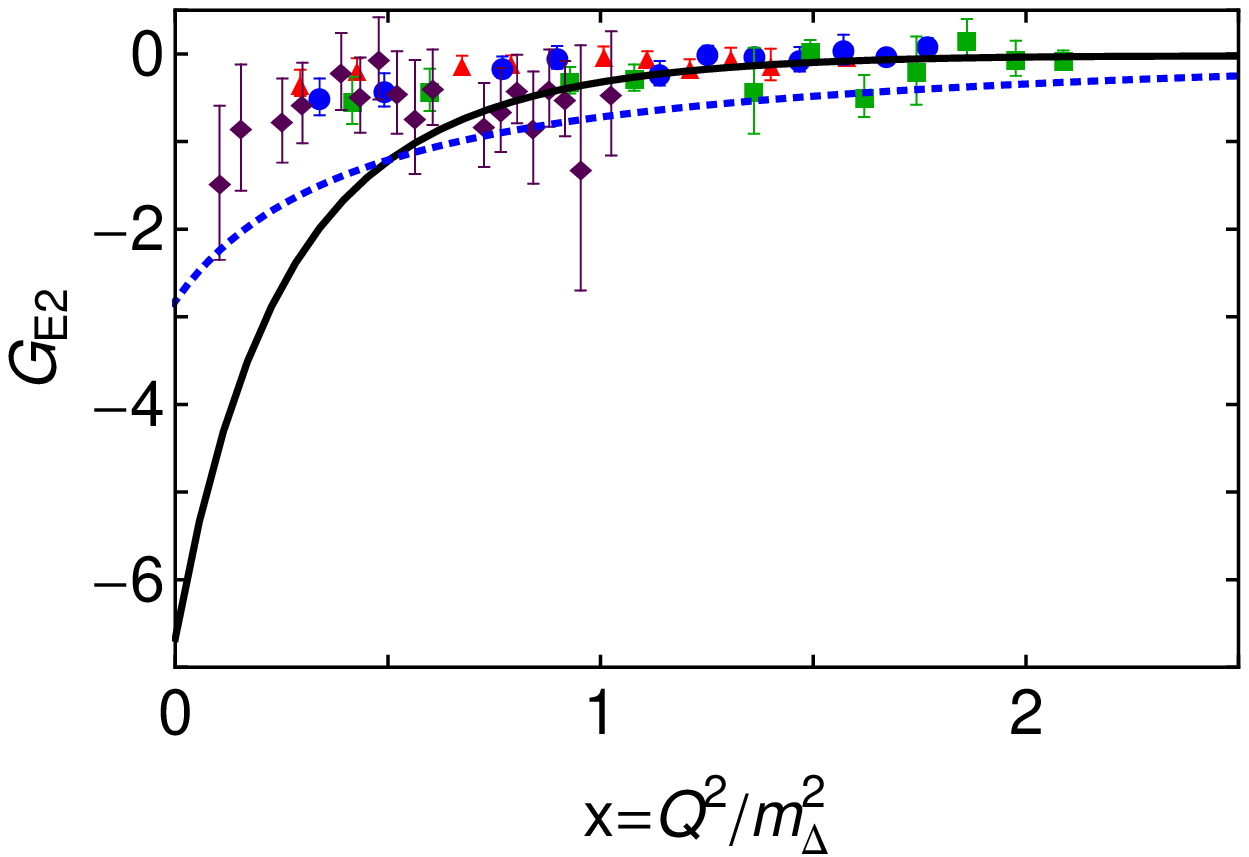}\vspace*
{-1ex} &
\includegraphics[clip,width=0.46\linewidth]{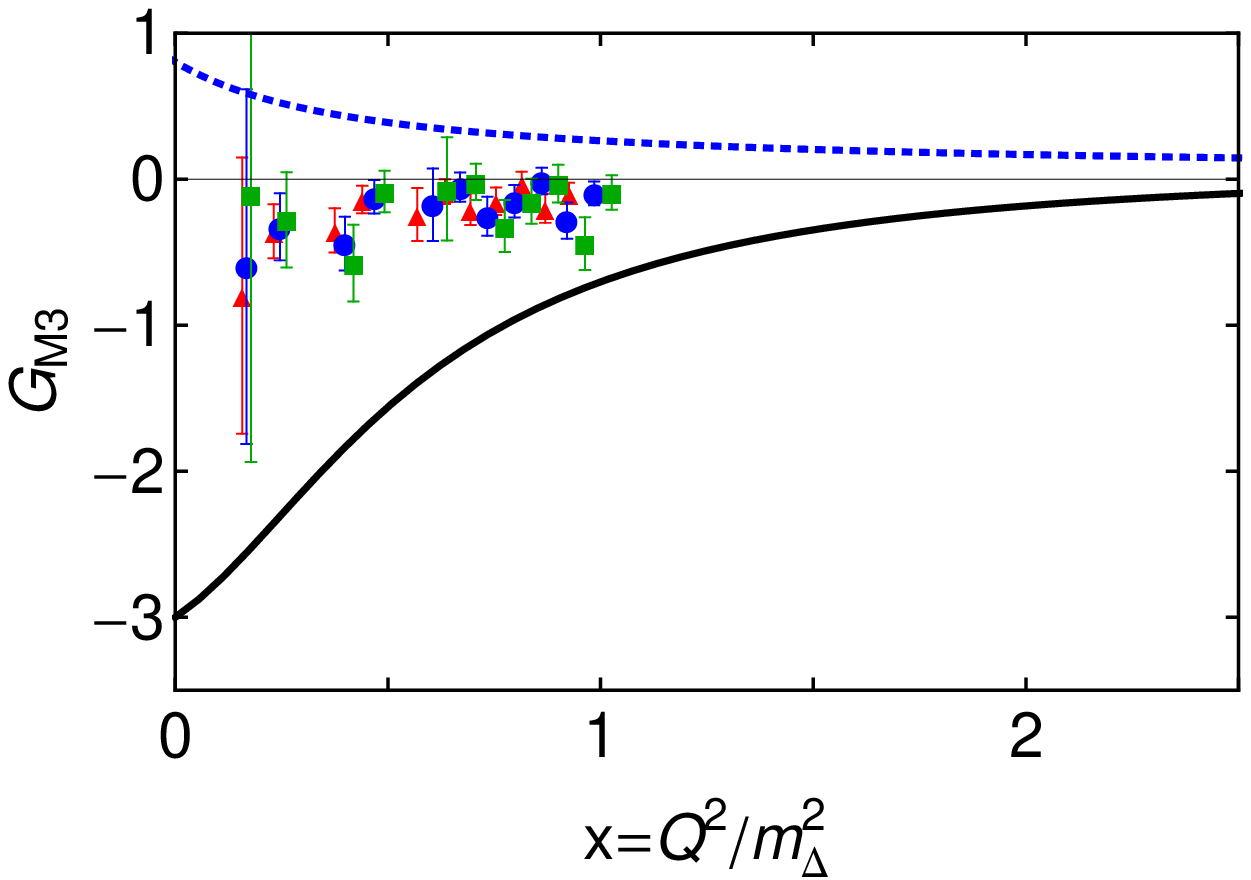}
\end{tabular}
\caption{\label{fig:elasticFFD}
Dressed-quark-core contributions to the $\Delta^{+}$ electromagnetic form factors:
$G_{E0}$, $G_{M1}$, $G_{E2}$ and $G_{M3}$.
Curves in all panels: {\it solid, black} -- result obtained herein, using the QCD-kindred framework; and {\it dotted, blue} -- CI result \cite{Segovia:2013uga}.
In the absence of precise experimental data, the points depict results from numerical
simulations of lattice-regularised QCD:
$G_{E0}$, $G_{M1}$ and $G_{E2}$, unquenched \protect\cite{Alexandrou:2009hs} (red
triangles -- $m_{\pi}=691\,{\rm MeV}$, blue circles -- $m_{\pi}=509\,{\rm MeV}$,
green squares -- $m_{\pi}=384\,{\rm MeV}$, and purple diamonds -- $m_{\pi}=353\,{\rm
MeV}$);
and $G_{M3}$, quenched \protect\cite{Alexandrou:2007we} (red triangles --
$m_{\pi}=563\,{\rm MeV}$, blue circles -- $m_{\pi}=490\,{\rm MeV}$,
green squares -- $m_{\pi}=411\,{\rm MeV}$).}
\end{center}
\end{figure}

\section{\mbox{\boldmath $\Delta$} Elastic Form Factors}
\label{subsec:FFdelta}
The $\Delta$-baryon electromagnetic current is
\begin{equation}
J_{\mu,\alpha\beta}(P_f,P_i) =  \sum_{n\neq 4}
\int\frac{d^4p}{(2\pi)^4} \frac{d^4k}{(2\pi)^4} \;
\bar\Psi_\alpha(-p;P_f) \, J_\mu^n(p,P_f,k,P_i) \, \Psi_\beta(k;P_i)\,,
\label{JDeltaExplicit}
\end{equation}
where $\Psi_\alpha$ is the $\Delta$ Faddeev amplitude and the sum ranges over each one of the six diagrams depicted in Fig.\,\ref{vertex} and detailed in App.\,\ref{NPVertex} \emph{except} Diagram~4, which contributes nothing because the $\Delta$-baryon does not contain a scalar diquark.  This fact also reduces the number of contributions from all the surviving diagrams.  The $\Delta$ elastic form factors are obtained from Eq.\,\eqref{JDeltaExplicit} via the projections in Eq.\,\eqref{eq:scalars}.  Notably, so long as charge symmetry is respected, the elastic electromagnetic form factors of all $\Delta$ charge states are identical up to an overall multiplicative factor that describes the electric charge of the system under consideration.  It follows that the $\Delta^0$ form factors vanish.

We depict our computed dressed-quark-core contributions to the $\Delta^{+}$ elastic electromagnetic form factors in Fig.\,\ref{fig:elasticFFD}.  They are compared with the contact interaction predictions from Ref.\,\cite{Segovia:2013uga} and, since there are no precise experimental data, results obtained via numerical simulations of lattice-regularised QCD: $G_{E0}$, $G_{M1}$ and $G_{E2}$, unquenched \cite{Alexandrou:2009hs}; and $G_{M3}$, quenched only \cite{Alexandrou:2007we}.  The lattice simulation parameters yield the masses listed in Table~\ref{tab:pionmasses} and may be characterised as producing a stable $\Delta(1232)$-baryon with a root-mean-square mass of 1.55\,GeV (unquenched) or 1.43\,GeV (quenched), which are noticeably larger than both the empirical value (1.232\,GeV in Ref.\,\cite{Beringer:1900zz}) and our computed result (1.33\,GeV in Table~\ref{tableNmass}).

It is worth examining separately each panel of Fig.~\ref{fig:elasticFFD}, so first
consider the top-left panel, which displays the $\Delta^+$ electric monopole form
factor.  The solid curve, obtained with QCD-like input, is consistent with the lattice-QCD results on the entire domain and displays a zero at $Q^2=2.1 \,m_\Delta^2$.  As seen with the proton elastic form factors, the contact interaction results \cite{Segovia:2013uga} are harder and exhibit a zero at a smaller value of $Q^2 = 1.3\,m_\Delta^2$.  Studies of the $\rho$-meson's electric form factor using similar interactions also produce a zero \cite{Hawes:1998bz,Bhagwat:2006pu,Roberts:2011wy}.  In all these cases; viz., $\rho^+$, proton, $\Delta^+$, the expressions for the electric form factors involve a destructive interference, with one or more negative contributions magnified by $Q^2$.  This interference does not guarantee a zero in the electric form factor of a charged $J\geq 1/2$ bound-state but it does suggest both that a zero might be difficult to avoid and, as highlighted above and elsewhere \cite{Wilson:2011aa,Cloet:2013gva}, that its appearance and location are a sensitive measure of the dynamics which underlies the bound-state's structure.

\begin{table}[t]
\begin{center}
\caption{\label{tab:pionmasses} Masses, in GeV, of the $\pi$, $\rho$ and $\Delta$, computed in the numerical simulations of lattice-regularised QCD that produced the points in Fig.\,\protect\ref{fig:elasticFFD}.}
\begin{tabular*}
{\hsize}
{
l@{\extracolsep{0ptplus1fil}}
c@{\extracolsep{0ptplus1fil}}
c@{\extracolsep{0ptplus1fil}}
c@{\extracolsep{0ptplus1fil}}}\hline
Approach & $m_{\pi}$ & $m_{\rho}$ & $m_{\Delta}$ \\\hline
Unquenched I   \protect\cite{Alexandrou:2009hs} & $0.691$ & $0.986$ & $1.687$ \\
Unquenched II  \protect\cite{Alexandrou:2009hs} & $0.509$ & $0.899$ & $1.559$ \\
Unquenched III \protect\cite{Alexandrou:2009hs}& $0.384$ & $0.848$ & $1.395$ \\
Hybrid         \protect\cite{Alexandrou:2009hs}& $0.353$ & $0.959$ & $1.533$ \\\hline
Quenched I    \protect\cite{Alexandrou:2007we} & $0.563$ & $0.873$ & $1.470$ \\
Quenched II   \protect\cite{Alexandrou:2007we} & $0.490$ & $0.835$ & $1.425$ \\
Quenched III  \protect\cite{Alexandrou:2007we} & $0.411$ & $0.817$ & $1.382$ \\
\hline
\end{tabular*}
\end{center}
\end{table}

Given the electric form factor, one can readily compute a $\Delta^+$ charge radius:
\begin{equation}
\left\langle\right.\!\!r_{E0}^{2}\!\!\left.\right\rangle =
- 6 \left.\frac{dG_{E0}}{dQ^{2}}\right|_{Q^{2}=0} \approx 1.5\, \langle r_p^2\rangle\,,
\label{DeltaRadius}
\end{equation}
where $r_p=0.61\,$fm when computed from the solid curve in Fig.\,\ref{fig:FFNucleon1} [see Eq.\,\eqref{nucleonstatic}].  Thus, the electromagnetic size of the $\Delta^+$-baryon's dressed-quark-core is greater than that of the proton.\footnote{N.B.\ The dimensionless product $m_\rho^2 r_{\Delta^+}^2$ computed using lattice-QCD is very sensitive to $m_\pi^2$: it grows rapidly as $m_\pi^2$ is decreased.  Therefore, in the absence of simulations at realistic masses, we choose not to report a lattice value for $r_{\Delta^+}$.}
To pursue this further, we recall the analysis in Refs.\,\cite{Buchmann:1996bd} and thus provide the following comparison:\footnote{Precise equality; viz., $r_{\Delta^+}^2 = \langle r_p^2\rangle - \langle r_n^2\rangle$, is a prediction of the nonrelativistic chiral constituent-quark model and associated current constructed from numerous ingredients in Ref.\,\cite{Buchmann:1996bd}.  Slightly modified relations have been obtained using a large-$N_c$ analysis \protect\cite{Buchmann:2000wf}.}
\begin{equation}
r_{\Delta^+}^2 = 1.5\,\langle r_p^2\rangle \quad \mbox{cf.} \quad
\langle r_p^2\rangle - \langle r_n^2\rangle = 1.2 \,\langle r_p^2\rangle
\end{equation}
viz., the $\Delta^+$ dressed-quark-core charge-radius-squared computed using our QCD-kindred framework differs by 25\% from the isovector combination of nucleon dressed-quark-core radii-squared calculated in the same approach.  Whilst this value for the difference is consistent with approximate equality, it is six-times greater than that obtained when the contact interaction is used.  The explanation for this lies in the far richer diquark correlation structures generated by QCD-like propagators in the Faddeev amplitudes and current.

The top-right panel of Fig.~\ref{fig:elasticFFD} depicts the $\Delta^+$ magnetic
dipole form factor.  The computed dimensionless magnetic moment $\hat\mu_{\Delta^+}=G_{M1}(Q^2=0)$ is listed in Table~\ref{tab:comparative}.  Notably, the value of $\hat\mu_{\Delta^+}$ is dynamical; i.e., it is not constrained by any symmetry, and whereas results obtained with the contact interaction lie uniformly above the lattice output, those obtained herein using QCD-like propagators and vertices agree with the lattice results.  With both continuum interactions, however, the level of agreement is influenced to some extent by the range of lattice-QCD masses for the pion, $\rho$-meson and $\Delta$-baryon in Table~\ref{tab:pionmasses}, which are too large.  We will return to this point.

\begin{table*}[t]
\begin{center}
\caption{\label{tab:comparative}
Static electromagnetic properties of the $\Delta^+(1232)$: the first row reports results obtained with QCD-like momentum-dependent quark dressing and the second row lists results obtained with the contact interaction  \cite{Segovia:2013uga}.
The experimental value for $G_{M1}(Q^{2}=0)$ is drawn from Ref.\,\protect\cite{Beringer:1900zz};
and the remaining rows report a representative selection of results from other calculations.  The $Q^2=0$ values associated with lattice simulations were obtained by fitting the available results and extrapolating (see Fig.\,\protect\ref{fig:elasticFFD}).
N.B.\ A symbol ``-'' in any location indicates that no result was reported for that quantity in the related reference.}
\end{center}
\begin{tabular*}
{\hsize}
{
l|@{\extracolsep{0ptplus1fil}}
c@{\extracolsep{0ptplus1fil}}
l@{\extracolsep{0ptplus1fil}}
l@{\extracolsep{0ptplus1fil}}
l@{\extracolsep{0ptplus1fil}}}\hline
Approach & Reference & $G_{M1}(0)$ & $G_{E2}(0)$ & $G_{M3}(0)$\\\hline
DSE-herein & & $+2.86$ & $-6.67$ & $-3.00$ \rule{-0.6em}{0ex} \rule{0em}{2.2ex} \\\hline
DSE-contact \cite{Segovia:2013uga}
 & & $+3.83$ & $-2.82$ &  $+0.80$ \rule{-0.6em}{0ex} \rule{0em}{2.2ex} \\\hline
%
%
Exp & \cite{Beringer:1900zz} & $+3.6^{+1.3}_{-1.7}\pm2.0\pm4$ & - & - \rule{-0.6em}{0ex} \rule{0em}{2.2ex}\\\hline
Lattice-QCD (hybrid) & \cite{Alexandrou:2009hs}
& $+3.0\pm0.2$ & $-2.06^{+1.27}_{-2.35}$ & $~0.00$ \rule{-0.6em}{0ex} \rule{0em}{2.2ex}\\
$1/N_{c}+ N\to \Delta$ &\cite{Alexandrou:2009hs}
& - & $-1.87 \pm 0.08$ & - \\
%
Faddeev equation & \cite{Maris:1999nt,Sanchis-Alepuz:2013iia} & $+2.38$ & $-0.67$ & $>0$\\
Covariant $\chi$PT  & \cite{Geng:2009ys}
& $+3.74\pm0.03$ & $-0.9\pm0.6$ & $-0.9\pm2.1$ \\
$+\,$qLQCD & \cite{Boinepalli:2009sq} & & & \\
QCD-SR                     & \cite{Aliev:2009np} & $+4.2\pm1.1$ & $-0.6\pm0.2$ & $-0.7\pm0.2$ \\
$\chi$QSM                  & \cite{Ledwig:2008es} & $+3.1$ & $-2.0$ & - \\
General Param. Method  & \cite{Buchmann:2002xq,Buchmann:2008zza}
& - & $-4.4$ & $-2.6$ \\
QM$+$exchange-currents & \cite{Buchmann:1996bd} & $+4.6$ & $-4.6$ & - \\
$1/N_{c}+m_s$-expansion              & \cite{Luty:1994ub} & $+3.8\pm0.3$ & - & - \\
RQM                        & \cite{Schlumpf:1993rm} & $+3.1$ & - & - \\
HB$\chi$PT                 & \cite{Butler:1993ej} & $+2.8\pm0.3$ & $-1.2\pm0.8$ & - \\
nrCQM                        & \cite{Krivoruchenko:1991pm} & $+3.6$ & $-1.8$ & - \\
\hline
\end{tabular*}
\end{table*}

The bottom-left panel in Fig.\,\ref{fig:elasticFFD} displays our calculated  $\Delta^+$ electric quadrupole form factor.  The prediction computed using QCD-kindred input agrees with the lattice results on the domain $x>0.5$ but is significantly larger in magnitude on $0<x<0.5$, a feature it shares with the CI result.
The dimensionless quadrupole moment $\hat{\mathpzc{Q}}_{\,\Delta}=G_{E2}(Q^2=0)$ is listed in Table~\ref{tab:comparative}.  The value is negative and possesses a magnitude that is roughly a factor of three larger than the quoted lattice result.  In the latter connection, the large lattice masses (see Table~\ref{tab:pionmasses}) are playing a role.

The bottom-right panel in Fig.\,\ref{fig:elasticFFD} displays the $\Delta^+$ magnetic octupole form factor.  In this case, only quenched lattice results are available \cite{Alexandrou:2007we}: they favour a negative form factor but are consistent with zero.  The prediction obtained with QCD-like input (solid curve) is negative, too, but much larger in magnitude.  As we shall see, the question posed by this mismatch is again answered by the unrealistically large value of the lattice results for $m_\pi$, $m_\rho$, $m_\Delta$ in Table~\ref{tab:pionmasses}.

Another conflict is also noticeable in the bottom-right panel of Fig.\,\ref{fig:elasticFFD}; namely, the contact-interaction produces a positive form factor.  This difference between the continuum results is readily understood.
The treatment of the contact interaction in Ref.\,\cite{Segovia:2013uga} produces a $\Delta$-baryon Faddeev amplitude that is independent of relative momentum.  In this case, of the eight possible Dirac structures in the $\Delta$ amplitude [Eq.\,\eqref{eq:DlP}], only one survives; viz., that which corresponds to an $S$-wave in the nonrelativistic limit.  If the $\Delta$ were truly a nonrelativistic $S$-wave state, then one would find $G_{M3}\equiv 0$.  Consequently, the nonzero value results from relativistic effects in the amplitude and current.  In contrast, the Faddeev equation constructed with QCD-like input supports all eight structures in the $\Delta$ Faddeev amplitude.  If one retains only that term which corresponds to an $S$-wave in the nonrelativistic limit, then this QCD-kindred framework also yields $G_{M3}> 0$.  The full result is negative owing to contributions from the $P$- and $D$-wave components in the $\Delta$ amplitude and interference effects in the coherent sum that describes the complete current, Eq.\,\eqref{JDeltaExplicit}.

It is now interesting to address the $m_\Delta$-dependence of the $\Delta(1232)$ elastic form factors.  The pattern of pion, $\rho$-meson and $\Delta$-baryon masses in Table~\ref{tab:pionmasses} matches that explained in Refs.\,\cite{Eichmann:2008ae,Nicmorus:2010sd}.  The momentum-dependent interaction therein, based on Refs.\,\cite{Eichmann:2008ae,Maris:1999nt}, produces $m_\Delta = 1.73\,$GeV; i.e., both type-1 and type-2 meson-cloud corrections are omitted.
We have therefore recomputed the $\Delta(1232)$-baryon elastic form factors by using $m_\Delta=1.73\,{\rm GeV}$.  We did this in two ways.  (M1): In an internally consistent computation, we increased $m_{1^+}$ to $1.1\,$GeV and solved the Faddeev equation, producing $m_\Delta=1.73\,$GeV and a dynamically altered Faddeev amplitude.  (M2): As an alternative, we kept the Faddeev amplitude unchanged and simply shifted $m_\Delta \to 1.73\,$GeV in the kinematic expressions that define the elastic form factors, Eqs.\,\eqref{eq:Gammamualbe}\,--\,\eqref{DEprojections}.  M1 allows the correlations that comprise the $\Delta(1232)$-baryon to evolve with a change in constituent masses whereas M2 assumes that the internal structure of the $\Delta(1232)$-baryon's dressed-quark-core is insensitive to changes in the masses of the constituent degrees-of-freedom.

\begin{figure}[t]
\begin{center}
\begin{tabular}{cc}
\includegraphics[clip,width=0.46\linewidth]{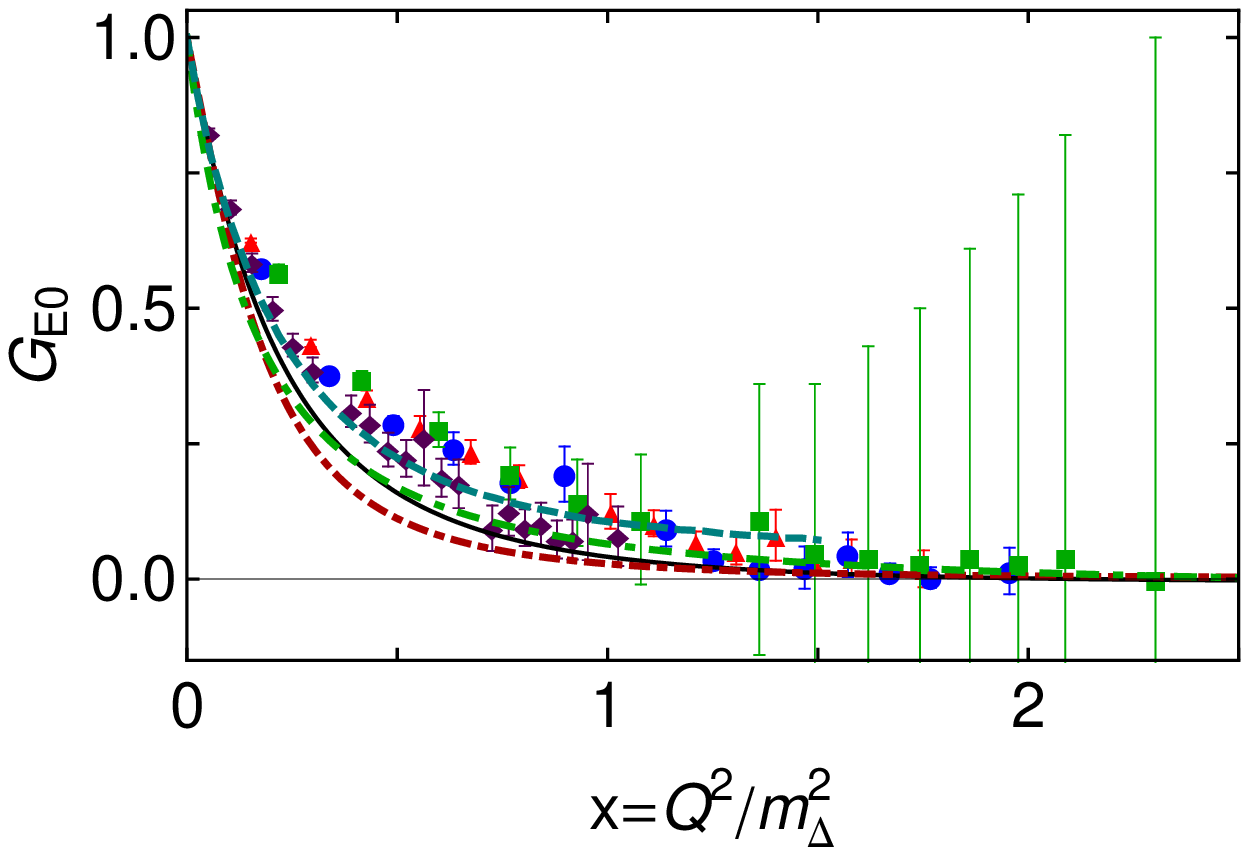}\vspace*
{-1ex } &
\includegraphics[clip,width=0.44\linewidth]{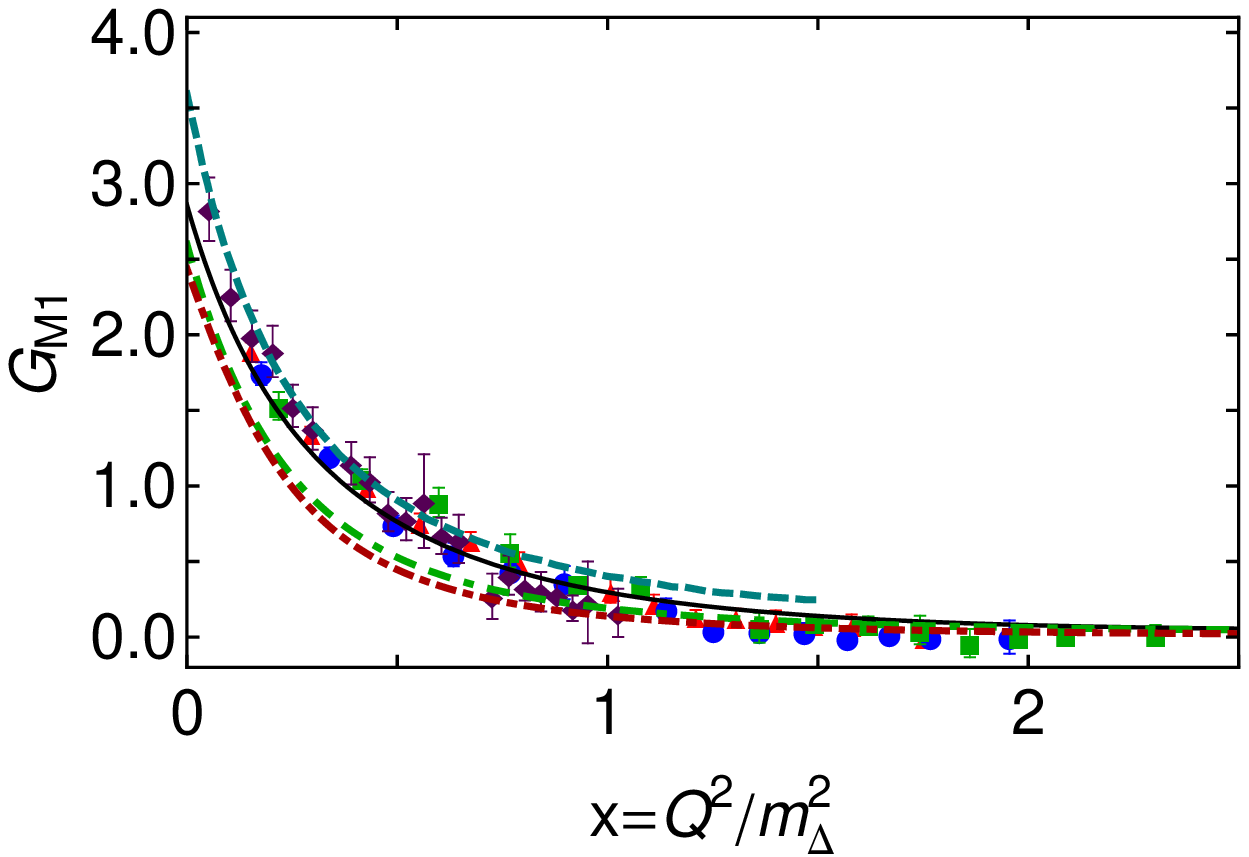}\vspace*
{-1ex} \\
\includegraphics[clip,width=0.44\linewidth]{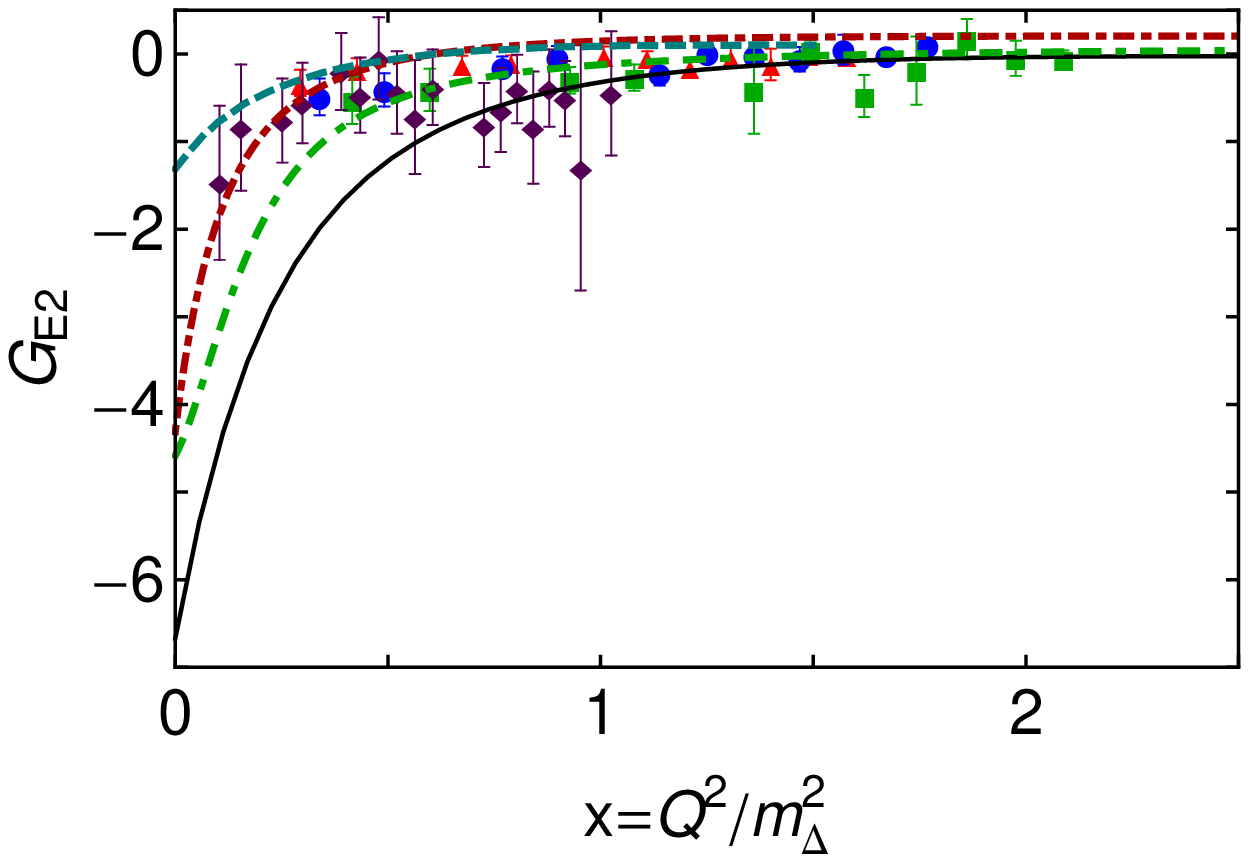}\vspace*
{-1ex} &
\includegraphics[clip,width=0.46\linewidth]{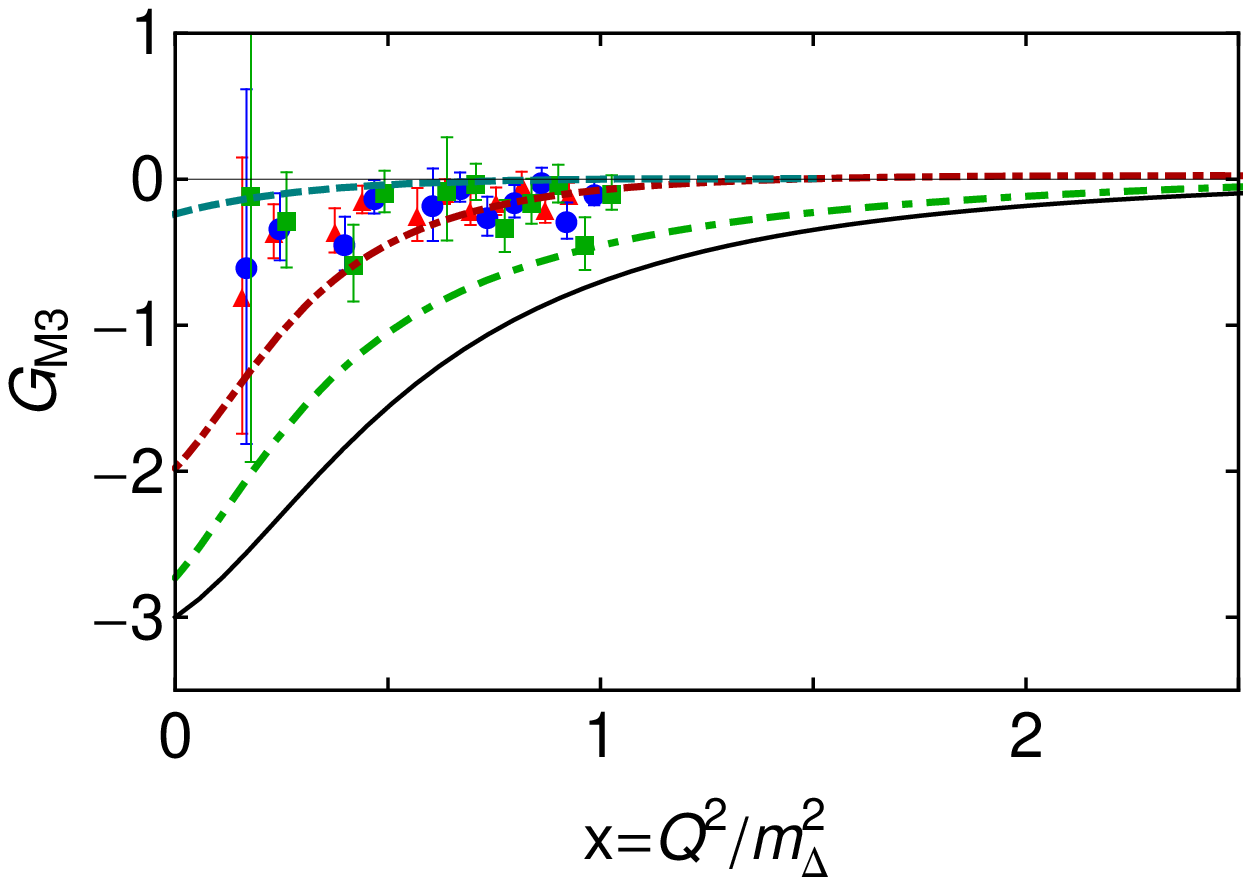}
\end{tabular}
\caption{\label{fig:hugemassFFD}
Dressed-quark-core contributions to the $\Delta^{+}$ electromagnetic form factors:
$G_{E0}$, $G_{M1}$, $G_{E2}$ and $G_{M3}$.
Curves in all panels: {\it solid, black} -- results depicted in Fig.\,\ref{fig:elasticFFD}, which were obtained herein using the QCD-kindred framework and $m_\Delta=1.33\,$GeV from Table~\ref{tableNmass}; {\it dot-dash-dashed, green} -- dynamically modified result, M1, obtained with same method but $m_\Delta=1.73\,$GeV; {\it dot-dashed, red} -- kinematically modified result, M2, obtained with $m_\Delta=1.73\,$GeV in Eqs.\,\eqref{eq:Gammamualbe}\,--\,\eqref{DEprojections} only; and \emph{dashed, cyan} -- computed results from Ref.\,\protect\cite{Nicmorus:2010sd}, which were obtained with $m_\Delta=1.73\,$GeV.  (The numerical algorithm employed in Ref.\,\protect\cite{Nicmorus:2010sd} limits the calculations to $x \lesssim 1.5$.)
The points depict results from numerical simulations of lattice-regularised QCD:
$G_{E0}$, $G_{M1}$ and $G_{E2}$, unquenched \protect\cite{Alexandrou:2009hs} (red triangles -- $m_{\pi}=691\,{\rm MeV}$, blue circles -- $m_{\pi}=509\,{\rm MeV}$, green squares -- $m_{\pi}=384\,{\rm MeV}$, and purple diamonds -- $m_{\pi}=353\,{\rm MeV}$);
and $G_{M3}$, quenched \protect\cite{Alexandrou:2007we} (red triangles -- $m_{\pi}=563\,{\rm MeV}$, blue circles -- $m_{\pi}=490\,{\rm MeV}$,
green squares -- $m_{\pi}=411\,{\rm MeV}$).}
\end{center}
\end{figure}

The results are depicted in Fig.\,\ref{fig:hugemassFFD}.  The top row shows that when QCD-like propagators and vertices are used to construct the Faddeev equation, then $G_{E0}$ and $G_{M1}$ are almost insensitive to increasing $m_\Delta$ by 30\%, independent of whether M1 or M2 is used to effect the change.  Using the contact interaction, on the other hand, $G_{M1}$ reacts rather more strongly.  Nevertheless, with both the contact interaction and QCD-like input, the biggest impact is on $G_{E2}$ and $G_{M3}$, the form factors that describe deformation of the $\Delta$-baryon: increasing $m_\Delta$ by 30\% has the effect of markedly suppressing the deformation, thereby shifting the continuum predictions toward the lattice-QCD results.  It appears, therefore, that the lattice results are materially affected by the kinematic impact of an unrealistically large mass for the $\Delta(1232)$ so that it is misleading to infer too much about the empirical $\Delta(1232)$ resonance from existing lattice studies.

The similarity between the form factors obtained using M1 and M2 is also informative: it highlights the simplicity of the $\Delta(1232)$-baryon's Faddeev amplitude.  This was already noted in Ref.\,\cite{Roberts:2011cf}, which showed that the $\Delta$-baryon's dressed-quark-core is accurately described as an almost non-interacting system of a dressed-quark and axial-vector diquark over a large range of current-quark masses.

An examination of Table~\ref{tab:comparative} is instructive.  Omitting our computations for the present, results from a diverse array of analyses are presented.  If they are weighted equally, then one obtains a mean value of $\hat \mu_{\Delta^+} = 3.5$, a median value of $3.7$ and a standard deviation of $0.7$.  Including our result, then one has a mean of $3.5$, a median of $3.6$ and a standard deviation of $0.7$.  So, there is fair agreement between the theoretical predictions.
With $\hat{\mathpzc{Q}}_{\;\Delta^+}$, on the other hand, one obtains a median value of $(-1.9)$ and a mean value of $(-2.5)$ with a standard deviation of $1.9$, so this quantity must be called uncertain.
There is plainly no consensus on the octupole moment.  However, the results we obtained with Faddeev equation solutions based on QCD-like propagators and vertices indicate that a negative value should be associated with the dressed-quark-core.  This differs from the CI prediction for the reasons described in connection with the bottom-right panel in Fig.\,\ref{fig:elasticFFD}.

It is common to attempt to interpret a nonzero electric quadrupole moment with deformation of the bound-state's charge distribution.  The robust indication from Table~\ref{tab:comparative} is $\hat{\mathpzc{Q}}_{\;\Delta^+}<0$: analyses with and without a meson-cloud agree on this sign.  If one supposes that for the $\Delta$-baryon the Fourier transform of a Breit-frame momentum-space form factor is, at least for small momentum transfers, a reasonable approximation to the configuration space charge distribution, then the negative value indicates an oblate deformation of the $\Delta^+$. It is notable that our QCD-kindred framework produces a strongly deformed $\Delta^+$.  This signals the presence of significant quark orbital-angular-momentum correlations in the $\Delta$ Faddeev amplitude that we have computed.
However, as indicated by Table~\ref{tab:comparative}, this is not a necessary outcome: rather, it depends sensitively on the structure of the Faddeev equation's kernel.  Consequently, at this stage we hesitate to draw any firm physical conclusions based on the size of $\hat{\mathpzc{Q}}_{\;\Delta^+}$.  Instead we merely observe that bound-state solutions obtained using rainbow-ladder (RL) truncation typically underestimate the size of DCSB-induced angular momentum correlations \cite{Chang:2009zb,Chang:2010hb,Chang:2011ei,Bashir:2011dp,Qin:2013mta,Qin:2014vya}.  In contrast, our formulation of the Faddeev equation and current, based on propagator parametrisations and vertices fitted to empirical information \cite{Burden:1995ve,Hecht:2000xa}, implicitly includes effects that transcend the RL truncation.


\section{\mbox{\boldmath $\gamma^\ast N \to \Delta$} Transition}
\label{subsec:FFnucdel}
One may write the $\gamma N\to \Delta$ transition current as
\begin{equation}
J_{\mu,\alpha}(P_f,P_i) =  \sum_{n=1}^6
\int\frac{d^4p}{(2\pi)^4} \frac{d^4k}{(2\pi)^4} \;
\bar\Psi_\alpha(-p;P_f) \, J_{\mu}^n(p,P_f,k,P_i) \, \Psi(k;P_i)\,,
\label{JDeltaNExplicit}
\end{equation}
where $\Psi_\alpha$, $\Psi$ are, respectively, the $\Delta$ and nucleon Faddeev amplitudes.  The sum in Eq.\,\eqref{JDeltaNExplicit} ranges over each one of the six diagrams depicted in Fig.\,\ref{vertex} and detailed in App.\,\ref{NPVertex}.  However, since the $\Delta$-baryon does not contain a scalar diquark correlation, the number of individual terms to be computed is much smaller than is the case for the nucleon elastic form factor. The transition form factors are obtained from Eq.\,\eqref{JDeltaNExplicit} via the projections in Eq.\,\eqref{GMGEGC}.

In considering the behaviour of the $\gamma^\ast N \to \Delta$ transition form factors, it is useful to begin by recapitulating upon a few facts.  Note then that in analyses of baryon electromagnetic properties, using a quark model framework which implements a current that transforms according to the adjoint representation of
spin-flavor $SU(6)$, one finds simple relations between magnetic-transition matrix
elements~\cite{Beg:1964nm,Buchmann:2004ia}:
\begin{equation}
\label{eqBeg}
 \langle p | \mu | \Delta^+\rangle = -\langle n | \mu | \Delta^0\rangle\,,\quad
 \langle p | \mu | \Delta^+\rangle = - \surd 2 \langle n | \mu | n \rangle\,;
\end{equation}
i.e., the magnetic components of the $\gamma^\ast p \to \Delta^+$ and $\gamma^\ast n
\to \Delta^0$ are equal in magnitude and, moreover, simply proportional to the
neutron's magnetic form factor.  Furthermore, both the nucleon and $\Delta$ are
$S$-wave states (neither is deformed) and hence $G_{E}^{\ast} \equiv 0 \equiv
G_{C}^{\ast}$~\cite{Alexandrou:2012da}.

The second entry in Eq.\,\eqref{eqBeg} is consistent with perturbative QCD (pQCD) \cite{Carlson:1985mm} in the following sense: both suggest that $G_{M}^{\ast p}(Q^2)$ should decay with $Q^2$ at the same rate as the neutron's magnetic form factor, which is dipole-like in QCD.  It is often suggested that this is not the case empirically
\cite{Aznauryan:2011ub,Aznauryan:2011qj}.  However, as argued elsewhere \cite{Segovia:2013rca,Segovia:2013uga}, such claims arise from a confusion between the form factors defined in the Ash \cite{Ash1967165} and Jones-Scadron \cite{Jones:1972ky} conventions:
\begin{equation}
\label{DefineAsh}
G_{M,Ash}^{\ast}(Q^2)= G_M^{\ast}(Q^2)/[1+Q^2/t_+ ]^{1/2},
\end{equation}
where $G_M^{\ast}(Q^2)$ is the Jones-Scadron form factor in Eq.\,\eqref{GMGEGC}.

In addition, helicity conservation arguments within the context of pQCD enable one to make \cite{Carlson:1985mm} the follow predictions for the ratios in Eq.\,\eqref{eqREMSM}:
\begin{equation}
\label{eqUVREMSM}
R_{EM} \stackrel{Q^2\to\infty}{=} 1 \,,\quad
R_{SM} \stackrel{Q^2\to\infty}{=} \,\mbox{\rm constant}\,,
\end{equation}
up to $\ln^2 Q^2$ corrections \cite{Idilbi:2003wj}.  These predictions are in marked disagreement with the outcomes produced by $SU(6)$-based quark models: $R_{EM} \equiv 0 \equiv R_{SM}$.  More importantly, they are inconsistent with available data \cite{Aznauryan:2011ub,Aznauryan:2011qj}.

\begin{figure}[t]
\begin{center}
\begin{tabular}{cc}
\includegraphics[clip,width=0.47\linewidth]{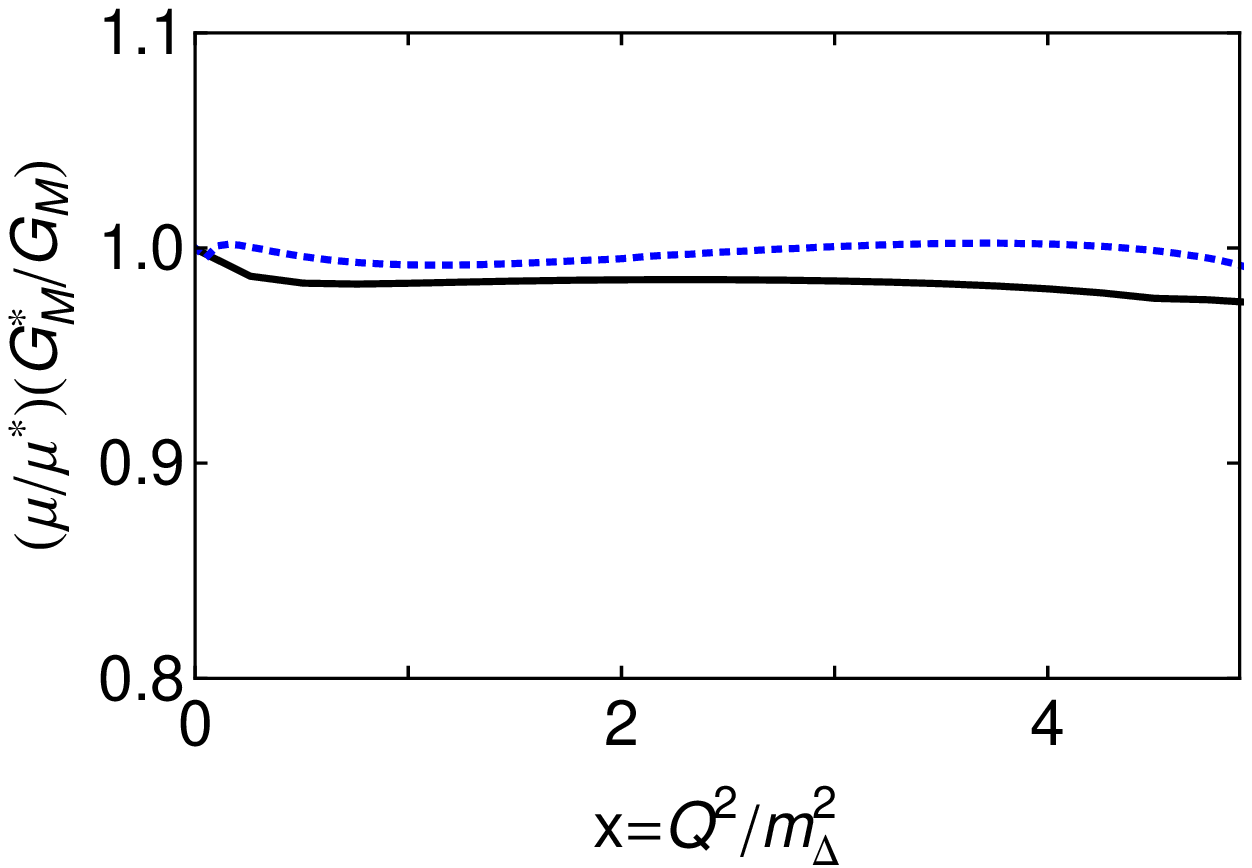} \vspace*{-1ex}
& \includegraphics[clip,width=0.47\linewidth]{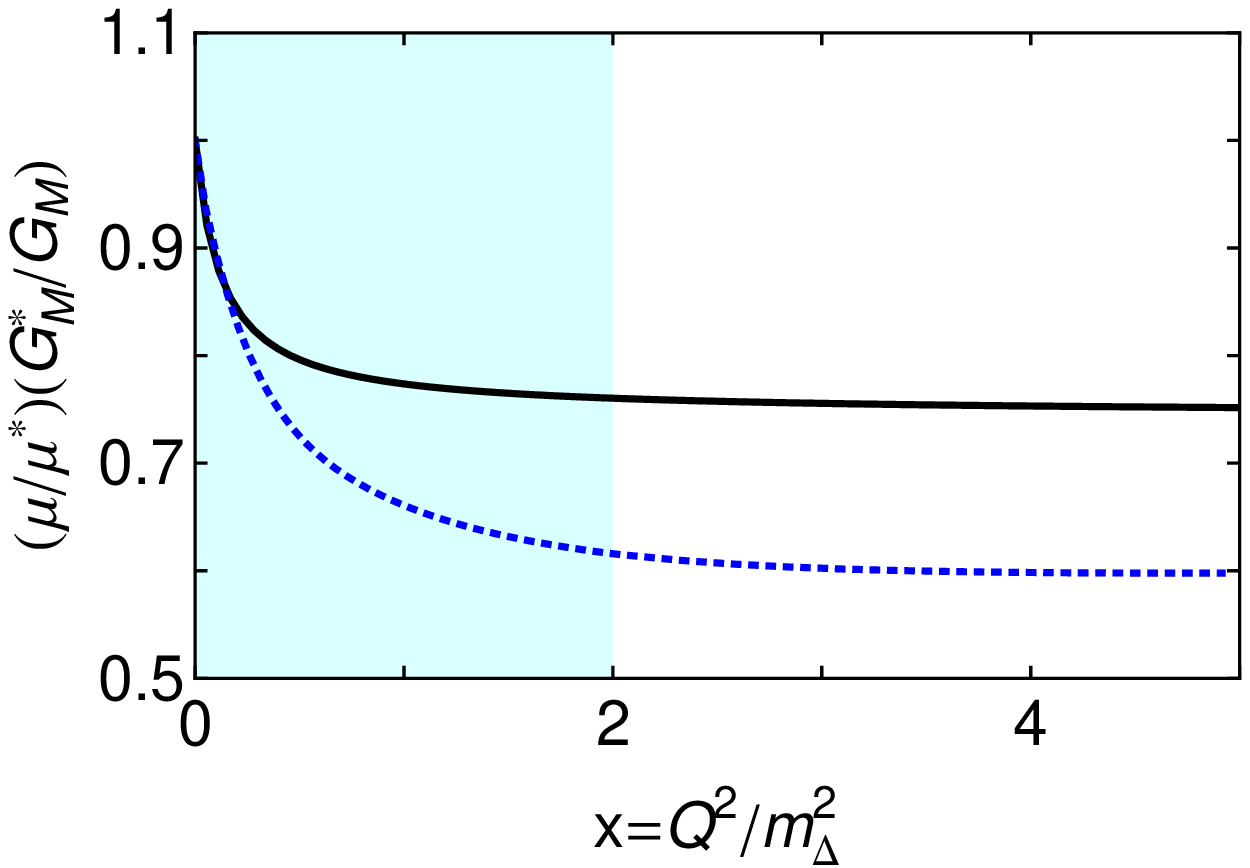} 
\end{tabular}
\caption{\label{figRatio}
Curves in both panels: \emph{Solid, black} -- dressed-quark core contribution to $\mu_n G_{M}^{\ast p}/\mu^\ast_p G_{M}^{n}$ as a function of $x=Q^2/m_{\Delta}^2$, obtained herein using our QCD-kindred framework; and \emph{dotted, blue} -- result for this ratio computed using a symmetry-preserving regularisation of a contact interaction \cite{Segovia:2013rca}.
%
\emph{Left panel}, DqAMM neglected; and \emph{right panel}, DqAMM included.
N.B.\,$\mu^\ast_p = G_M^{\ast N}(Q^2=0)$; and $\mu_n = G_M^n(Q^2=0)$.  The \emph{blue-shaded region} indicates the domain -- domain upon which meson cloud effects are certainly important.}
\end{center}
\end{figure}

In Fig.\,\ref{figRatio} we depict the ratio $\mu_n G_{M}^{\ast p}/\mu^\ast_p G_{M}^{n}$.  The DqAMM was omitted in preparing the left panel, which thus represents the cleanest expression of the dressed-quark core contribution: it is almost identically one, irrespective of the Faddeev equation kernel.  With isospin symmetry, the first entry in Eq.\,\eqref{eqBeg} is valid, so the same is true of the ratio constructed from the $\gamma^{\ast}n \to\Delta^{0}$ magnetic form factor.  It therefore appears that the second entry in Eq.\,\eqref{eqBeg} is also an excellent approximation, even in the absence of SU(6) symmetry.  This may be explained by a straightforward generalisation of the argument presented in association with Eqs.\,(19)\,--\,(22) in Ref.\,\cite{Segovia:2013uga}, which capitalises on the following features of the transition current: only photon-quark interactions are hard and each diagram describing such an interaction produces the same asymptotic behaviour; the axial-vector diquark contribution from Diagram~1 vanishes when computing proton's elastic form factor but survives in the neutron; and the $\Delta(1232)$-baryon does not contain scalar-diquark correlations.  Note that these are statements about the dressed-quark-core contributions to the transitions.  In general, one should only expect them to be valid empirically outside the domain upon which meson-cloud effects are certainly important; i.e., for $x \gtrsim 2\,$ \cite{Sato:2000jf,JuliaDiaz:2006xt}.  To illustrate this, the right-panel in Fig.\,\ref{figRatio} was prepared using form factors computed with an active DqAMM.  Notably, irrespective of the Faddeev equation kernel, the ratio is almost constant on $x\gtrsim 2$.

\begin{figure}[t]
\begin{center}
\begin{tabular}{cc}
\includegraphics[clip,width=0.43\linewidth]{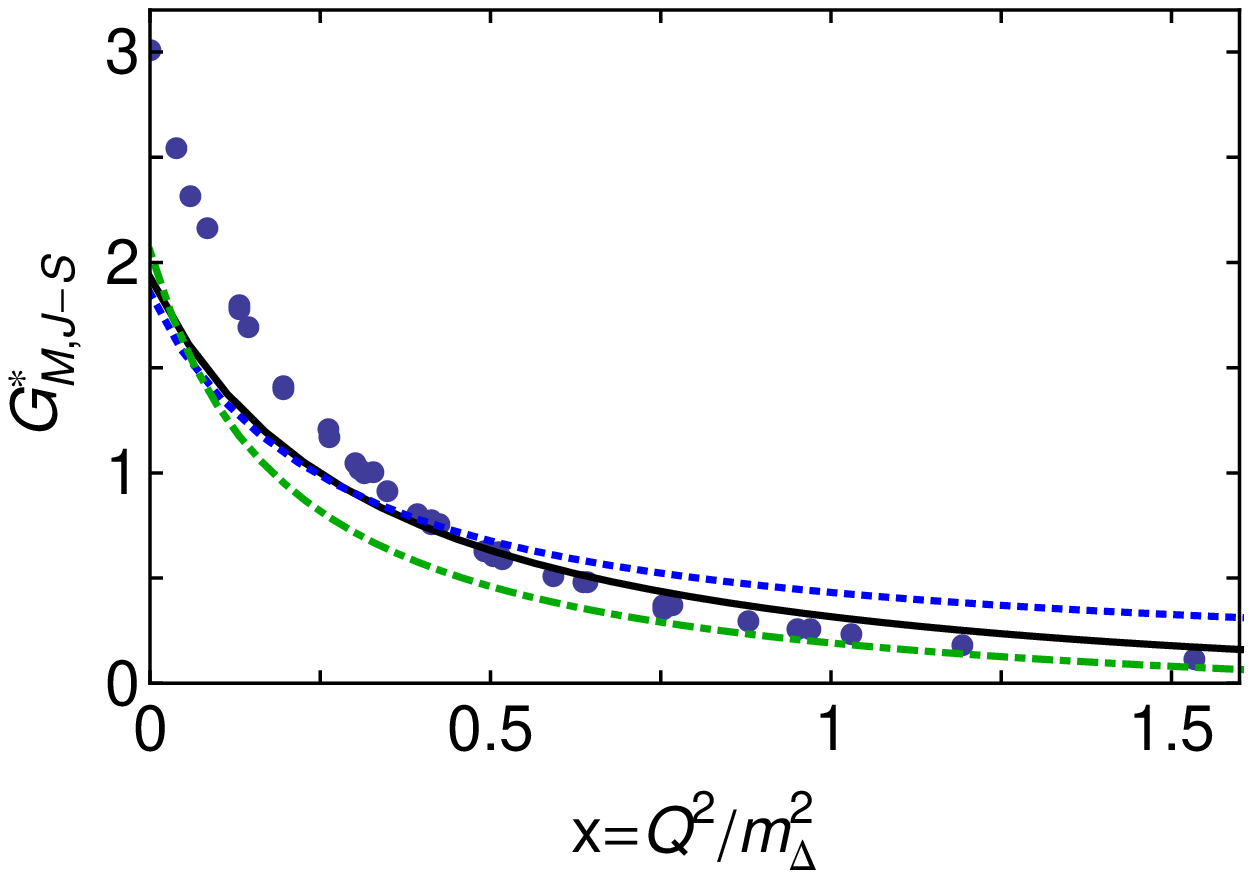}
\vspace*{-1ex} &
\includegraphics[clip,width=0.45\linewidth]{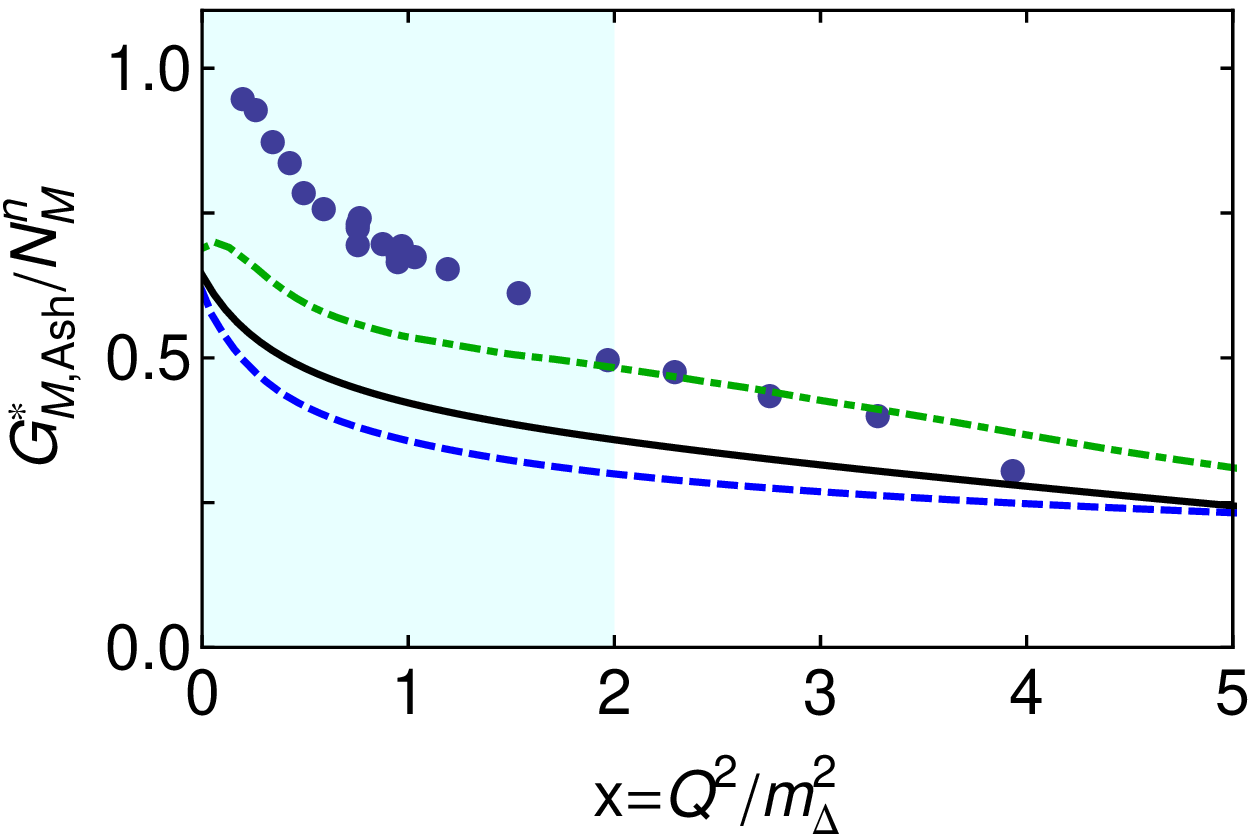}
\end{tabular}
\caption{\label{figGMast}
\emph{Left panel}. $G_{M}^{\ast}(Q^2)$: result obtained herein, with QCD-based Faddeev kernel (solid, black); CI result \cite{Segovia:2013rca,Segovia:2013uga} (dotted, blue); dressed-quark core contribution inferred using SL-model \protect\cite{JuliaDiaz:2006xt} (dot-dashed, green); and data from Refs.\,\protect\cite{Beringer:1900zz,Aznauryan:2009mx,Bartel:1968tw,Stein:1975yy,
Sparveris:2004jn,Stave:2008aa}, whose errors are commensurate with the point size.
\emph{Right panel}.  $G_{M,Ash}^{\ast}(Q^2)/N^n_M(Q^2)$.
In this panel the empirical results are from Ref.\,\cite{Aznauryan:2009mx}.  In all cases, $N(Q^2)=3 G_M^n(Q^2)/\mu_n$, where $G_M^n$ is the neutron magnetic form factor computed in the associated framework (solid and dashed curves) or the empirical result.}
\end{center}
\end{figure}

The left panel of Fig.\,\ref{figGMast} displays the $\gamma^\ast p \to \Delta^+$ magnetic transition form factor:
\begin{equation}
\tilde \mu_{N\Delta}^\ast:= (\sqrt{m_\Delta/m_N}) G_M^{\ast}(0) = 2.05\,.
\end{equation}
Our prediction, obtained with a QCD-based kernel, agrees with the data on $x\gtrsim 0.4$ and, allowing for the natural hardness of the contact interaction result, it is also consistent with data on that domain.  On the other hand, both curves disagree markedly with the data at infrared momenta.  This is explained by the similarity between these predictions and the ``bare'' or dressed-quark-core result determined using the Sato-Lee (SL) dynamical meson-exchange model (dot-dashed curve) \cite{JuliaDiaz:2006xt}.  The SL result supports a view that the discrepancy owes to omission of meson-cloud effects in the DSE computations.
Looking closer, it is worth reiterating that the difference between form factors obtained with the QCD-kindred and CI frameworks increases with $x=Q^2/m_\Delta^2$.  Consequently, future  measurements of the magnetic $N\to\Delta$ transition form factor, such as those using CLAS12 at JLab on $3<x<8$ \cite{Gothe:1209003}, can serve as a sensitive probe of the running coupling and masses in QCD.  In large part this is because meson-cloud effects are suppressed on $x>3$ and hence the behaviour of $G_M^{\ast}$ is thereupon determined primarily by the dressed-quark cores of the hadrons involved.  This is also true of many other elastic and transition form factors, which will be measured on $x>3$ at JLab \cite{Dudek:2012vr}.  As emphasised elsewhere (e.g., Refs.\,\cite{Roberts:2011rr,Aznauryan:2012baS,Bashir:2012fs,Chang:2011vu,Cloet:2013gva}), the domain $3<x<8$ is particularly important because it delivers momentum transfers to a dressed-quark within the hadron that covers the region of transition between nonperturbative and perturbative behaviour of the running coupling and masses.

In contrast to the left panel of Fig.\,\ref{figGMast}, presentations of experimental
data typically use the Ash form factor, Eq.\,\eqref{DefineAsh}.  This comparison is depicted in Fig.\,\ref{figGMast}, right panel.  (The DSE dressed-quark
core results are quantitatively similar to Fig.\,3 of Ref.\,\cite{Aznauryan:2012ec}).
Plainly, $G_{M,Ash}^{\ast}(Q^2)$ falls faster than a dipole.  This was historically viewed by many as a conundrum.  However, as observed previously \cite{Carlson:1985mm} and elucidated elsewhere \cite{Segovia:2013rca,Segovia:2013uga} there is no sound reason to expect $G_{M,Ash}^{\ast}(Q^2)/G_M^n(Q^2) \approx\,$constant.  Instead, the Jones-Scadron form factor should exhibit $G_{M}^{\ast}(Q^2)/G_M^n(Q^2) \approx\,$constant.  The empirical Ash form factor falls rapidly for two reasons.  First: meson-cloud effects provide up-to $35\%$ of the form factor for $x \lesssim 2$; these contributions are very soft; and hence they disappear rapidly.  Second: the additional kinematic factor $\sim 1/\sqrt{Q^2}$ in Eq.\,\eqref{DefineAsh} provides material damping for $x\gtrsim 2$.

\begin{figure}[t]
\begin{center}
\begin{tabular}{cc}
\includegraphics[clip,width=0.46\linewidth]{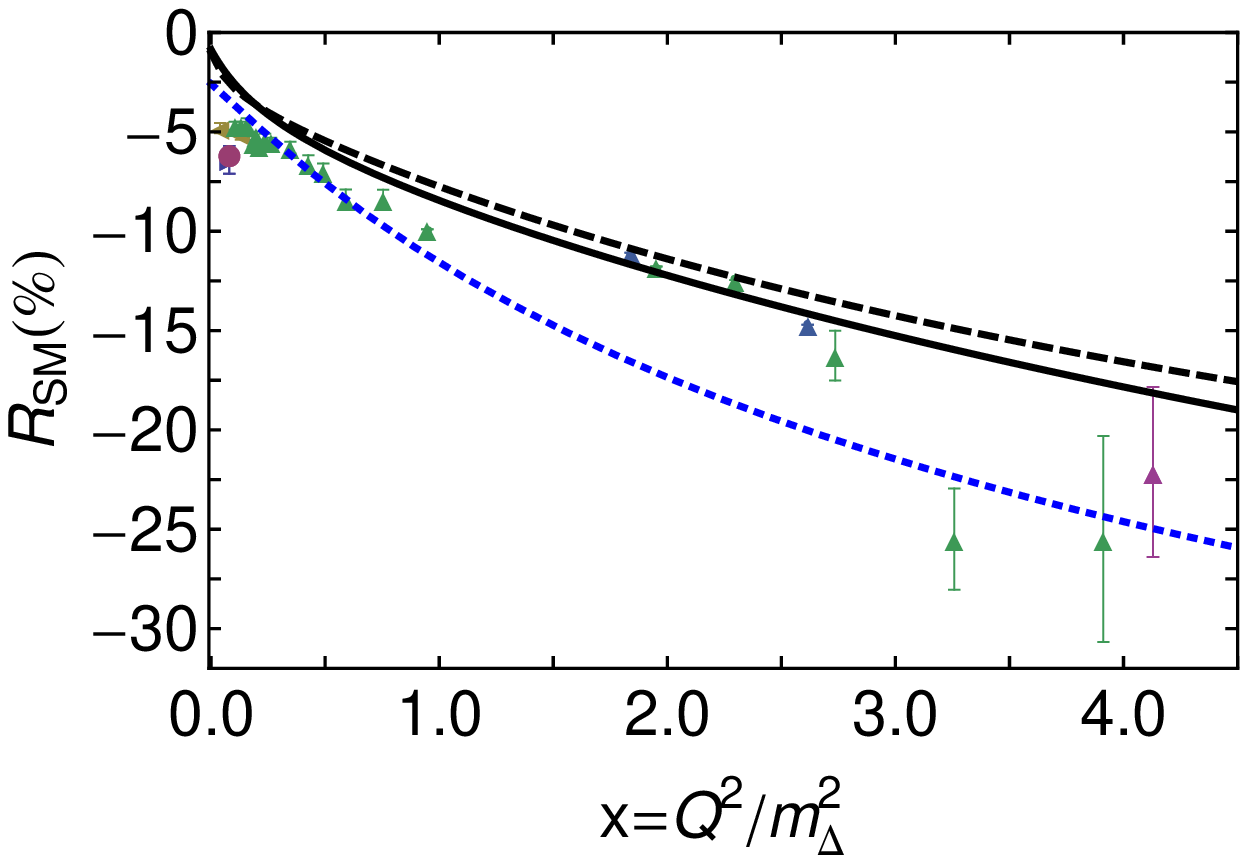}
\vspace*{-1ex} &
\includegraphics[clip,width=0.45\linewidth]{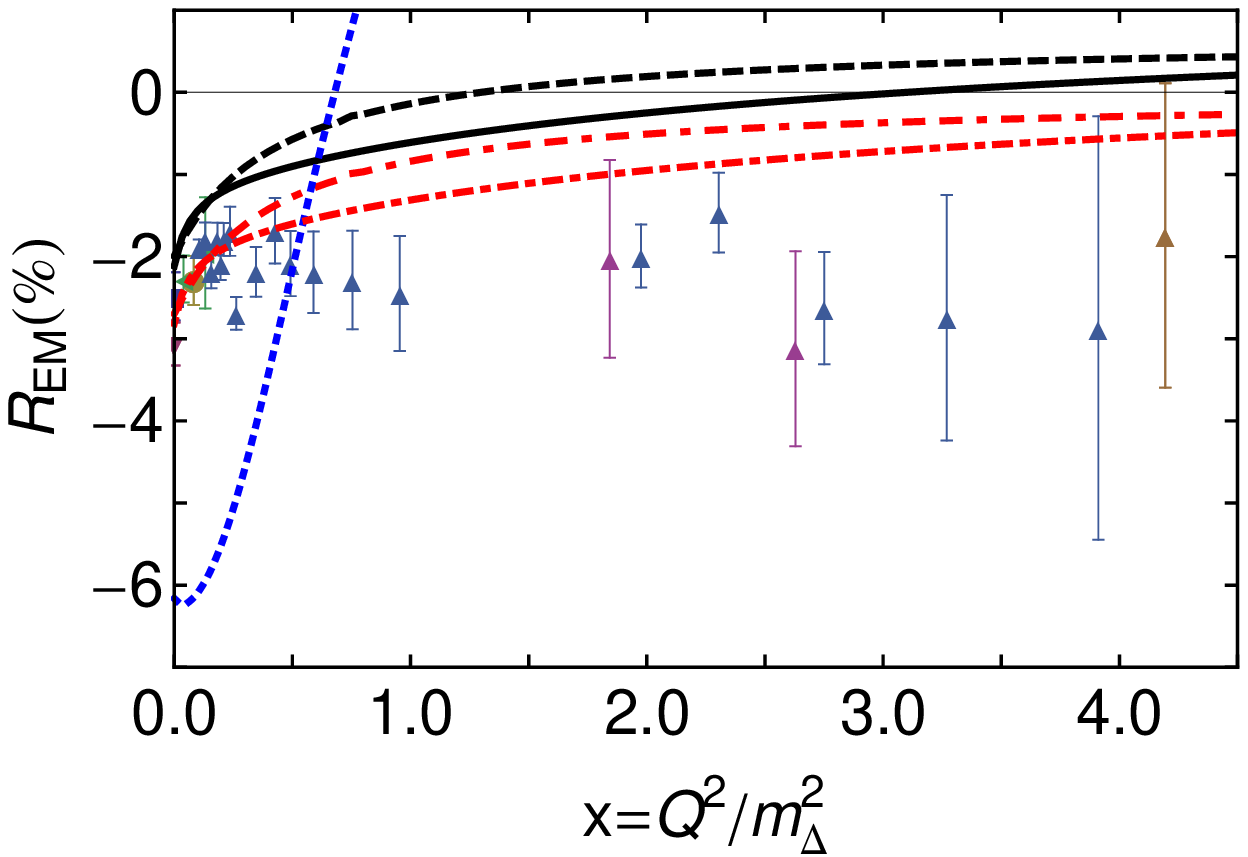}
\end{tabular}
\caption{\label{figREMSM} Ratios in Eq.\,\protect\eqref{eqREMSM}.
\emph{Left panel} -- $R_{SM}$: prediction of QCD-based kernel, including DqAMM (black, solid); no DqAMM (black, dashed); and CI result (dotted, blue).
\emph{Right panel} -- $R_{EM}$:
prediction obtained with QCD-kindred framework (solid, black); 
same input but without DqAMM (dashed, black);
these results renormalised (by a factor of 1.34) to agree with experiment at $x=0$ (dot-dashed, red - zero at $x\approx 14$; and dot-dash-dashed, red, zero at $x\approx 6$); and contact-interaction result (dotted, blue).
The data in both panels are drawn from Refs.\,\protect\cite{Aznauryan:2009mx,Sparveris:2004jn,Stave:2008aa,Beck:1999ge,
Pospischil:2000ad,Blanpied:2001ae}.
}
\end{center}
\end{figure}

Our predictions for the ratios in Eqs.\,\eqref{eqREMSM} are depicted in Fig.\,\ref{figREMSM}.  These quantities are commonly read as measures of deformation in one or both of the hadrons involved because they are zero in $SU(6)$-symmetric constituent-quark models.  However, the ratios also measure the way in which such
deformation influences the structure of the transition current.

It is worth examining separately each panel in Fig.~\ref{figREMSM} so first consider the left figure, which displays the Coulomb quadrupole ratio.  Both the prediction obtained with QCD-like propagators and vertices and the contact-interaction result are broadly consistent with available data.  This shows that even a contact-interaction can produce correlations between dressed-quarks within Faddeev wave-functions and related
features in the current that are comparable in size with those observed empirically.  Moreover, suppressing the DqAMM in the transition current has little impact.  These remarks highlight that $R_{SM}$ is quite robust; i.e., as might have been anticipated from the simple structure of the relevant projections in Eq.\,\eqref{GMGEGC}, $R_{SM}$ is not particularly sensitive to details of the Faddeev kernel and transition current.

As emphasised by the right panel in Fig.\,\ref{figREMSM}, this is certainly not the case with $R_{\rm EM}$, which involves an electric quadrupole form factor.  The differences between the curves displayed show that this ratio is a particularly sensitive measure of diquark and orbital angular momentum correlations, both within the hadrons involved and in the excitation current.  The contact-interaction result is inconsistent with data, possessing a zero that appears at a rather small value of $x$, just as is the case for the proton's electric form factor, Fig.\,\ref{fig:FFNucleon2}.  On the other hand, predictions obtained with QCD-like propagators and vertices can be viable.  We have presented four variants, which differ primarily in the location of the zero that is a feature of this ratio in all cases we have considered.  The inclusion of a DqAMM shifts the zero to a larger value of $x$.\footnote{N.B.\, Changing the dressed-quark propagator in precisely the manner which eliminates a zero in $G_E^p$ has only a limited effect on $R_{EM}$: it shifts the zero to a value of $Q^2$ that is $\lesssim 20$\% larger than that in the figure.}
Given the uniformly small value of this ratio and its sensitivity to the DqAMM, we judge that meson-cloud affects must play a large role on the entire domain that is currently accessible to experiment.  It is plausible that their contribution is serving to displace the zero in $R_{\rm EM}$ to a point beyond the domain that is currently accessible empirically.

In connection with Eqs.\,\eqref{eqUVREMSM} one may readily show that the helicity conservation arguments in Ref.\,\cite{Carlson:1985mm} apply equally to both the results obtained within our QCD-kindred framework and those produced by an internally-consistent symmetry-preserving treatment of a contact interaction.  As a consequence, Eqs.\,\eqref{eqUVREMSM} are certainly valid.  However, the predicted behaviour will probably not become apparent until $x\gtrsim 20$ \cite{Segovia:2013uga}.  Notwithstanding that, empirical discovery of a zero in $R_{\rm EM}$ will certainly serve as an harbinger for the transition to the domain upon which helicity conservation begins to play a role in determining the transition form factors.  That cannot otherwise be true.


\section{Epilogue}
\label{sec:summary}
We described a unified study of nucleon and $\Delta$ elastic and transition form factors that compares predictions made by a QCD-kindred framework, built upon a Faddeev equation kernel and interaction vertices that possess QCD-like momentum dependence, with results obtained using a symmetry-preserving treatment of a vector$\,\otimes\,$vector contact-interaction (CI).  The comparison established clearly that experiment is sensitive to the momentum dependence of the running couplings and masses in the strong interaction sector of the Standard Model and hence an experiment-theory collaboration can effectively constrain the evolution to infrared momenta of the $\beta$ function in QCD.  Indeed, the difference between form factors obtained with these different interaction kernels grows with increasing momentum transfer so that new experiments using upgraded facilities at JLab will gain access to the region of transition between nonperturbative and perturbative behaviour of QCD's running coupling and masses.

In order to achieve a satisfactory unification of all the form factors, we found it necessary to make small changes to the interaction current used in an earlier, wide-ranging analysis of nucleon elastic form factors alone \cite{Cloet:2008re}.  These modifications had little impact on the nucleon form factors, but the effects they did have were improvements.  For example, in the revised calculation, with no fine tuning, the momentum-dependence of the ratio $G_E^p(Q^2)/G_M^p(Q^2)$ agrees better with available high-$Q^2$ data and possesses a zero at $Q^2=9.5\,$GeV$^2$ [Fig.\,\ref{fig:FFNucleon2}].

In revisiting the analysis in Ref.\,\cite{Cloet:2008re} we were led to reemphasise that the possible existence and location of a zero in $G_E^p(Q^2)/G_M^p(Q^2)$ are a fairly direct measure of the nature of the quark-quark interaction in the Standard Model.  They are amongst a growing number of quantities that can serve as a cumulative gauge of the momentum dependence of the interaction, the transition between the associated theory's nonperturbative and perturbative domains, and the width of the transition region.  We also produced some interesting corollaries.
For instance, owing to the presence of strong diquark correlations in the nucleon Faddeev amplitudes and approximate charge symmetry, the neutron ratio reacts oppositely to $G_E^p(Q^2)/G_M^p(Q^2)$; viz., any change in the interaction which acts to shift a zero in the proton ratio to larger $Q^2$ relocates a zero in $G_E^n(Q^2)/G_M^n(Q^2)$ to smaller $Q^2$ [Fig.\,\ref{fig:GEnGEp}].  Furthermore, as a consequence of the faster-than-dipole decrease of the proton's electric form factor (and possible appearance of a zero), there will probably be a domain of $Q^2$ upon which the magnitude of the neutron's electric form factor exceeds that of the proton's.  This being so, then at some value of momentum transfer the electric form factor of Nature's most basic neutral composite fermion becomes larger than that of its positively charged counterpart.

We also reviewed and reanalysed available experimental and theoretical information pertaining to the nucleons' flavour-separated Dirac and Pauli form factors.  This led us to reiterate observations made elsewhere \cite{Roberts:2010hu,Cloet:2011qu,Wilson:2011aa}.  Namely, whilst the inclusion of meson-cloud contributions to the Faddeev kernels and interaction currents may quantitatively affect a comparison between data and predictions made within this framework, such contributions are not key to understanding the data.  The presence of strong diquark correlations within the nucleon is sufficient to explain the empirically verified features of the flavour-separated form factors [Fig.\,\ref{fig:flavoursepF1}, \ref{fig:flavoursepF2}].

It was natural, too, to update predictions for the ratio of nucleon spin-averaged and longitudinal-spin-dependent parton distribution functions on Bjorken-$x\simeq 1$ using simple formulae derived elsewhere \cite{Wilson:2011aa,Roberts:2013mja} and results obtained with our modified nucleon elastic electromagnetic current.  These values are important because they do not change under DGLAP evolution \cite{Holt:2010vj} and hence are scale invariant, nonperturbative features of QCD that serve as keen discriminators between frameworks that claim to explain nucleon structure.  Our revised values are practically indistinguishable from those produced by the original current [Table~\ref{tab:a}].

Our analysis of $\Delta(1232)$-baryon elastic form factors also produced observations of interest.  For example, the electric monopole, $G_{E0}$, and magnetic dipole form factors obtained using our QCD-kindred framework agree well with extant results from numerical simulations of lattice-regularised QCD.  However, there is marked disagreement between the results of these two approaches for the electric quadrupole, $G_{E2}$, and magnetic octupole, $G_{M3}$, form factors [Fig.\,\ref{fig:elasticFFD}].  This mismatch is at least partly explained by the unrealistically large values of $m_\pi$, $m_\rho$, $m_\Delta$ used in the lattice simulations, as we showed by recomputing the form factors using an inflated value of $m_\Delta$ [Fig.\,\ref{fig:hugemassFFD}].  We judge, therefore, that it would be a mistake to infer too much about $\Delta$-baryon elastic form factors from existing lattice results.  That part of our study also reemphasised the simple structure of the $\Delta$-baryon, whose Faddeev amplitude is dominated by axial-vector diquark correlations.

Our comparison between QCD-kindred and CI results also pointed to a number of robust predictions about the $\Delta$-baryon: $G_{E0}(Q^2)$ possesses a zero at $Q^2 \sim 2\,$GeV$^2$ [Fig.\,\ref{fig:elasticFFD}]; $G_{E2}(Q^2=0)$ is negative but its magnitude is very sensitive to the nature and strength of correlations in the $\Delta(1232)$ baryon's Faddeev amplitude [Table~\ref{tab:comparative}]; and $G_{M3}$ is negative, so long as one includes those correlations in the Faddeev amplitude which correspond to $P$- and $D$-waves in the $\Delta(1232)$ baryon's rest frame [Fig.\,\ref{fig:elasticFFD}].

Turning to the $N \to \Delta$ transition, we found that, independent of the Faddeev kernel and interaction current, the momentum-dependence of the magnetic transition form factor in the Jones-Scadron convention, $G_M^\ast$, matches that of the neutron's magnetic form factor once the momentum transfer enters the domain upon which meson-cloud contributions are negligible [Fig.\,\ref{figRatio}].  Moreover, the  prediction for $G_M^\ast$ obtained using the QCD-kindred framework is in almost pointwise agreement with the dressed-quark core contribution inferred from a dynamical coupled channels analysis of the transition [Fig.\,\ref{figGMast}].  It follows naturally that the Ash form factor connected with the $\gamma^\ast N \to \Delta$ transition should fall faster than the neutron's magnetic form factor.

We also considered the quadrupole ratios associated with the $\gamma^\ast N \to \Delta$ transition and found that the Coulomb ratio does not depend strongly on the precise structure of the Faddeev kernel and interaction current: the presence of diquark correlations is sufficient to qualitatively explain the behaviour.  On the other hand, the electric quadrupole ratio, $R_{\rm EM}$, presented quite a different picture.  It is a particularly keen measure of diquark and orbital angular momentum correlations, both within the hadrons involved and in the excitation current; and results obtained using the QCD-kindred framework can be viable.  Our analysis predicts that $R_{\rm EM}$ should possess a zero.  However, we argued that meson-cloud effects obscure this feature on that domain which is currently accessible to experiment.

In closing we would like to emphasise that the key element in our unification of elastic and transition form factors involving the nucleon and/or $\Delta(1232)$-baryon is a veracious expression of dynamical chiral symmetry breaking (DCSB) in hadrons.  DCSB is very plainly expressed in the dressed-quark mass-function; but it is also evident in the strength of diquark correlations in the scalar and axial-vector channels within the nucleon, and in the axial-vector channel within the $\Delta$.  As our analysis showed, numerous empirically observed baryon properties can be explained by the presence of diquark correlations and the overlaps and interferences between them.  It should be stressed here that the diquark correlations predicted to exist within baryons are not the static, pointlike degrees-of-freedom which were historically introduced in order to simplify the study of systems comprised from three constituent-quarks.  The modern dynamical diquark correlation is soft and possesses nontrivial electromagnetic structure.  It is unsound to consider diquarks as inert and structureless, and any expectations grounded in such a picture should be discarded.

In this study we used simple parametrisations of propagators and vertices, an expedient which enabled us to compute elastic and transition form factors for arbitrary spacelike momenta using QCD-like input.  A natural next step is to replicate these calculations using propagators and vertices computed from a realistic interaction \cite{Qin:2011dd} in conjunction with a numerical algorithm that facilitates computation of form factors to arbitrarily large momentum transfers with such input.  The approach in Ref.\,\cite{Chang:2013nia} is an obvious candidate.  Progress in that direction will materially enhance the ability of experiment and theory together to be used effectively to chart the infrared evolution of QCD's $\beta$-function.


\section*{Acknowledgments}
JS thanks G.~Eichmann for informative discussions and invaluable help when requested.  We also thank the following people for useful input: M.~Diehl, R.~Gothe and V.~Mokeev.
JS, ICC and CDR are grateful for the opportunity to participate in the workshops ``Many Manifestations of Nonperturbative QCD under the Southern Cross'', Ubatuba, and the ``2$^{\rm nd}$ Workshop on Perspectives in Nonperturbative QCD'' at IFT-UNESP, S\~ao Paulo, during both of which substantial fractions of this work were completed.
CDR acknowledges support from an \emph{International Fellow Award} from the Helmholtz Association.
Work otherwise supported by:
U.S.\ Department of Energy, Office of Science, Office of Nuclear Physics, under contract no.~DE-AC02-06CH11357;
GAUSTEQ (Germany and U.S.\ Nuclear Theory Exchange Program for QCD Studies of Hadrons and Nuclei) under contract number DE-SC0006758;
and For\-schungs\-zentrum J\"ulich GmbH.


\appendix
\setcounter{equation}{0}
\renewcommand{\theequation}{\Alph{section}.\arabic{equation}}
\setcounter{figure}{0}
\renewcommand{\thefigure}{\Alph{section}.\arabic{figure}}

\section{Nucleon and $\Delta$ Faddeev Equations}
\label{app:FE}

\subsection{General Structure}

The nucleon is represented by a Faddeev amplitude
\begin{equation}
\label{PsiNucleon}
\Psi_{N} = \Psi_{1} + \Psi_{2} + \Psi_{3}  \,,
\end{equation}
where the subscript identifies the bystander quark and, e.g., $\Psi_{1,2}$ are
obtained from $\Psi_{3}$ by a cyclic permutation of all the quark labels. The spin-
and isospin-$1/2$ nucleon is a sum of scalar and axial-vector diquark correlations:
\begin{equation}
\label{Psi}
\Psi_3(p_i,\alpha_i,\tau_i) = {\cal N}_{\,\,3}^{0^+} + {\cal N}_{\,\,3}^{1^+},
\end{equation}
with $(p_i,\alpha_i,\tau_i)$ the momentum, spin and isospin labels of the quarks
constituting the bound state, and $P=p_1+p_2+p_3$ the total momentum of the system.

The scalar diquark piece in Eq.\,(\ref{Psi}) is
\begin{eqnarray}
{\cal N}_{\,\,3}^{0^+}(p_i,\alpha_i,\tau_i)&=&
[\Gamma^{0^+}(\sfrac{1}{2}p_{[12]};K)]_{\alpha_1
\alpha_2}^{\tau_1 \tau_2}\, \Delta^{0^+}(K) \,[{\cal S}(\ell;P)
u(P)]_{\alpha_3}^{\tau_3}\,,%
\label{calS}
\end{eqnarray}
where: the spinor satisfies (App.\,\ref{App:EM}) 
\begin{equation}
(i\gamma\cdot P + M)\, u(P) =0= \bar u(P)\, (i\gamma\cdot P + M)\,,
\end{equation}
with $M$ the mass obtained by solving the Faddeev equation, and it is also a spinor
in isospin space with $\varphi_+= {\rm col}(1,0)$ for the proton and $\varphi_-= {\rm
col}(0,1)$ for the neutron; $K= p_1+p_2=: p_{\{12\}}$, $p_{[12]}= p_1 - p_2$, $\ell
:= (-p_{\{12\}} + 2 p_3)/3$; $\Delta^{0^+}$ is a pseudoparticle propagator for the
scalar diquark formed from quarks $1$ and $2$, and $\Gamma^{0^+}\!$ is a
Bethe-Salpeter-like amplitude describing their relative momentum correlation; and
${\cal S}$, a $4\times 4$ Dirac matrix, describes the relative quark-diquark momentum
correlation (${\cal S}$, $\Gamma^{0^+}$ and $\Delta^{0^+}$ are discussed in
Sect.\,\ref{completing}.). The colour antisymmetry of $\Psi_3$ is implicit in
$\Gamma^{J^P}\!\!$, with the Levi-Civita tensor, $\epsilon_{c_1 c_2 c_3}$, expressed
via the antisymmetric Gell-Mann matrices; viz., defining
\begin{equation}
\{H^1=i\lambda^7,H^2=-i\lambda^5,H^3=i\lambda^2\}\,,
\end{equation}
then $\epsilon_{c_1 c_2 c_3}= (H^{c_3})_{c_1 c_2}$.  [See Eqs.\,(\ref{Gammaqq}).]

The axial-vector component in Eq.\,(\ref{Psi}) is
\begin{eqnarray}
{\cal N}_{\,\,3}^{1^+}(p_i,\alpha_i,\tau_i) & =&
[{\tt t}^i\,\Gamma_\mu^{1^+}(\sfrac{1}{2}p_{[12]};K)]_{\alpha_1 \alpha_2}^{\tau_1
\tau_2}\,\Delta_{\mu\nu}^{1^+}(K)\, [{\cal A}^{i}_\nu(\ell;P)
u(P)]_{\alpha_3}^{\tau_3}\,,
\label{calA}
\end{eqnarray}
where the symmetric isospin-triplet matrices are
\begin{equation}
{\tt t}^+ = \frac{1}{\surd 2}(\tau^0+\tau^3) \,,\;
{\tt t}^0 = \tau^1\,,\;
{\tt t}^- = \frac{1}{\surd 2}(\tau^0-\tau^3)\,,
\end{equation}
and the other elements in Eq.\,(\ref{calA}) are straightforward generalisations of
those in Eq.\,(\ref{calS}).

Since it is not possible to combine an isospin-$0$ diquark with an isospin-$1/2$ dressed-quark to obtain isospin-$3/2$, the spin- and isospin-$3/2$ $\Delta$-baryon contains only an axial-vector diquark component
\begin{equation}
\label{DeltaA}
\Psi_{3}(p_{i},\alpha_{i},\tau_{i}) = {\cal D}_{3}^{1^{+}}.
\end{equation}
Understanding the $\Delta$'s structure is plainly far simpler than is the case of the nucleon since, whilst the general form of the Faddeev amplitude for a spin- and isospin-$3/2$ can be complicated, isospin symmetry means that one can focus on the $\Delta^{++}$, with its simple flavour structure, because all the charge states
are degenerate:
\begin{equation}
{\cal D}_{3}^{1^{+}}(p_{i},\alpha_{i},\tau_{i}) = [{\tt
t}^{+}\Gamma^{1^{+}}_{\mu}(\frac{1}
{2}p_{[12]};K)]_{\alpha_{1}\alpha_{2}}^{\tau_{1}\tau_{2}} \Delta_{\mu\nu}^{1^{+}}(K)
[{\cal D}_{\nu\rho}(\ell;P) u_{\rho}(P) \varphi_{+}]_{\alpha_{3}}^{\tau_{3}} \,,
\label{eq:D31p}
\end{equation}
where $u_{\rho}(P)$ is a Rarita-Schwinger spinor defined via Eq.\,\eqref{eq:rarita}.

The general forms of the matrices ${\cal S}(\ell;P)$, ${\cal A}^i_\nu(\ell;P)$ and ${\cal D}_{\nu\rho}(\ell;P)$, which describe the momentum space correlation between the quark and diquark in the nucleon and $\Delta$-baryon are described in Refs.\,\cite{Cloet:2007pi,Oettel:1998bk}.  The requirement that ${\cal S}(\ell;P)$
represent a positive energy nucleon entails
\begin{equation}
\label{Sexp}
{\cal S}(\ell;P) = s_1(\ell;P)\,I_{\rm D} + \left(i\gamma\cdot \hat\ell - \hat\ell
\cdot \hat P\, I_{\rm D}\right)\,s_2(\ell;P)\,,
\end{equation}
where $(I_{\rm D})_{rs}= \delta_{rs}$, $\hat \ell^2=1$, $\hat P^2= - 1$.  In the
nucleon rest frame, $s_{1,2}$ describe, respectively, the upper, lower component of
the bound-state nucleon's spinor.  Placing the same constraint on the axial-vector
component, one has
\begin{equation}
\label{Aexp}
 {\cal A}^i_\nu(\ell;P) = \sum_{n=1}^6 \,
p_n^i(\ell;P)\,\gamma_5\,A^n_{\nu}(\ell;P)\,,\; i=+,0,-\,,
\end{equation}
where ($ \hat \ell^\perp_\nu = \hat \ell_\nu + \hat \ell\cdot\hat P\, \hat P_\nu$, $
\gamma^\perp_\nu = \gamma_\nu + \gamma\cdot\hat P\, \hat P_\nu$)
\begin{equation}
\label{Afunctions}
\begin{array}{lll}
A^1_\nu= \gamma\cdot \hat \ell^\perp\, \hat P_\nu \,,\; &
A^2_\nu= -i \hat P_\nu \,,\; &
A^3_\nu= \gamma\cdot\hat \ell^\perp\,\hat \ell^\perp_{\nu}\,,\\
A^4_\nu= i \,\hat \ell_\mu^\perp\,,\; &
A^5_\nu= \gamma^\perp_\nu - A^3_\nu \,,\; &
A^6_\nu= i \gamma^\perp_\nu \gamma\cdot\hat \ell^\perp - A^4_\nu\,.
\end{array}
\end{equation}
Finally, requiring also that ${\cal D}_{\nu\rho}(\ell;P)$ be an eigenfunction of the
positive-energy projection operator, one obtains
\begin{equation}
{\cal D}_{\nu\rho}(\ell;P) = {\cal S}^{\Delta}(\ell;P) \delta_{\nu\rho} +
\gamma_{5}{\cal
A}_{\nu}^{i\,\Delta}(\ell;P) \ell_{\rho}^{\bot},
\label{eq:DlP}
\end{equation}
with ${\cal S}^{\Delta}$ and ${\cal A}_{\nu}^{i\,\Delta}$ given by obvious analogues
of Eqs.~(\ref{Sexp}) and~(\ref{Aexp}), respectively.

One can now write the Faddeev equation satisfied by $\Psi_3$ in the case of the
nucleon as
\begin{equation}
 \left[ \begin{array}{r}
{\cal S}(k;P)\, u(P)\\
{\cal A}^i_\mu(k;P)\, u(P)
\end{array}\right]\\
 = -\,4\,\int\frac{d^4\ell}{(2\pi)^4}\,{\cal M}(k,\ell;P)
\left[
\begin{array}{r}
{\cal S}(\ell;P)\, u(P)\\
{\cal A}^j_\nu(\ell;P)\, u(P)
\end{array}\right]\,.
\label{FEone}
\end{equation}

The kernel in Eq.~(\ref{FEone}) is
\begin{equation}
\label{calM} {\cal M}(k,\ell;P) = \left[\begin{array}{cc}
{\cal M}_{00} & ({\cal M}_{01})^j_\nu \\
({\cal M}_{10})^i_\mu & ({\cal M}_{11})^{ij}_{\mu\nu}\rule{0mm}{3ex}
\end{array}
\right] ,
\end{equation}
with
\begin{equation}
 {\cal M}_{00} = \Gamma^{0^+}\!(k_q-\ell_{qq}/2;\ell_{qq})\,
S^{\rm T}(\ell_{qq}-k_q) \,\bar\Gamma^{0^+}\!(\ell_q-k_{qq}/2;-k_{qq})\,
S(\ell_q)\,\Delta^{0^+}(\ell_{qq}) \,,
\end{equation}
where: $\ell_q=\ell+P/3$, $k_q=k+P/3$, $\ell_{qq}=-\ell+ 2P/3$,
$k_{qq}=-k+2P/3$ and the superscript ``T'' denotes matrix transpose; and
\begin{subequations}
\begin{eqnarray}
({\cal M}_{01})^j_\nu &=& {\tt t}^j \,
\Gamma_\mu^{1^+}\!(k_q-\ell_{qq}/2;\ell_{qq})
S^{\rm T}(\ell_{qq}-k_q)\,\bar\Gamma^{0^+}\!(\ell_q-k_{qq}/2;-k_{qq})\,
S(\ell_q)\,\Delta^{1^+}_{\mu\nu}(\ell_{qq}) \,, \label{calM01} \\
 ({\cal M}_{10})^i_\mu &=&
\Gamma^{0^+}\!(k_q-\ell_{qq}/2;\ell_{qq})\, S^{\rm T}(\ell_{qq}-k_q)\,{\tt t}^i\,
\bar\Gamma_\mu^{1^+}\!(\ell_q-k_{qq}/2;-k_{qq})\,
S(\ell_q)\,\Delta^{0^+}(\ell_{qq}) \,,\\
 ({\cal M}_{11})^{ij}_{\mu\nu} &=& {\tt t}^j\,
\Gamma_\rho^{1^+}\!(k_q-\ell_{qq}/2;\ell_{qq}) S^{\rm T}(\ell_{qq}-k_q)\,{\tt t}^i\,
\bar\Gamma^{1^+}_\mu\!(\ell_q-k_{qq}/2;-k_{qq})\,
S(\ell_q)\,\Delta^{1^+}_{\rho\nu}(\ell_{qq}) \,.
\label{calM11}
\end{eqnarray}
\end{subequations}

The $\Delta$'s Faddeev equation is
\begin{equation}
\begin{split}
{\cal D}_{\lambda\rho}(k;P)u_{\rho}(P) = -4 \int \frac{d^{4}\ell}{(2\pi)^{4}} {\cal
M}_{\lambda\mu}^{\Delta}(k,\ell;P){\cal D}_{\mu\sigma}(\ell;P)u_{\sigma}(P)\,,
\label{FEdelta}
\end{split}
\end{equation}
with
\begin{equation}
{\cal M}_{\lambda\mu}^{\Delta}={\tt t}^{+}
\Gamma_{\sigma}^{1^{+}}(k_q-\ell_{qq}/2;\ell_{qq}) S^{T}(\ell_{qq}-k_q) {\tt t}^{+}
\bar{\Gamma}_{\lambda}^{1^{+}}(\ell_{q}-k_{qq}/2;-k_{qq}) S(\ell_q)
\Delta^{1^{+}}_{\sigma\mu}(\ell_{qq})\,.
\end{equation}

\subsection{Kernel of the Faddeev Equation}
\label{completing}

To complete the Faddeev equations, Eqs.~(\ref{FEone}) and~(\ref{FEdelta}), one must specify the dressed-quark propagator, the diquark Bethe-Salpeter amplitudes and the diquark propagators.

\subsubsection{Dressed-quark propagator}
\label{subsubsec:S}

The dressed-quark propagator has the general form
\begin{eqnarray}
S(p) & = & -i \gamma\cdot p\, \sigma_V(p^2) + \sigma_S(p^2) = 1/[i\gamma\cdot p\,
A(p^2) + B(p^2)]\label{SpAB}
\end{eqnarray}
and can be obtained from QCD's gap equation.  It is a longstanding prediction of DSE studies in QCD that for light-quarks the wave function renormalisation and dressed-quark mass:
\begin{equation}
\label{ZMdef}
Z(p^2)=1/A(p^2)\,,\;M(p^2)=B(p^2)/A(p^2)\,,
\end{equation}
respectively, receive strong momentum-dependent corrections at infrared momenta \cite{Lane:1974he,Politzer:1976tv,Chang:2011vu,Bashir:2012fs,Cloet:2013jya}: $Z(p^2)$ is suppressed and $M(p^2)$ enhanced.  These features are an expression of dynamical chiral symmetry breaking (DCSB) and, plausibly, of confinement \cite{Cloet:2013jya}.  The enhancement of $M(p^2)$ is central to the appearance of a constituent-quark mass-scale and an existential prerequisite for Goldstone modes.  These DSE predictions are confirmed in numerical simulations of lattice-regularised QCD \cite{Zhang:2004gv}, and the conditions have been explored under which pointwise agreement between DSE results and lattice simulations may be obtained \cite{Bhagwat:2003vw,Bhagwat:2004kj,Bhagwat:2006tu}.

The impact of this infrared dressing on hadron phenomena has long been emphasised \cite{Roberts:1994hh} and, while numerical solutions of the quark DSE are now readily obtained, the utility of an algebraic form for $S(p)$ when calculations require the evaluation of numerous multidimensional integrals is self-evident.  An efficacious
parametrisation of $S(p)$, which exhibits the features described above, has been used
extensively in hadron studies \cite{Roberts:2007jh}.  It is expressed via
\begin{subequations}
\begin{eqnarray}
\bar\sigma_S(x) & =&  2\,\bar m \,{\cal F}(2 (x+\bar m^2)) + {\cal
F}(b_1 x) \,{\cal F}(b_3 x) \,
\left[b_0 + b_2 {\cal F}(\epsilon x)\right]\,,\label{ssm} \\
\label{svm} \bar\sigma_V(x) & = & \frac{1}{x+\bar m^2}\, \left[ 1 - {\cal F}(2
(x+\bar m^2))\right]\,,
\end{eqnarray}
\end{subequations}
with $x=p^2/\lambda^2$, $\bar m$ = $m/\lambda$,
\begin{equation}
\label{defcalF}
{\cal F}(x)= \frac{1-\mbox{\rm e}^{-x}}{x}  \,,
\end{equation}
$\bar\sigma_S(x) = \lambda\,\sigma_S(p^2)$ and $\bar\sigma_V(x) =
\lambda^2\,\sigma_V(p^2)$.  The mass-scale, $\lambda=0.566\,$GeV, and
parameter values\footnote{$\epsilon=10^{-4}$ in Eq.\ (\ref{ssm}) acts only to
decouple the large- and intermediate-$p^2$ domains.}
\begin{equation}
\label{tableA}
\begin{array}{ccccc}
   \bar m& b_0 & b_1 & b_2 & b_3 \\\hline
   0.00897 & 0.131 & 2.90 & 0.603 & 0.185
\end{array}\;,
\end{equation}
were fixed in a least-squares fit to light-meson observables \cite{Burden:1995ve,Hecht:2000xa}.  The dimensionless $u=d$ current-quark mass in
Eq.~(\ref{tableA}) corresponds to
\begin{equation}
\label{mcq}
m=5.08\,{\rm MeV} = :m^{\rm phys}\,.
\end{equation}
The parametrisation yields the following Euclidean constituent-quark mass, defined as the solution of $p^2=M^2(p^2)$:
\begin{equation}
\label{MEq} M_{u,d}^E = 0.33\,{\rm GeV}.
\end{equation}

The ratio $M^E/m = 65$ is one expression of DCSB in the parametrisation of $S(p)$.
It emphasises the dramatic enhancement of the dressed-quark mass function at infrared
momenta. Another is the chiral-limit in-pion condensate
\begin{equation}
\label{qbqparam}
-\langle \bar q q \rangle_\zeta^0 = \lambda^3 \, \frac{3}{4\pi^2}\, \frac{b_0}{b_1
b_3} \, \ln \frac{\zeta^2}{\Lambda_{\rm QCD}^2},
\end{equation}
which assumes the value ($\Lambda_{\rm QCD} = 0.2\,$GeV)
\begin{equation}
-\langle \bar q q \rangle_{\zeta=1\,{\rm GeV}}^0 = (0.221 \,{\rm GeV})^3.
\end{equation}
Detailed discussions of the in-pion condensate in QCD can be found, e.g., in
Refs.\,\cite{Brodsky:2012ku,Cloet:2013jya}.

\subsubsection{Diquark Bethe-Salpeter amplitudes}
\label{qqBSA}

The rainbow-ladder DSE truncation yields asymptotic diquark states in the strong interaction spectrum.  Such states are not observed and their appearance is an artefact of the truncation.  Higher-order terms in the quark-quark scattering kernel, whose analogue in the quark-antiquark channel do not materially affect the properties of vector and flavour non-singlet pseudoscalar mesons, ensure that QCD's quark-quark scattering matrix does not exhibit singularities which correspond to asymptotic
diquark states \cite{Bender:2002as,Bhagwat:2004hn}.  Nevertheless, studies with kernels that don't generate diquark bound states, do support a physical interpretation of the masses, $m_{(qq)_{J^P}}$, obtained using the rainbow-ladder truncation: the quantity $l_{(qq)_{J^P}}=1/m_{(qq)_{J^P}}$ may be interpreted as a range over which the
diquark correlation can propagate within a baryon.  These observations motivate an
{\it Ansatz} for the quark-quark scattering matrix that is employed in deriving the
Faddeev equation:
\begin{equation}
[M_{qq}(k,q;K)]_{rs}^{tu} = \sum_{J^P=0^+,1^+,\ldots} \bar\Gamma^{J^P}\!(k;-K)\,
\Delta^{J^P}\!(K) \, \Gamma^{J^P}\!(q;K)\,. \label{AnsatzMqq}
\end{equation}

One manner of specifying the $\Gamma^{J^P}\!\!$ in Eq.\,(\ref{AnsatzMqq}) is to employ solutions of a rainbow-ladder quark-quark Bethe-Salpeter equation (BSE), as, e.g., in Refs.\,\cite{Maris:2002yu,Maris:2004bp,Eichmann:2008ef}.  Using the properties of the Gell-Mann matrices one finds easily that $\Gamma^{J^P}_C:= \Gamma^{J^P}C^\dagger$ satisfies exactly the same equation as the $J^{-P}$ colour-singlet meson {\it but} for a halving of the coupling strength \cite{Cahill:1987qr}.  This makes clear that the interaction in the ${\bar 3_c}$ $(qq)$ channel is strong and attractive.

A solution of the BSE equation requires a simultaneous solution of the quark-DSE.
However, since we chose to simplify the calculations by parametrising $S(p)$, we
also employ that expedient with $\Gamma^{J^P}\!$, using the following one-parameter
forms:
\begin{subequations}
\label{Gammaqq}
\begin{eqnarray}
\label{Gamma0p} \Gamma^{0^+}(k;K) &=& \frac{1}{{\cal N}^{0^+}} \,
H^a\,C i\gamma_5\, i\tau_2\, {\cal F}(k^2/\omega_{0^+}^2) \,, \\
\label{Gamma1p} {\tt t}^i \Gamma^{1^+}_\mu (k;K) &=& \frac{1}{{\cal N}^{1^+}}\,
H^a\,i\gamma_\mu C\,{\tt t}^i\, {\cal F}(k^2/\omega_{1^+}^2)\,,
\end{eqnarray}
\end{subequations}
with the normalisation, ${\cal N}^{J^P}\!$, fixed by requiring
\begin{subequations}
\begin{eqnarray}
\label{BSEnorm}
2 \,K_\mu & = &
\left[ \frac{\partial}{\partial Q_\mu} \Pi(K,Q) \right]_{Q=K}^{{K^2=-m_{J^P}^2}},\\
\Pi(K,Q) & = & tr\!\! \int\!\!
\frac{d^4 q}{(2\pi)^4}\, \bar\Gamma(q;-K) \, S(q+Q/2) \, \Gamma(q;K) \, S^{\rm
T}(-q+Q/2) .
\end{eqnarray}
\end{subequations}

The {\it Ans\"atze} of Eqs.\,\eqref{Gammaqq} retain only that single Dirac-amplitude which would represent a point particle with the given quantum numbers in a local Lagrangian density.  They are usually the dominant amplitudes in a solution of the rainbow-ladder BSE for the lowest mass $J^P$ diquarks \cite{Burden:1996nh,Maris:2002yu} and mesons \cite{Maris:1997tm,Maris:1999nt,Maris:1999bh,Qin:2011xq}.

\subsubsection{Diquark propagators}
\label{qqprop}
Solving for the quark-quark scattering matrix using the rainbow-ladder truncation
yields free particle propagators for $\Delta^{J^P}$ in Eq.~(\ref{AnsatzMqq}).  As already noted, however, higher-order contributions remedy that defect, eliminating asymptotic diquark states from the spectrum.  The attendant modification of $\Delta^{J^P}$ can be modeled efficiently by simple functions that are free-particle-like at spacelike momenta but pole-free on the timelike axis \cite{Bender:2002as,Bhagwat:2004hn}; namely,\footnote{These forms satisfy a sufficient condition for confinement because of the associated violation of reflection positivity.  See Sect.~2.2 of Ref.\,\cite{Cloet:2013jya} for a brief discussion.}
\begin{subequations}
\begin{eqnarray}
\Delta^{0^+}(K) & = & \frac{1}{m_{0^+}^2}\,{\cal F}(K^2/\omega_{0^+}^2)\,,\\
\Delta^{1^+}_{\mu\nu}(K) & = &
\left(\delta_{\mu\nu} + \frac{K_\mu K_\nu}{m_{1^+}^2}\right) \, \frac{1}{m_{1^+}^2}\,
{\cal F}(K^2/\omega_{1^+}^2) \,,
\end{eqnarray}
\end{subequations}
where the two parameters $m_{J^P}$ are diquark pseudoparticle masses and
$\omega_{J^P}$ are widths characterising $\Gamma^{J^P}\!$.  Herein we require
additionally that
\begin{equation}
\label{DQPropConstr}
\left. \frac{d}{d K^2}\,\left(\frac{1}{m_{J^P}^2}\,{\cal
F}(K^2/\omega_{J^P}^2)\right)^{-1} \right|_{K^2=0}\! = 1 \; \Rightarrow \;
\omega_{J^P}^2 = \sfrac{1}{2}\,m_{J^P}^2\,,
\end{equation}
which is a normalisation that accentuates the free-particle-like propagation
characteristics of the diquarks {\it within} the hadron.


\setcounter{equation}{0}
\section{Euclidean Conventions}
\label{App:EM}
In our Euclidean formulation:
\begin{equation}
p\cdot q=\sum_{i=1}^4 p_i q_i\,;
\end{equation}
\begin{equation}
\{\gamma_\mu,\gamma_\nu\}=2\,\delta_{\mu\nu}\,;\;
\gamma_\mu^\dagger = \gamma_\mu\,;\;
\sigma_{\mu\nu}= \sfrac{i}{2}[\gamma_\mu,\gamma_\nu]\,; \;
{\rm tr}\,[\gamma_5\gamma_\mu\gamma_\nu\gamma_\rho\gamma_\sigma]=
-4\,\epsilon_{\mu\nu\rho\sigma}\,, \epsilon_{1234}= 1\,.
\end{equation}

A positive energy spinor satisfies
\begin{equation}
\bar u(P,s)\, (i \gamma\cdot P + M) = 0 = (i\gamma\cdot P + M)\, u(P,s)\,,
\end{equation}
where $s=\pm$ is the spin label.  It is normalised:
\begin{equation}
\bar u(P,s) \, u(P,s) = 2 M \,,
\end{equation}
and may be expressed explicitly:
\begin{equation}
u(P,s) = \sqrt{M- i {\cal E}}\left(
\begin{array}{l}
\chi_s\\
\displaystyle \frac{\vec{\sigma}\cdot \vec{P}}{M - i {\cal E}} \chi_s
\end{array}
\right)\,,
\end{equation}
with ${\cal E} = i \sqrt{\vec{P}^2 + M^2}$,
\begin{equation}
\chi_+ = \left( \begin{array}{c} 1 \\ 0  \end{array}\right)\,,\;
\chi_- = \left( \begin{array}{c} 0\\ 1  \end{array}\right)\,.
\end{equation}
For the free-particle spinor, $\bar u(P,s)= u(P,s)^\dagger \gamma_4$.

The spinor can be used to construct a positive energy projection operator:
\begin{equation}
\label{Lplus} \Lambda_+(P):= \frac{1}{2 M}\,\sum_{s=\pm} \, u(P,s) \, \bar
u(P,s) = \frac{1}{2M} \left( -i \gamma\cdot P + M\right).
\end{equation}

A negative energy spinor satisfies
\begin{equation}
\bar v(P,s)\,(i\gamma\cdot P - M) = 0 = (i\gamma\cdot P - M) \, v(P,s)\,,
\end{equation}
and possesses properties and satisfies constraints obtained via obvious analogy
with $u(P,s)$.

A charge-conjugated Bethe-Salpeter amplitude is obtained via
\begin{equation}
\label{chargec}
\bar\Gamma(k;P) = C^\dagger \, \Gamma(-k;P)^{\rm T}\,C\,,
\end{equation}
where ``T'' denotes a transposing of all matrix indices and
$C=\gamma_2\gamma_4$ is the charge conjugation matrix, $C^\dagger=-C$. We note that
\begin{equation}
C^{\dagger}\gamma_{\mu}^{T}C = -\gamma_{\mu}, [C,\gamma_{5}] =0.
\end{equation}

In describing the $\Delta$ resonance we employ a Rarita-Schwinger spinor to
unambiguously represent a covariant spin-$3/2$ field. The positive energy spinor is
defined by the following equations:
\begin{equation}
(i\gamma\cdot P + M) u_{\mu}(P;r) = 0,\quad \gamma_{\mu}u_{\mu}(P;r) = 0,\quad
P_{\mu}u_{\mu}(P;r) = 0,
\label{eq:rarita}
\end{equation}
where $r=-3/2,-1/2,1/2,3/2$. It is normalised:
\begin{equation}
\bar{u}_{\mu}(P;r) \, u_{\mu}(P,r) = 2 M \,,
\end{equation}
and satisfies a completeness relation
\begin{equation}
\frac{1}{2M} \sum_{r=-3/2}^{3/2} u_{\mu}(P;r) \bar{u}_{\nu}(P;r) =
\Lambda_{+}(P)R_{\mu\nu},
\end{equation}
where
\begin{equation}
R_{\mu\nu} = \delta_{\mu\nu} I_{D} - \frac{1}{3}\gamma_{\mu}\gamma_{\nu} +
\frac{2}{3} \hat{P}_{\mu}\hat{P}_{\nu} I_{D} - i \frac{1}{3}
\left[\hat{P}_{\mu}\gamma_{\nu} - \hat{P}_{\nu}\gamma_{\mu} \right]\,,
\end{equation}
with $\hat{P}^{2}=-1$, which is very useful in simplifying the positive energy
$\Delta$'s Faddeev equation.


\setcounter{equation}{0}
\setcounter{figure}{0}
\section{Baryon-Photon Vertex}
\label{NPVertex}
In order to compute the electromagnetic vertices one must specify how the photon couples to the constituents within the composite hadrons.  In the present context this amounts to specifying the nature of the couplings of the photon to the dressed quarks and the diquark correlations, since the incoming and outgoing baryons are described by the quark-diquark Faddeev amplitudes.

\begin{figure}[t]
\begin{minipage}[t]{\textwidth}
\begin{minipage}[t]{0.45\textwidth}
\centerline{\includegraphics[width=0.70\textwidth]{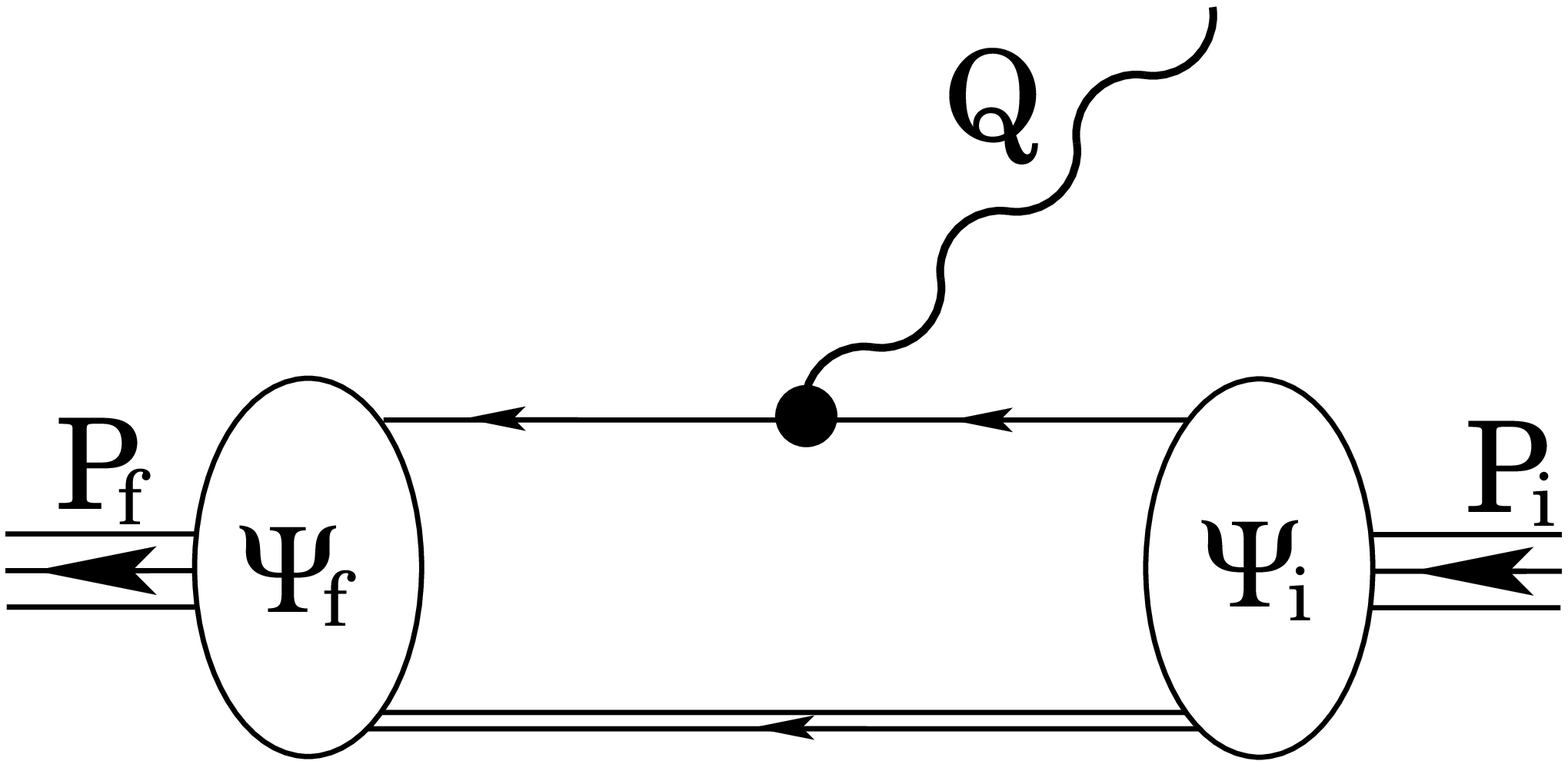}}
\end{minipage}
\begin{minipage}[t]{0.45\textwidth}
\centerline{\includegraphics[width=0.70\textwidth]{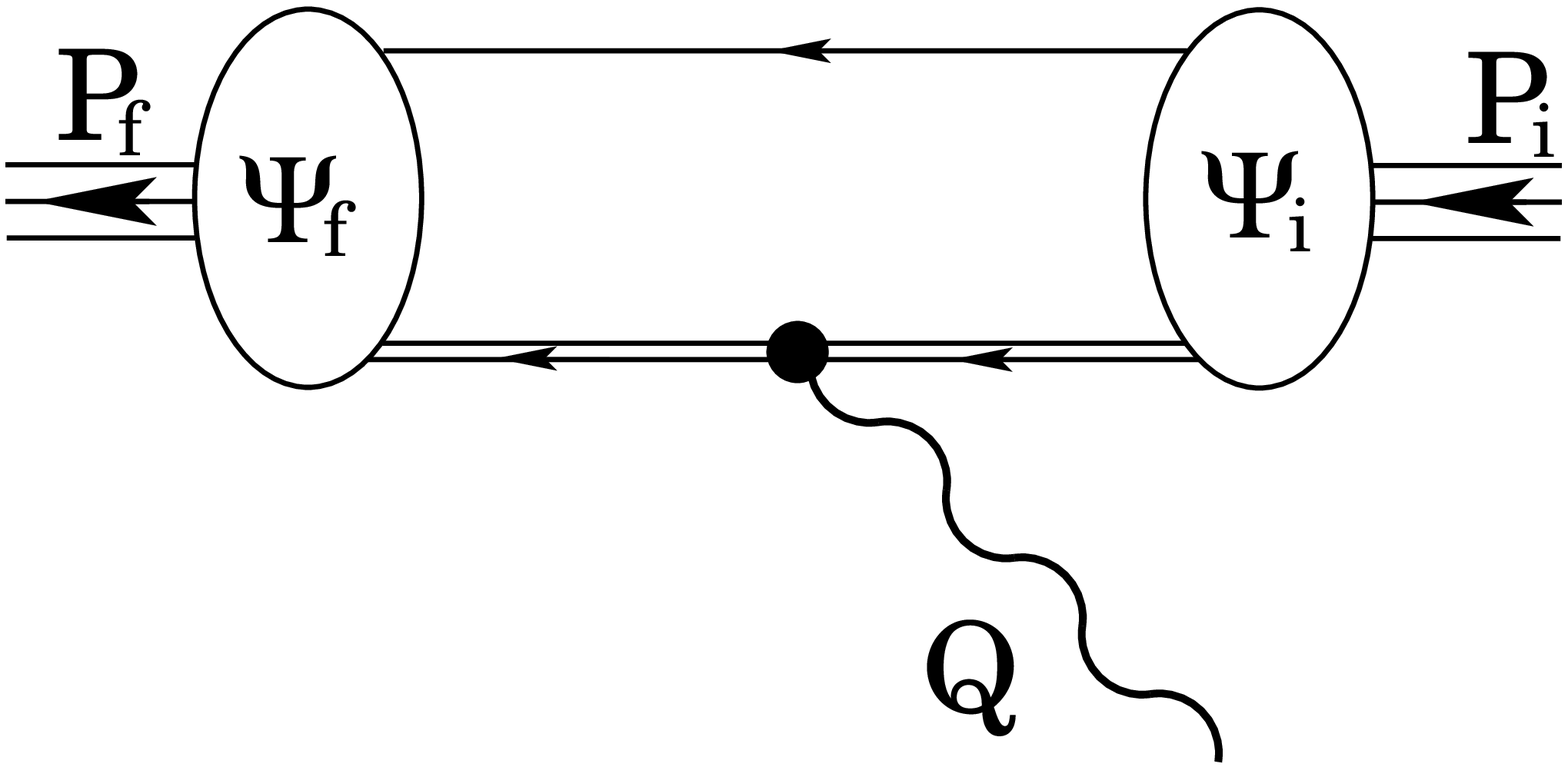}}
\end{minipage}\vspace*{3ex}

\begin{minipage}[t]{0.45\textwidth}
\centerline{\includegraphics[width=0.70\textwidth]{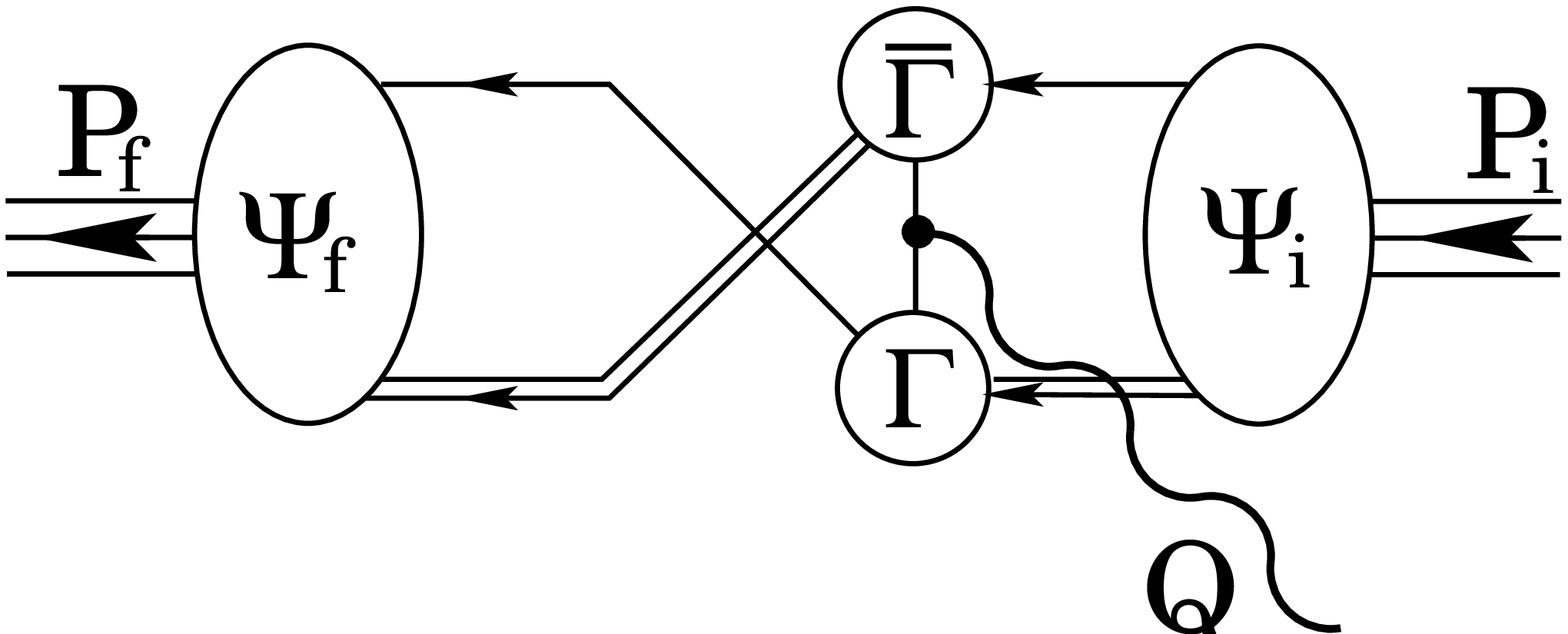}}
\end{minipage}
\begin{minipage}[t]{0.45\textwidth}
\centerline{\includegraphics[width=0.70\textwidth]{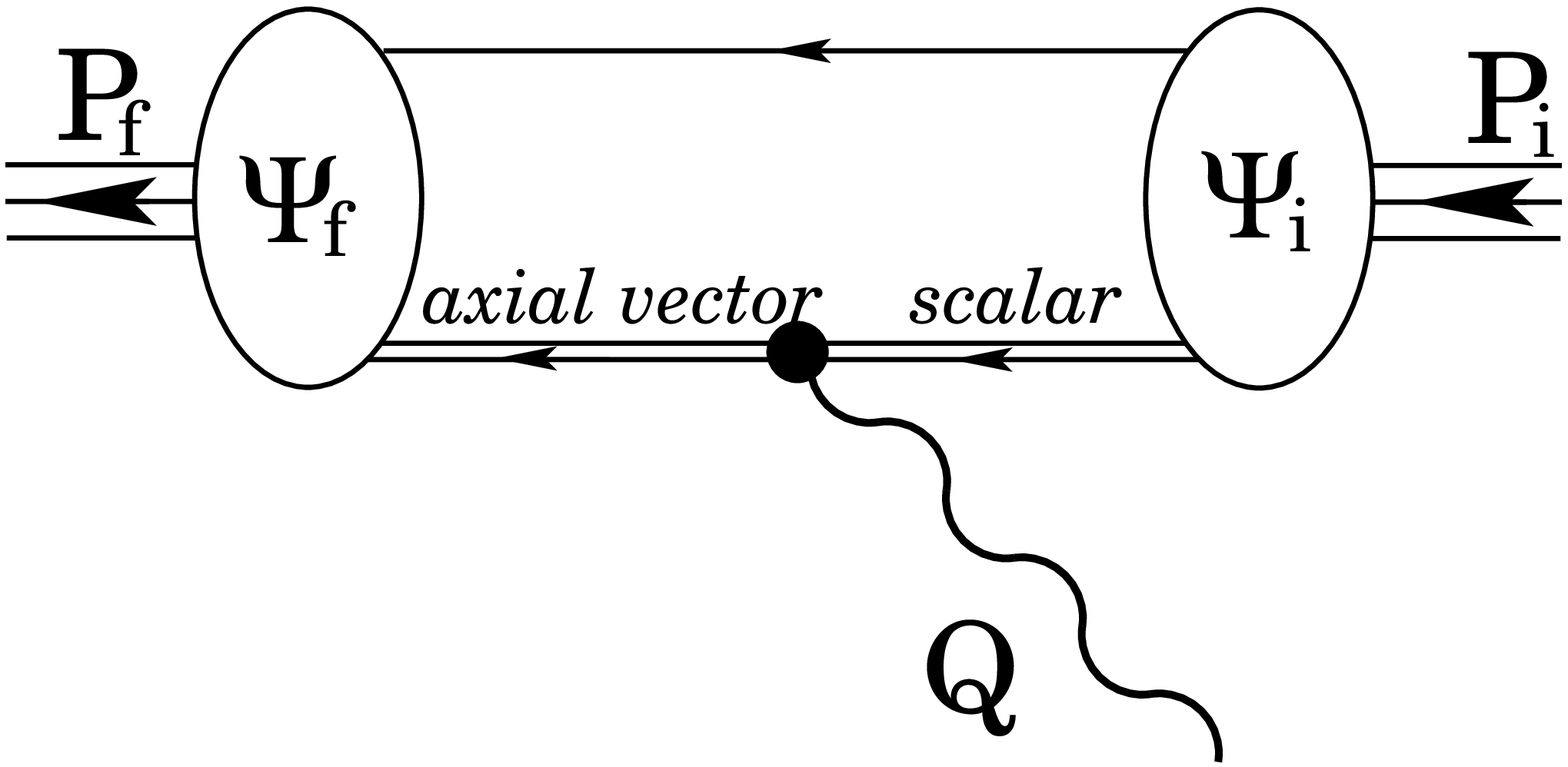}}
\end{minipage}\vspace*{3ex}

\begin{minipage}[t]{0.45\textwidth}
\centerline{\includegraphics[width=0.70\textwidth]{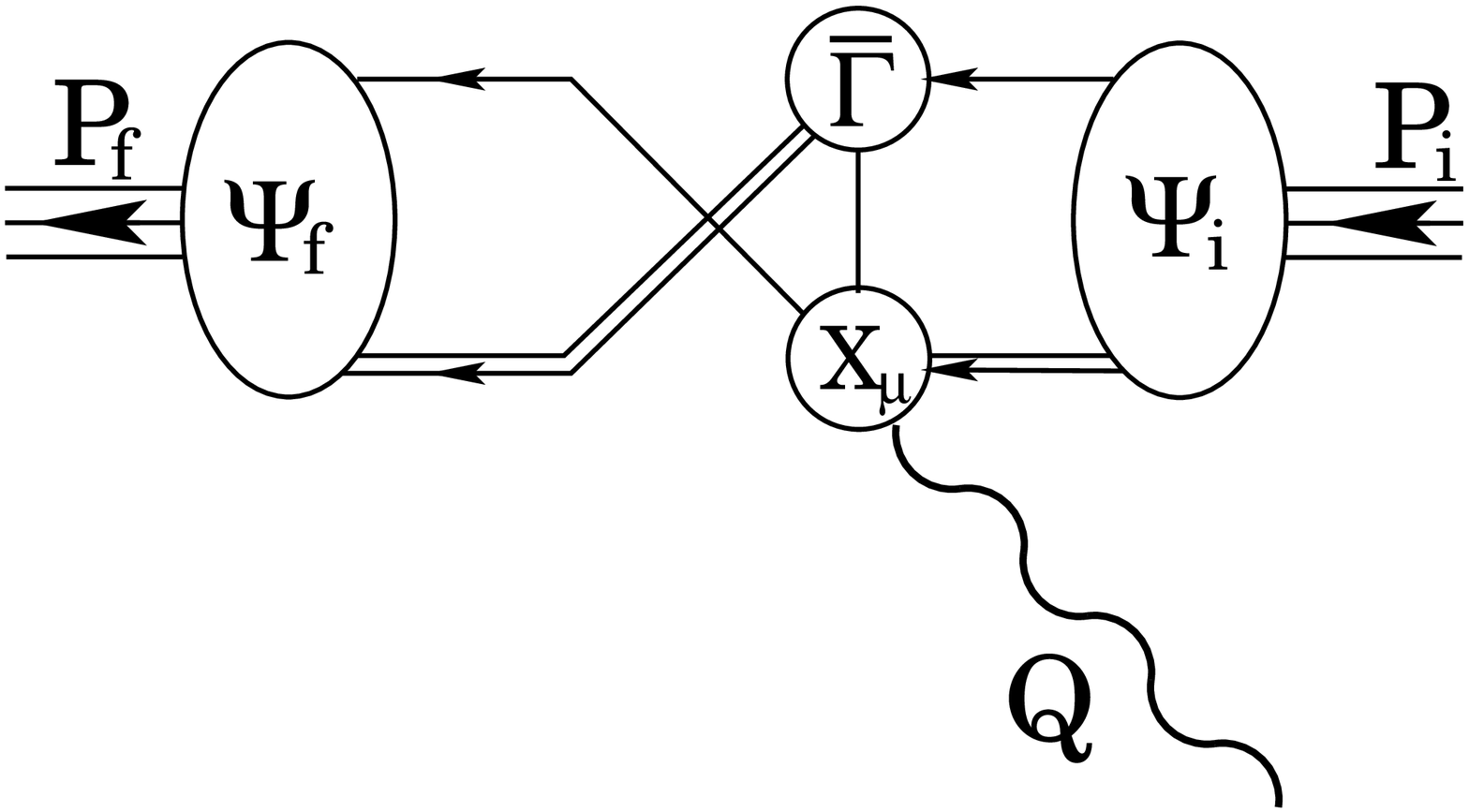}}
\end{minipage}
\begin{minipage}[t]{0.45\textwidth}
\centerline{\includegraphics[width=0.70\textwidth]{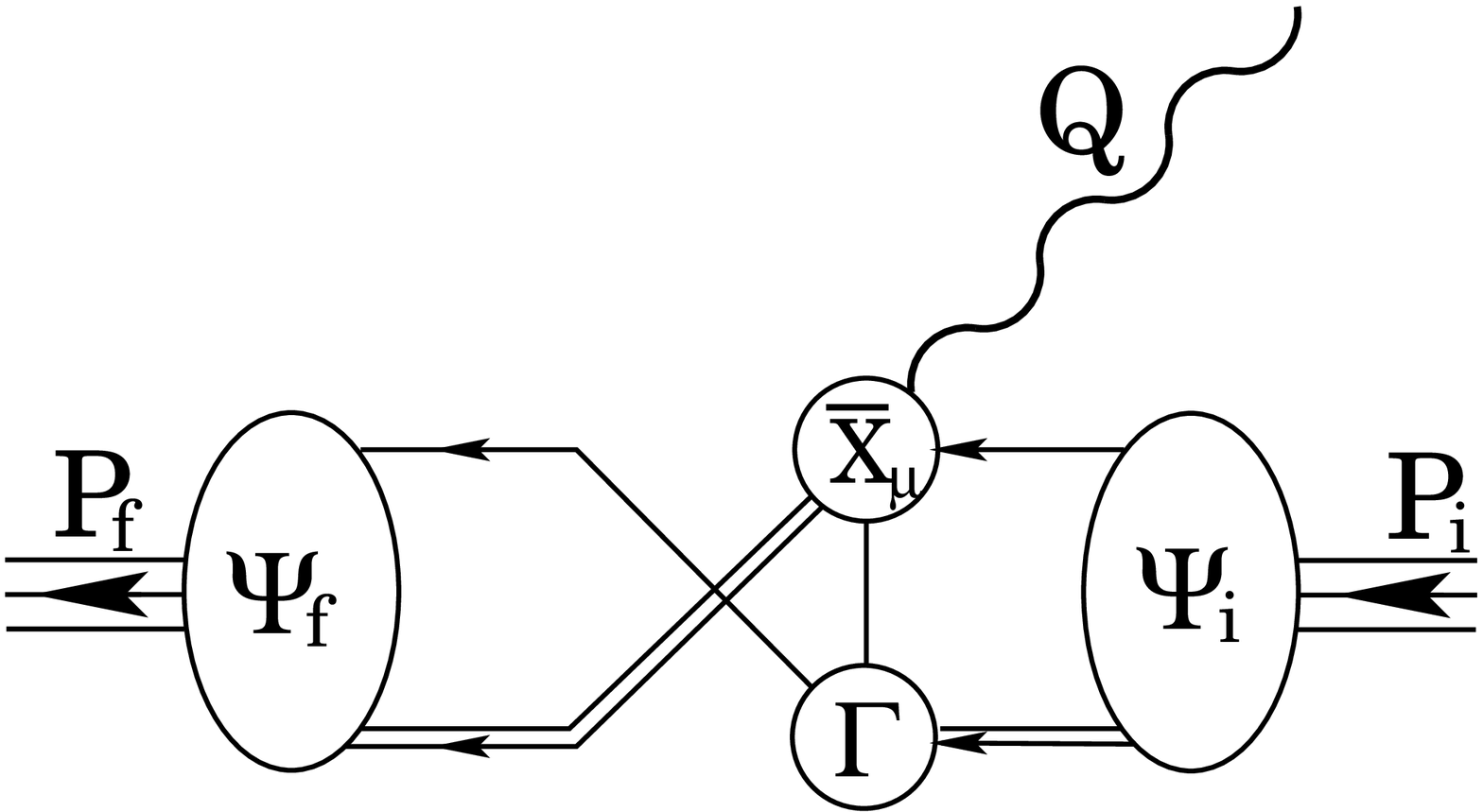}}
\end{minipage}
\end{minipage}
\caption{\label{vertex} Vertex which ensures a conserved current for on-shell baryons described by the Faddeev amplitudes, $\Psi_{i,f}$, described in Sect.\,\protect\ref{sec:BaryonModel} and App.\,\protect\ref{app:FE}.  The single line represents $S(p)$, the dressed-quark propagator, Sec.\,\protect\ref{subsubsec:S}, and the double line, the diquark propagator, Sec.\,\protect\ref{qqprop}; $\Gamma$ is the diquark Bethe-Salpeter amplitude, Sec.\,\protect\ref{qqBSA}; and the remaining vertices are described in App.\,\ref{NPVertex} -- the top-left image is Diagram~1; the top-right, Diagram~2; and so on, with the bottom-right image, Diagram~6.}
\end{figure}

In Fig.\,\ref{vertex} we have separated the different contributions to the currents into six terms, each of which we subsequently make precise. N.B.\,Diagrams~1, 2 and 4 are one-loop integrals, which we evaluate by Gau{\ss}ian quadrature.  The remainder, Diagrams 3, 5 and 6, are two-loop integrals, for whose evaluation Monte-Carlo methods are employed.  A technical aspect concerning the computation is
described in App.\,\ref{CHexpand}.

For explicit calculations, we work in the Breit frame: $P_\mu=P_\mu^{BF}-Q_\mu /2$,
$P'_\mu=P_\mu^{BF}+Q_\mu /2$ and $P_\mu^{BF}=(0,0,0,i\sqrt{M_n^2+Q^2/4})$.  There is no scalar diquark correlation inside the $\Delta(1232)$-baryon and so appropriate simplifications of the general formulae below should be used in that case.

\subsection{Diagram~1}
\label{Diag1}

This represents the photon coupling directly to the bystander quark. It is expressed
as \begin{eqnarray}
\label{B1}
J_{\mu}^{qu} &=&  S(p_q) \hat{\Gamma}_{\mu}^{qu}(p_q;k_q) S(k_q)
\left(\Delta^{0^+}(k_s) + \Delta^{1^+}(k_s) \right)
(2\pi)^4 \delta^4(p-k-\hat{\eta}Q)\,,
\end{eqnarray}
where $\hat\Gamma_{\mu}^{qu}(p_q;k_q)= Q_q \, \Gamma_{\mu}(p_q;k_q)$, with $Q_q={\rm
diag}[2/3,-1/3]$ being the quark electric charge matrix, and $\Gamma_{\mu}(p_q;k_q)$
is the dressed-quark-photon vertex.  In Eq.\,(\ref{B1}) the momenta are
\begin{eqnarray}
\label{etavalue}
\begin{array}{lc@{\qquad}l}
k_q=\eta P+k\,, & & p_q=\eta P'+p\,, \\
k_d=\hat{\eta}P-k\,, & & p_d=\hat{\eta}P'-p\,,
\end{array}
\end{eqnarray}
with $\eta + \hat{\eta}=1$.  The results reported herein were obtained with $\eta=1/3$ for the nucleon and $\eta=2/5$ for the $\Delta$-baryon.  The value $\eta=1/3$ provides a single quark with one-third of the baryon's total momentum and is thus the natural choice.  
Notably, in a manifestly Poincar\'e covariant approach the precise value of $\eta$ is immaterial so long as the numerical methods preserve that covariance.  However, we have retained only the leading-order piece of the diquark Bethe-Salpeter amplitudes.  This leads to a subleading violation of Poincar\'e covariance, so our results exhibit some sensitivity to $\eta$.  That sensitivity is least  with the choices mentioned above.

It is a necessary condition for current conservation that the quark-photon vertex
satisfy the Ward-Green-Takahashi (WGT) identity:
\begin{equation}
\label{vwti}
Q_\mu \, i\Gamma_\mu(\ell_1,\ell_2) = S^{-1}(\ell_1) - S^{-1}(\ell_2)\,,
\end{equation}
where $Q=\ell_1-\ell_2$ is the photon momentum flowing into the vertex.  The vertex can be obtained by solving an inhomogeneous Bethe-Salpeter equation.  However, since we have parametrised $S(p)$, we follow Ref.~\cite{Roberts:1994hh} and write \cite{Ball:1980ay}
\begin{equation}
\label{bcvtx}
i\Gamma_\mu^{BC}(\ell_1,\ell_2)  =
i\Sigma_A(\ell_1^2,\ell_2^2)\,\gamma_\mu +
2 k_\mu \left[i\gamma\cdot k_\mu \,
\Delta_A(\ell_1^2,\ell_2^2) + \Delta_B(\ell_1^2,\ell_2^2)\right] \!;
\end{equation}
with $k= (\ell_1+\ell_2)/2$, $Q=(\ell_1-\ell_2)$ and
\begin{equation}
\Sigma_F(\ell_1^2,\ell_2^2) = \sfrac{1}{2}\,[F(\ell_1^2)+F(\ell_2^2)]\,,\;
\Delta_F(\ell_1^2,\ell_2^2) =
\frac{F(\ell_1^2)-F(\ell_2^2)}{\ell_1^2-\ell_2^2}\,,
\label{DeltaF}
\end{equation}
where $F= A, B$; viz., the scalar functions in Eq.\,(\ref{SpAB}).  It is critical that $\Gamma_\mu$ in Eq.\ (\ref{bcvtx}) satisfies Eq.\ (\ref{vwti}) and very useful that it is completely determined by the dressed-quark propagator.  Notably, an analysis of the coupled longitudinal and transverse WGT identities shows that Eq.\,\eqref{bcvtx} is uniquely determined as the ``longitudinal part'' of the dressed-vertex \cite{Qin:2013mta}.

In the presence of dynamical chiral symmetry breaking, a dressed light-quark possesses a large anomalous electromagnetic moment \cite{Singh:1985sg,Bicudo:1998qb,Kochelev:1996pv,Chang:2010hb,%
Bashir:2011dp,Qin:2013mta}.  To illustrate the effect on form factors that one might expect from this phenomenon, we often present results obtained with the dressed-quark--photon interaction modified as follows:
\begin{equation}
\Gamma_{\mu}^f(\ell_1,\ell_2) = \Gamma_{\mu}^{f,BC}(\ell_1,\ell_2) - \kappa_f
\sigma_{\mu\nu}q_{\nu} \Delta_{B}^f(\ell_1^{2},\ell_2^{2})\,,
\label{eq:DQAMM}
\end{equation}
where $f$ labels the quark flavour and $\kappa_f=0.4$ is the modulating magnitude.\footnote{$\kappa_f=0.4$ was the value used in Ref.\,\cite{Chang:2011tx}.  As explained in Refs.\,\cite{Chang:2010hb,Bashir:2011dp,Qin:2013mta}, any value on the order of $1/2$ is justified.}  We ignore the effect this modification would have on electromagnetic form factors involving the axial-vector diquark correlation.  Such modifications are noticeable in the magnetic and quadrupole form factors \cite{Wilson:2011aa} but this has no bearing on the points we wish to highlight.

In this connection, we note that the two-flavour quark charge operator $\hat{Q} = \frac{1}{6} \boldmath{I} + \frac{1}{2} \tau_3$ has both isoscalar and isovector components.  Therefore the anomalous electromagnetic moment (AMM) of the dressed $u$- and $d$-quarks can differ.  This may be illustrated by considering the contribution that pion loops can conceivably produce \cite{Horikawa:2005dh,Cloet:2014rja}, in which case it was found that $\zeta_u - \zeta_d \approx 1/2$; i.e., the isovector combination is large, and $\zeta_d + 2 \zeta_u \approx 0$.

It follows that when considering form factors involving the spin-$1/2$, isospin-$1/2$ neutron and proton, a reliable estimate of the effect produced by dressed-quark AMMs can be obtained by ignoring the flavour dependence in Eq.\,\eqref{eq:DQAMM} and using a common value of $\varsigma$.

On the other hand, the $J=3/2$, $I=3/2$ $\Delta^+$ baryon is predominantly a quark$\,+\,$axial-vector diquark in a relative $S$-wave, so that the spin-flavour wave function for the $\Delta^+$ may be represented as $ \sqrt{2} \, u_\uparrow\{u_\uparrow d_\uparrow\} + d_\uparrow \{u_\uparrow u_\uparrow\}$.  The dressed-quark AMM contribution must largely cancel within a bound-state with this spin-flavour structure.  This expectation is supported by an analysis of lattice-QCD results for $\Delta$-baryon form factors \cite{Cloet:2003jm}.  We therefore ignore the dressed-quark AMM when computing elastic form factors of the decuplet baryons.

\subsection{Diagram~2}
\label{dqff}

This figure depicts the photon coupling directly to a diquark correlation.  It is
expressed as
\begin{eqnarray}
\label{B2}
J_{\mu}^{dq} &=& \Delta^i(p_{d})
\left[ \hat{\Gamma}_{\mu}^{dq}(p_{d};k_{d}) \right]^{i j}
\Delta^{j}(k_{d}) S(k_q)
(2\pi)^4 \delta^4(p-k+\eta Q)\,,
\end{eqnarray}
with $ [\hat{\Gamma}_{\mu}^{dq}(p_{d};k_{d})]^{i j}={\rm diag}[Q_{0^+}
\Gamma_\mu^{0^+},Q_{1^+}\Gamma_\mu^{1^+}] $, where $Q_{0^+}=1/3$ and
$\Gamma_\mu^{0^+}$ is given in Eq.\,(\ref{Gamma0plus}), and $Q_{1^+}={\rm
diag}[4/3,1/3,-2/3]$ with $\Gamma_\mu^{1^+}$ given in Eq.\,(\ref{AXDQGam}).
Naturally, the diquark propagators match the line to which they are attached.

In the case of a scalar correlation, the general form of the diquark-photon vertex is
\begin{equation}
\Gamma_\mu^{0^+}(\ell_1,\ell_2) = 2\, k_\mu\, f_+(k^2,k\cdot Q,Q^2) + Q_\mu  \,
f_-(k^2,k\cdot Q,Q^2)\,.
\end{equation}
If one is dealing with an elementary scalar correlation, then the WGT identity reads:
\begin{equation}
\label{VWTI0}
Q_\mu \,\Gamma_\mu^{0^+}(\ell_1,\ell_2) = \Pi^{0^+}(\ell_1^2)  -
\Pi^{0^+}(\ell_2^2)\,,\; \Pi^{J^P}(\ell^2) = \{\Delta^{J^P}(\ell^2)\}^{-1}.
\end{equation}
However, for a composite system of the type we consider this identity is modified;
viz., \cite{Frank:1993ye}:
\begin{equation}
\label{VWTI0Comp}
Q_\mu \,\Gamma_\mu^{0^+}(\ell_1,\ell_2) = \left[\Pi^{0^+}(\ell_1^2)  -
\Pi^{0^+}(\ell_2^2)\right] F_{qq}(Q^2)\,,
\end{equation}
where
\begin{equation}
\label{Fqqform}
F_{qq}(Q^2) = \frac{1}{1 + \frac{1}{6} r_{qq}^2 Q^2}
\end{equation}
is a form factor describing the distribution of charge within the correlation.

The evaluation of scalar diquark elastic electromagnetic form factors in Ref.\,\cite{Maris:2004bp} is a first step toward calculating $\Gamma_\mu^{0^+}(\ell_1,\ell_2)$.  However, in providing only an on-shell component,
it is insufficient for our requirements.  We choose to adapt Eq.\,(\ref{bcvtx}) to
our needs and employ
\begin{equation}
\label{Gamma0plus}
\Gamma_\mu^{0^+}(\ell_1,\ell_2) =  k_\mu\,
\Delta_{\Pi^{0^+}}(\ell_1^2,\ell_2^2)\,F_{qq}(Q^2)\,,
%
\end{equation}
with the definition of $\Delta_{\Pi^{0^+}}(\ell_1^2,\ell_2^2)$ apparent from
Eq.\,(\ref{DeltaF}) and $r_{qq}=0.80\,{\rm fm}$, Eq.\,\eqref{radiusqq}.

Equation~(\ref{Gamma0plus}) is an \textit{Ansatz} that satisfies Eq.\,(\ref{VWTI0Comp}), is completely determined by quantities introduced already and is free of kinematic singularities on the relevant domain.  It implements $f_- \equiv 0$, which is a requirement for elastic form factors, and guarantees a valid normalisation of electric charge; viz.,
\begin{equation}
\lim_{\ell^\prime\to \ell} \Gamma_\mu^{0^+}(\ell^\prime,\ell) = 2 \, \ell_{\mu} \,
\frac{d}{d\ell^2}\, \Pi^{0^+}(\ell^2) \stackrel{\ell^2\sim 0}{=} 2 \, \ell_{\mu}\,,
\end{equation}
owing to Eq.\,(\ref{DQPropConstr}).  N.B.\ We have factored the fractional diquark
charge, which therefore appears subsequently in our calculations as a simple multiplicative factor.

For the case in which the struck diquark correlation is axial-vector and the
scattering is elastic, the vertex assumes the form
\cite{Hawes:1998bz}:\,\footnote{If the scattering is inelastic the general form
of the vertex involves eight scalar functions \protect\cite{Salam:1964zk}.  For
simplicity, we ignore the additional structure in this \textit{Ansatz}.}
\begin{equation}
\label{AXDQGam}
\Gamma^{1^+}_{\mu\alpha\beta}(\ell_1,\ell_2)
= -\sum_{i=1}^{3} \Gamma^{\rm [i]}_{\mu\alpha\beta}(\ell_1,\ell_2)\,,
\end{equation}
with ($T_{\alpha\beta}(\ell) = \delta_{\alpha\beta} - \ell_\alpha \ell_\beta/\ell^2$)
\begin{subequations}
\label{AXDQGamT}
\begin{eqnarray}
\label{AXDQGam1}
\Gamma^{\rm [1]}_{\mu\alpha\beta}(\ell_1,\ell_2)
&=& (\ell_1+\ell_2)_\mu \, T_{\alpha\lambda}(\ell_1) \, T_{\lambda\beta}(\ell_2)\;
F_1(\ell_1^2,\ell_2^2)\,,
\\
\label{AXDQGam2}
\Gamma^{\rm [2]}_{\mu\alpha\beta}(\ell_1,\ell_2)
&=& \left[ T_{\mu\alpha}(\ell_1)\, T_{\beta\rho}(\ell_2) \, \ell_{1 \rho}
+ T_{\mu\beta}(\ell_2) \, T_{\alpha\rho}(\ell_1) \, \ell_{2\rho} \right]
F_{2}(\ell_1^2,\ell_2^2) \,,
\\
\label{AXDQGam3}
\Gamma^{\rm [3]}_{\mu\alpha\beta}(\ell_1,\ell_2)
&=& -\frac{1}{2 m_{1^+}^2}\, (\ell_1+\ell_2)_\mu\, T_{\alpha\rho}(\ell_1)\, \ell_{2
\rho}
\, T_{\beta\lambda}(\ell_2)\, \ell_{1 \lambda}\; F_{3}(\ell_1^2,\ell_2^2) \,.
\end{eqnarray}
\end{subequations}
This vertex satisfies:
\begin{equation}
\ell_{1\alpha} \, \Gamma^{1^+}_{\mu\alpha\beta}(\ell_1,\ell_2) = 0 =
\Gamma^{1^+}_{\mu\alpha\beta}(\ell_1,\ell_2) \, \ell_{2\beta} \,,
\end{equation}
which is a general requirement of the elastic electromagnetic vertex of axial-vector
bound states and guarantees that the interaction does not induce a pseudoscalar
component in the axial-vector correlation.  We note that the electric, magnetic and
quadrupole form factors of an axial-vector bound state are expressed
\cite{Hawes:1998bz}
\begin{subequations}
\begin{eqnarray}
\label{GEDQ}
& &
G_{\cal E}^{1^+}(Q^2) = F_1 + \sfrac{2}{3}\, \tau_{1^+}\,
G_{\cal Q}^{1^+}(Q^2) \,, \; \tau_{1^+} = \frac{Q^2}{ 4 \,m_{1^{+}}^{2}}\,,
\\
\label{GMDQ}
& &
G_{\cal M}^{1^+}(Q^2) = - F_2(Q^2) ~,
\\
& &
\label{GQDQ}
G_{\cal Q}^{1^+}(Q^2) = F_1(Q^2) + F_2(Q^2) + \left( 1 + \tau_{1^+}\right) F_3(Q^2)
\,.
\end{eqnarray}
\end{subequations}

Owing to the fact that $\Gamma^{J^P}_C:= \Gamma^{J^P}C^\dagger$ satisfies exactly the
same Bethe-Salpeter equation as the $J^{-P}$ colour-singlet meson {\it but} for a
halving of the coupling strength, the vector meson form factor calculation in
Ref.\,\cite{Bhagwat:2006pu} might become useful as a guide in understanding the form
factors in Eqs.\,(\ref{AXDQGam}), (\ref{AXDQGamT}).  However, in providing only an
on-shell component, that information is insufficient for our requirements.  Hence we
employ the following \textit{Ans\"atze}:
\begin{subequations}
\label{AnsatzFT}
\begin{eqnarray}
\label{AnsatzF1}
F_{1}(\ell_1^2,\ell_2^2) &=& \Delta_{\Pi^{1^+}}(\ell_1^2,\ell_2^2)\,F_{qq}(Q^2)\,, \\
\label{AnsatzF2}
F_{2}(\ell_1^2,\ell_2^2) &=& -\, F_{1}  +
(1-\tau_{1^+}) \,( \tau_{1^+} F_{1}+1 - \mu_{1^+})\, d(\tau_{1^+})\,, \\
\label{AnsatzF3}
F_{3}(\ell_1^2,\ell_2^2) &=& -\,(\chi_{1^+}\,(1- \tau_{1^+})\,d(\tau_{1^+})+F_1 +
F_2)\, d(\tau_{1^+})\,,
\end{eqnarray}
\end{subequations}
with $d(x)=1/(1+x)^2$.  This construction ensures a valid electric charge normalisation for the axial-vector correlation; viz.,
\begin{equation}
\lim_{\ell^\prime \to\ell} \, \Gamma^{1^+}_{\mu\alpha\beta}(\ell^\prime,\ell) =
T_{\alpha\beta}(\ell) \,\frac{d}{d\ell^2}\, \Pi^{1^+}(\ell^2)
\stackrel{\ell^2\sim 0}{=}  T_{\alpha\beta}(\ell) \,2 \,\ell_{ \mu}\,,
\end{equation}
owing to Eq.\,(\ref{DQPropConstr}), and current conservation
\begin{equation}
\lim_{\ell_2\to\ell_1} \, Q_\mu \Gamma^{1^+}_{\mu\alpha\beta}(\ell_1,\ell_2) = 0\,.
\end{equation}
The diquark's static electromagnetic properties follow:
\begin{equation}
\label{pointp}
G_{\cal E}^{1^+}(0) = 1\,,\;
G_{\cal M}^{1^+}(0) = \mu_{1^+}\,,\;
G_{\cal Q}^{1^+}(0) = -\chi_{1^+}\,.
\end{equation}
For an on-shell or pointlike axial-vector: $\mu_{1^+}=2$; and $\chi_{1^+}=1$.  In
addition, Eqs.\,(\ref{AXDQGam}), (\ref{AXDQGamT}) with Eqs.\,\eqref{AnsatzFT} realise the constraints of Ref.\,\cite{Brodsky:1992px}; namely, independent of the values of $\mu_{1^+}$ \& $\chi_{1^+}$, the form factors assume the ratios
\begin{equation}
\label{pQCDavdq}
G_{\cal E}^{1^+}(Q^2): G_{\cal M}^{1^+}(Q^2): G_{\cal Q}^{1^+}(Q^2)
\stackrel{Q^2\to \infty}{=} (1 - \sfrac{2}{3} \tau_{1^+}) : 2 : - 1 \,.
\end{equation}

It is noteworthy that within a nucleon the diquark correlation is not on-shell. Hence, in contrast to Ref.\,\cite{Alkofer:2004yf}, we do not assume herein that a point-particle value for the magnetic moment in Eq.\,(\ref{pointp}) serves as a good
reference point.  Instead we employ the value:
\begin{equation}
\mu_{1^+}=1.0\,,
\end{equation}
which may be compared with $\mu_{1^+}=0.37$ in Ref.~\cite{Cloet:2008wg}.  The elastic electromagnetic form factors of the nucleon are not particularly sensitive to the magnetic properties of the axial vector diquark because, amongst other things,  axial-vector correlations inside the nucleon are less probable than scalar correlations, owing to DCSB \cite{Chen:2012qr} (see Table~\ref{tab:probabilities}).  On the other hand, axial-vector diquarks are the only quark-quark correlations within the $\Delta$ and hence their properties have a large influence on $\Delta$-baryon properties.  It follows that changes may be made to $\mu_{1^+}$ in order to improve a description of the $\Delta$ without much affecting nucleon properties.  N.B.\ Whilst one need not employ the point-particle value for $\chi_{1^+}$, changing to $\chi_{1^+}=0$ has little impact on the results \protect\cite{Alkofer:2004yf}.  We therefore retain $\chi_{1^+}=1$.

\subsection{Diagram~3}

This image depicts a photon coupling to the quark that is exchanged as one diquark
breaks up and another is formed.  It is expressed as
\begin{equation}
\label{B3}
J_{\mu}^{ex} = -\frac{1}{2} S(k_{q}) \Delta^{i}(k_{d})
\Gamma^{i}(p_1,k_{d})
S^T(q) \hat{\Gamma}_{\mu}^{quT}(q',q) S^T(q')
\bar{\Gamma}^{jT}(p'_2,p_{d})
\Delta^j(p_{d}) S(p_{q})\,,
\end{equation}
wherein the vertex $\hat{\Gamma}_{\mu}^{qu}$ appeared in Eq.\,(\ref{B1}).  While this
is the first two-loop diagram we have described, no new elements appear in its
specification: the dressed-quark-photon vertex was discussed in App.\,\ref{Diag1}.  In
Eq.\,(\ref{B3}) the momenta are
\begin{eqnarray}
\begin{array}{lc@{\qquad}l}
q = \hat{\eta}P-\eta P'-p-k\,, & & q' = \hat{\eta}P'-\eta P-p-k \,,\\
p_1 = (p_q-q)/2\,,& & p'_2 = (-k_q+q')/2 \,,\\
p'_1 = (p_q-q')/2\,, & & p_2 = (-k_q+q)/2 \,.
\end{array}
\end{eqnarray}

It is noteworthy that the process of quark exchange provides the attraction necessary
in the Faddeev equation to bind the nucleon.  It also guarantees that the Faddeev
amplitude has the correct antisymmetry under the exchange of any two dressed-quarks.
This key feature is absent in models with elementary (noncomposite) diquarks.  The
full contribution is obtained by summing over the superscripts $i,j$, which can each
take the values $0^+$, $1^+$.

\subsection{Diagram~4}
\label{kTsec}
This differs from Diagram~2 in expressing the contribution to baryon form factors owing to an electromagnetically induced transition between scalar and axial-vector diquarks.  The explicit expression is given by Eq.\,(\ref{B2}) with
$[\hat{\Gamma}_{\mu}^{dq}(p_{d};k_{d})]^{i= j}=0$, and
$[\hat{\Gamma}_{\mu}^{dq}(p_{d};k_{d})]^{1,2}=\Gamma_{SA}$ and
$[\hat{\Gamma}_{\mu}^{dq}(p_{d};k_{d})]^{2,1}=\Gamma_{AS}$.
This transition vertex is a rank-2 pseudotensor, kindred to the matrix element describing the $\rho\,\gamma^\ast \pi^0$ transition \cite{Maris:2002mz}, and can therefore b expressed
\begin{equation}
\label{SAPhotVertex}
\Gamma_{SA}^{\gamma\alpha}(\ell_1,\ell_2) =
-\Gamma_{AS}^{\gamma\alpha}(\ell_1,\ell_2)
= \frac{i}{M_N} \, {\cal T}(\ell_1,\ell_2) \,
\varepsilon_{\gamma\alpha\rho\lambda}\ell_{1\rho} \ell_{2 \lambda}\,,
\end{equation}
where $\gamma$, $\alpha$ are, respectively, the vector indices of the photon and
axial-vector diquark. For simplicity we proceed under the assumption
\begin{equation}
\label{calTvalue}
{\cal T}(\ell_1,\ell_2) = \kappa_{\cal T} F_{qq}(Q^2) \,.
\end{equation}
A typical on-shell value for the dimensionless normalisation is
$\kappa_{\cal T} \sim 2$ \cite{Oettel:2000jj}.
However, as with $\mu_{1^+}$, we recognise herein that this value is not a useful reference point because, for the processes described by Fig.\,\ref{vertex}, $\kappa_{\cal T}$ can be smaller in magnitude.  We use the value:
\begin{equation}
\label{kappaTfit}
\kappa_{\cal T} = 1.30\,,
\end{equation}
which may be compared with $\kappa_{\cal T}=0.12$ in Ref.~\cite{Cloet:2008wg}.

In conducting this study we found that Diagram~4 plays an important role in describing the electromagnetic transition form factors in the $\gamma^{\ast}N\Delta$ reaction but contributes little to nucleon elastic form factors \cite{Cloet:2007pi,Alkofer:2004yf,Cloet:2008wg}.  This explains why we have revised the expression for the transition coupling.  Notably, the form factor, $F_{qq}(Q^2)$, is required in order to ensure that this diagram's contribution is concentrated in the infrared.

\subsection{Diagrams~5 \& 6}
\label{X56}
These two-loop diagrams are the so-called ``seagull'' terms, which appear as partners
to Diagram~3 and arise because binding in the baryons' Faddeev equations is effected
by the exchange of a dressed-quark between \textit{nonpointlike} diquark correlations
\cite{Oettel:1999gc}.  The explicit expression for their contribution to the
nucleons' form factors is
\begin{eqnarray}
\label{B5}
J_{\mu}^{sg} &=& \frac{1}{2} S(k_{q}) \Delta^{i}(k_{d})
\left( X_{\mu}^{i}(p_q,q',k_d) S^T(q')
\bar{\Gamma}^{jT}(p'_2,p_{d})
\right.
\nonumber\\
& & -
\left.
\Gamma^{i}(p_1,k_{d}) S^T(q)
\bar{X}_{\mu}^{j}(-k_q,-q,p_d)
\right) \Delta^{j}(p_{d}) S(p_{q})\,,
\end{eqnarray}
where, again, the superscripts are summed.

The new elements in these diagrams are the couplings of a photon to two
dressed-quarks as they either separate from (Diagram~5) or combine to form
(Diagram~6) a diquark correlation.  As such they are components of the five point
Schwinger function which describes the coupling of a photon to the quark-quark
scattering kernel.  This Schwinger function could be calculated, as is evident from
the computation of analogous Schwinger functions relevant to meson observables
\cite{Cotanch:2002vj}.  However, such a calculation provides relevant input only when a
uniform truncation of the DSEs has been employed to calculate each of the elements
described hitherto.  We must instead employ an algebraic parametrisation
\cite{Oettel:1999gc}, which for Diagram~5 reads
\begin{eqnarray}
\nonumber
X^{J^P}_\mu(k,Q) & =&  e_{\rm by}\,\frac{4 k_\mu- Q_\mu}{4 k\cdot Q -
Q^2}\,\left[\Gamma^{J^P}\!(k-Q/2)-\Gamma^{J^P}\!(k)\right]\\
& +& e_{\rm ex}\,\frac{4 k_\mu+ Q_\mu}{4 k\cdot Q +
Q^2}\,\left[\Gamma^{J^P}\!(k+Q/2)-\Gamma^{J^P}\!(k)\right], \label{X5}
\end{eqnarray}
with $k$ the relative momentum between the quarks in the initial diquark, $e_{\rm
by}$ the electric charge of the quark which becomes the bystander, and $e_{\rm ex}$
the charge of the quark that is reabsorbed into the final diquark.  Diagram~6 has
\begin{eqnarray}
\nonumber
\bar X^{J^P}_\mu(k,Q) & =&  e_{\rm by}\,\frac{4 k_\mu+ Q_\mu}{4 k\cdot Q +
Q^2}\,\left[\bar\Gamma^{J^P}\!(k+Q/2)-\bar\Gamma^{J^P}\!(k)\right]\\
& +& e_{\rm ex}\,\frac{4 k_\mu-Q_\mu}{4 k\cdot Q -
Q^2}\,\left[\bar\Gamma^{J^P}\!(k-Q/2)-\bar\Gamma^{J^P}\!(k)\right], \label{X6}
\end{eqnarray}
where $\bar\Gamma^{J^P}\!(\ell)$ is the charge-conjugated amplitude,
Eq.\,(\ref{chargec}).  Plainly, these terms vanish if the diquark correlation is
represented by a momentum-independent Bethe-Salpeter-like amplitude; i.e., the
diquark is pointlike.

It is naturally possible to use more complicated \textit{Ans\"atze} \cite{Eichmann:2008ef}.  However, like Eq.\,(\ref{Gamma0plus}), Eqs.\,(\ref{X5}) \&
(\ref{X6}) are simple forms, free of kinematic singularities and sufficient to ensure
the nucleon-photon vertex satisfies the WGT identity when the composite nucleon is obtained from the Faddeev equation.


\setcounter{equation}{0}
\section{Chebyshev Expansion}
\label{CHexpand}

In solving the Faddeev equation we employ a Chebyshev expansion of the scalar
functions appearing in the Faddeev amplitude and wave function in order to restrain
the use of computer memory.  (See, e.g., Ref.\,\cite{Maris:1997tm}.)  The results
herein were obtained with twelve terms in both.  The Chebyshev-expanded functions
then define the Faddeev amplitude that appears and is evaluated in the expressions
for the form factors.  Without due care, this can lead to a problem; namely, with
increasing $Q^2$ a function can be computed outside the expansion's domain of
convergence.

Consider a function $F(k^2,k\cdot P; P^2)$, which represents a term in the Faddeev
amplitude.  It is a function of only two variables: $k^2$ and $k\cdot P$, where $k$
is the relative quark-diquark momentum, because the total momentum always satisfies
$P^2=-M^2$, where $M$ is the bound-state's mass.  In the bound-state's rest frame one
can define an angle $\alpha$ through
\begin{equation}
\label{cosalpha}
i |k| M \cos \alpha := k\cdot P\,.
\end{equation}
Then, with $\{U_i(x),\,j=1\,\ldots\infty\}$ being Chebyshev polynomials of the second
kind,
\begin{equation}
\label{cheb}
F(k^2,k\cdot P; -M^2) = \lim_{N_m\to \infty} \sum_{j=0}^{N_m} \,^j\! F(|k|,i M;-M^2)
\,  U_j(\cos\alpha)\,.
\end{equation}
For any finite $N_m$ the expansion in Eq.\,(\ref{cheb}) is a true approximation to
the $k\cdot P$-dependence of the function $F$ in the sense that, with increasing
$N_m$, the right-hand-side (rhs) is uniformly pointwise an increasingly accurate
representation of the function.  The lhs of Eq.\,(\ref{cheb}) is Poincar\'e
invariant.  Hence, in the limit $N_m \to \infty$, so is the rhs.  These statements
are true so long as $\cos\alpha$ defined in Eq.\,(\ref{cosalpha}) satisfies $-1\leq
\cos\alpha \leq 1$.

In calculating a form factor one must compute the Faddeev amplitude of a bound-state
that is not at rest.  In the Breit frame, e.g., the total momentum can be written as
$P= (0,0,\pm Q/2,i E(Q/2))$, where $E^2(Q/2) = M^2 + Q^2/4$, the bound-state is
moving with three momentum $\pm Q/2$ and
\begin{equation}
\label{kcdotPo}
k\cdot P = \pm \frac{1}{2} |k| Q \cos\theta \sin\beta + i |k| E(Q) \cos\beta\,,
\end{equation}
with $k$ expressed using the standard definition of hyperspherical coordinates. In
principle, as demonstrated in Ref.\,\cite{Bhagwat:2006pu}, this is not a problem in a
Poincar\'e covariant framework.  However, it can consume large amounts of computer
memory and time.  We therefore proceed by writing
\begin{equation}
\label{kcdotP}
k\cdot P = i |k| E(Q) \left[\mp  \frac{iQ}{2E(Q)} \cos\theta \sin\beta +
\cos\beta\right] =: i |k| E(Q) z\,,
\end{equation}
in which case the real and imaginary parts of $z$ are bounded in magnitude by one,
and then define
\begin{equation}
F(k^2,k\cdot (P\pm Q/2); -M^2) =  \sum_{j=0}^{N_m} \,^j\! F(|k|,i E(Q);-M^2)\,
U_j(z)\,.
\end{equation}

\section{Interaction Probabilities}
\label{FFN}

In this appendix we define the probabilities expressed in Table~\ref{tab:probabilities}.
\begin{itemize}
\item $P_1^{p,q}$ -- \label{nota}{Sum} of all contributions to $G_E^p(Q^2=0)$ that can be represented by Diagram~1 in Fig.\,\ref{vertex}; i.e., in which the photon interacts with a bystander quark, either $u$ or $d$.  This quantity gauges the probability that the photon interacts with a bystander quark.

\item $P_1^{p,c}$ -- Sum of all contributions to $G_E^p(Q^2=0)$ that can be represented by either Diagram~2 or 4; i.e., in which the photon interacts with a diquark correlation, either scalar or axial-vector, or excites a transition between them.  This quantity  gauges the probability that the photon interacts with a diquark.

\item $P_1^{p,e}$ -- Sum of all contributions to $G_E^p(Q^2=0)$ that can be represented by one of Diagrams~3, 5 or 6; i.e., in which the photon interacts with a diquark in association with its breakup. $P_1^{p,e}=F_1^{p,e}(Q^2=0)$ gauges the probability that the photon acts in association with diquark breakup.

    N.B.\ $P_1^{p,q} + P_1^{p,c} + P_1^{p,e} = G_E^p(Q^2=0)=1$.

\item $G_E^{p,u}$ -- Sum of all contributions to $G_E^p(Q^2)$ in Fig.\,\ref{vertex} that are proportional to the charge of a $u$-quark, $e_u$; i.e., the total $u$-quark contribution $G_1^p$.

\item $G_E^{p,q,u}$ -- Sum of all contributions to $G_E^{p,u}(Q^2)$ that can be represented by Diagram~1 in Fig.\,\ref{vertex}; i.e., in which the photon interacts with a bystander $u$-quark.

\item $G_E^{p,c,u}$ -- Sum of all contributions to $G_E^p$ that can be represented by either Diagram~2 or 4 and are proportional to $e_u$; i.e., in which the photon resolves a $u$-quark within a diquark correlation.

\item $G_E^{p,e,u}$ -- Sum of all contributions to $G_E^{p,u}(Q^2)$ that can be represented by one of Diagrams~3, 5 or 6 and are proportional to $e_u$; i.e., in which the photon interacts with a $u$-quark in association with the breakup of a diquark.

    N.B.\ $G_E^{p,q,u} + G_E^{p,c,u} + G_E^{p,e,u} = G_E^{p,u}$; $G_E^{p,u}(0) = 2$; $2 P_1^{p,\alpha,u}:=G_E^{p,\alpha,u}(Q^2=0)$, $\alpha=q,d,e$.

\item $G_E^{p,d}$ and related functions are defined in direct analogy with those connected to $G_E^{p,u}$.

    N.B.\ $G_E^{p,q,d} + G_E^{p,c,d} + G_E^{p,e,d} = G_E^{p,d}$; $G_E^{p,d}(0) =  1$; $ P_1^{p,\alpha,d}:=G_E^{p,\alpha,d}(Q^2=0)$, $\alpha=q,d,e$.

\end{itemize}

\begin{table}[!t]
\begin{center}
\caption{\label{tab:probabilities} Probabilities derived from the proton's electric form factor at $Q^{2}=0$.}
\begin{tabular*}
{\hsize}
{
l@{\extracolsep{0ptplus1fil}}
l@{\extracolsep{0ptplus1fil}}
l@{\extracolsep{0ptplus1fil}}
c|@{\extracolsep{0ptplus1fil}}
l@{\extracolsep{0ptplus1fil}}
l@{\extracolsep{0ptplus1fil}}
l@{\extracolsep{0ptplus1fil}}}\hline
$P_{1}^{p,q}$ & $P_{1}^{p,d}$ & $P_{1}^{p,e}$ && $P_{1}^{p,s}$ & $P_{1}^{p,a}$ &
$P_{1}^{p,m}$ \\
$0.490$ & $0.358$ & $0.152$ && $0.621$ & $0.287$ & $0.092$ \\ 
\hline
$P_{1}^{p,q,u}$ & $P_{1}^{p,c,u}$ & $P_{1}^{p,e,u}$ && $P_{1}^{p,s,u}$ &
$P_{1}^{p,a,u}$ & $P_{1}^{p,m,u}$ \\
$0.457$ & $0.383$ & $0.160$ && $0.575$ & $0.338$ & $0.087$ \\ 
\hline
$P_{1}^{p,q,d}$ & $P_{1}^{p,c,d}$ & $P_{1}^{p,e,d}$ && $P_{1}^{p,s,d}$ &
$P_{1}^{p,a,e}$ & $P_{1}^{p,m,d}$ \\
$0.356$ & $0.459$ & $0.185$ && $0.439$ & $0.493$ & $0.068$ \\ 
\hline
\end{tabular*}
\end{center}
\end{table}

\vspace*{-1ex}

\begin{itemize}
\item $G_E^{p,s}$ -- Sum of all contributions to $G_E^p$ in Fig.\,\ref{vertex} that involve a scalar diquark component in both $\Psi_i$ and $\Psi_f$.  $P_1^{p,s}=G_E^{p,s}(Q^2=0)$ gauges the probability that the photon interacts with a scalar diquark component of the nucleon.

\item $G_E^{p,a}$ -- Sum of all contributions to $G_E^p$ that involve an axial-vector diquark component in both $\Psi_i$ and $\Psi_f$.  $P_1^{p,a}=G_E^{p,a}(Q^2=0)$ gauges the probability that the photon interacts with an axial-vector diquark component of the nucleon.

\item $G_E^{p,m}$ -- Sum of all contributions to $G_E^p$ in which the diquark component of $\Psi_i$ is different to that in $\Psi_f$. $P_1^{p,m}=G_E^{p,m}(Q^2=0)$ gauges the probability that the photon induces a transition between diquark components of the incoming and outgoing nucleon.

    N.B.\ $G_E^{p,s} + G_E^{p,a} + G_E^{p,m} = G_E^p$.

\item $G_E^{p,s,u}$ -- Sum of all contributions to $G_E^p$ in Fig.\,\ref{vertex} that involve a scalar diquark component in both $\Psi_i$ and $\Psi_f$, and are proportional to $e_u$; i.e., in which a $u$-quark is resolved in the presence of a scalar diquark.

\item $G_E^{p,a,u}$ -- Sum of all contributions to $G_E^p$ that involve an axial-vector diquark component in both $\Psi_i$ and $\Psi_f$, and are proportional to $e_u$.

\item $G_E^{p,m,u}$ -- Sum of all contributions to $G_E^p$ that are proportional to $e_u$ and in which the diquark component of $\Psi_i$ is different to that in $\Psi_f$.

    N.B.\ $G_E^{p,s,u} + G_E^{p,a,u} + G_E^{p,m,u} = G_E^{p,u}$;
    $2 P_1^{p,\alpha,u}:=G_E^{p,\alpha,u}(Q^2=0)$, $\alpha=s,a,m$.

\item $G_E^{p,s,d}$ \label{notb}{and} similar functions are defined in direct analogy with those connected to $G_E^{p,s,u}$.

    N.B.\ $G_E^{p,s,d} + G_E^{p,a,d} + G_E^{p,m,d} = G_E^{p,d}$;
    $P_1^{p,\alpha,d}:=G_E^{p,\alpha,d}(Q^2=0)$, $\alpha=s,a,m$.

\end{itemize}
With these definitions, $G_E^p = e_u G_E^{p,u} + e_d G_E^{p,d}$, which is the first entry in Eq.\,\eqref{eqFlavourSep}.



\end{document}